\documentclass[showpacs,twocolumn,superscriptaddress]{revtex4-1}
\usepackage{doi}%<----------
\usepackage{hyperref}
\hypersetup{
  colorlinks=true,        % false: boxed links; true: colored links
  linkcolor=blue,         % color of internal links
  citecolor=cyan,         % color of links to bibliography
}

\def\cblue#1{\textcolor{black}{#1}}
\usepackage{graphicx}% Include figure files
\usepackage{dcolumn}% Align table columns on decimal point
\usepackage{bm}% bold math
\usepackage{color}
\usepackage{enumitem}
\usepackage{amsmath}
\usepackage{amssymb}
\usepackage{orcidlink}
\usepackage{subcaption}
\usepackage{graphicx} % Required for inserting images
\begin{document}
\title{Gravitational waveforms from periodic orbits around a Schwarzschild black hole embedded in a Dehnen-type dark matter halo}

\author{Mirzabek Alloqulov}
\email{malloqulov@gmail.com}
\affiliation{University of Tashkent for Applied Sciences, Str. Gavhar 1, Tashkent 100149, Uzbekistan}
\affiliation{Shahrisabz State Pedagogical Institute, Shahrisabz Str. 10, Shahrisabz 181301, Uzbekistan}

\author{Tursunali Xamidov}
\email{xamidovtursunali@gmail.com}
\affiliation{Institute of Fundamental and Applied Research, National Research University TIIAME, Kori Niyoziy 39, Tashkent 100000, Uzbekistan}

\author{Sanjar Shaymatov}
\email{sanjar@astrin.uz}
\affiliation{Institute of Fundamental and Applied Research, National Research University TIIAME, Kori Niyoziy 39, Tashkent 100000, Uzbekistan}
\affiliation{Institute for Theoretical Physics and Cosmology,
Zhejiang University of Technology, Hangzhou 310023, China}
\affiliation{University of Tashkent for Applied Sciences, Str. Gavhar 1, Tashkent 100149, Uzbekistan}

\author{Bobomurat Ahmedov}
\email{ahmedov@astrin.uz}
\affiliation{Institute for Advanced Studies, New Uzbekistan University, Movarounnahr str. 1, Tashkent 100000, Uzbekistan}
\affiliation{Institute of Theoretical Physics, National University of Uzbekistan, Tashkent 100174, Uzbekistan}

\date{\today}

\begin{abstract}

In this paper, we study the periodic orbits, characterized by zoom-whirl behavior, around a Schwarzschild-like black hole (BH) embedded within a Dehnen-type dark matter (DM) halo. We demonstrate how the DM halo modifies the gravitational dynamics of the black hole, influencing the energy and angular momentum of timelike particle geodesics and enhancing their interaction with the BH. We determine the radii of the marginally bound orbits (MBOs) and innermost stable circular orbits (ISCOs), showing that the DM halo increases both. This provides a deeper understanding of how the DM alters the behavior, energy, and angular momentum of timelike particle geodesics. Furthermore, we explore the gravitational waveforms emitted by a timelike particle in periodic orbits around a supermassive black hole (SMBH) within this BH-DM system. Using a semi-analytical approach, we calculate particle trajectories and derive the corresponding waveforms, demonstrating that the DM halo modifies the zoom-whirl orbital behavior, leading to distinct changes in the waveform structure. Our findings suggest that future gravitational wave (GW) observations could constrain the properties of DM halos surrounding BHs, providing new insights into the gravitational wave signatures arising from the interaction between BH gravity and DM.

\end{abstract}

\maketitle

\section{Introduction}

Within the framework of general relativity (GR), black holes (BHs) emerge naturally as exact solutions to Einstein's field equations. The Schwarzschild~\cite{1916SPAW.......189S,2015arXiv151202061B} and Kerr BHs~\cite{1963PhRvL..11..237K} are the vacuum solutions of the field equation. However, astrophysical BHs are considered surrounded by matter distributions like dark matter halos in many scenarios. Because only around $5 \%$ of the Universe is known to us. According to the cosmic microwave background observation, the Universe contains $27\%$ dark matter (DM) and $68\%$ dark energy. Inspired by this, researchers are trying to understand the interaction between the DM halo and BHs. Investigating the influence of dark matter (DM) halos on supermassive black holes (SMBHs) is crucial to understanding the interaction between DM and BHs. DM halo structures significantly affect galactic rotation curves \cite{Rubin70ApJ,Bertone18Nature,Corbelli00MNRAS} and matter motion observed in events like the Bullet Cluster collision \cite{Clowe06ApJL}. While DM is currently understood to interact only gravitationally, compelling evidence supports its existence (see, e.g., \cite{Bertone05,deSwart17Nat,Wechsler18}). 

Observations confirm that star formation concentrates near galactic centers, leading to the formation of DM halos in the surrounding regions. Consequently, host galaxies (such as spiral or giant elliptical galaxies) develop within these DM halos (e.g., \cite{Valluri04ApJ,Akiyama19L1,Akiyama19L6,Akiyama22L12}). Although understanding the fundamental nature of DM remains a challenge within general relativity, analyzing DM models can provide valuable insights. Observational data suggest that the DM halo primarily explains the rotational velocities of stars orbiting their host galaxies \cite{Persic96}. Consequently, several analytical models have been developed that describe BH solutions within DM halos. These include the Einasto \cite{Merritt_2006,Dutton_2014}, Navarro-Frenk-White \cite{Navarro_1996}, and Burkert \cite{Burkert_1995} models, as well as the Dehnen model \cite{Dehnen93,refId0,Gohain_2024,Pantig_2022}. Further models incorporate a DM profile associated with a phantom scalar field \cite{Li-Yang12,Shaymatov21d,Shaymatov21pdu,Shaymatov22a}. 
Additionally, analytical models depicting a supermassive BH immersed in a DM halo have also been explored in Refs. \cite{Cardoso22DM,Hou18-dm,Shen24PLB}.

BHs, while subject to intense theoretical and observational scrutiny, remain among the most fascinating and intriguing astrophysical objects. GR, by describing gravity as the curvature of spacetime \cite{1916SPAW.......189S}, revolutionized our understanding of the universe. This framework has successfully explained phenomena such as gravitational waves (GWs) \cite{Abbott16a,Abbott16b}. The direct detection of GWs illuminates astrophysical phenomena in strong gravitational field regimes. GW detection and analysis are crucial to understanding the dynamics of compact objects such as BHs. Current ground-based detectors, including LISA \cite{Amaro-Seoane2017LISA} and Taiji \cite{10.1093/nsr/nwx116}, primarily focus on stellar-mass binary systems with comparable masses. However, extreme mass-ratio inspirals (EMRIs) \cite{Hughes_2001,Amaro-Seoane18LRR,Babak17PRD} represent another important source of GW. These systems, comprising a stellar-mass object orbiting a SMBH, emit low-frequency GWs. The resulting waveforms offer precise insights into the dynamics of test particles around black holes. Furthermore, gravitational-wave radiation from EMRIs offers a wealth of information about the phenomena near BH event horizons and also help elucidate the nature of astrophysical dark matter surrounding SMBHs \cite{Yue19ApJ,Duque24PRL,Dai24PRD}, making them a powerful tool for probing the nature of these objects.

It is important to note that the gravitational waveforms from the EMRI system are influenced by the motion of the smaller object. According to \cite{Levin_2008,Grossman_2009,Misra_2010}, both Schwarzschild and Kerr BH backgrounds exhibit periodic geodesic orbits. A key characteristic of EMRI dynamics is the presence of these periodic orbits, which are especially important as they capture the fundamental characteristics of orbital dynamics around BHs, together with typical bound trajectories. For these periodic orbits, a test particle must return to its starting point, and it should have zoom-whirl behavior, completing an integer number of radial and angular oscillations. Therefore, periodic orbits are defined by three integers: $z$, the zoom number; $w$, the whirl number; and $v$, the vertex number \cite{Levin_2008,Levin_2009}. The shape of these orbits is determined by these integers, with the ratio of angular to radial frequencies forming a rational number for each periodic orbit that can play an important role in explaining GW radiation \cite{Glampedakis02PRD}. Additionally, the properties of these periodic orbits are related to the black hole background geometry \cite{Levin2010,Babar17PRD,Liu_2019,Tu23,Mustapha2020,wei2019,Deng20,yang2024}. The Schwarzschild black hole surrounded by a Dehnen-type dark matter halo may serve as a prominent candidate for studying the interaction between the BH and the DM halo. Consequently, the DM surrounding the SMBHs can affect the motion of particles through dynamical orbits, resulting in a change in the EMRI waveform and background spacetime \cite{Sadeghian13PRD,Barausse14PRD,Cardoso22PRD,Haroon2025}. Hence, it is also important to analyze the EMRI signals to explore the DM halo distribution surrounding the SMBHs \cite{Navarro96ApJ,Gondolo99PRL}.

This paper extends the study of periodic orbits to BH spacetime surrounded by a Dehnen-type dark matter (DM) halo (see details in Ref.~\cite{Al-Badawi_2025}), examining its impact on both periodic orbits and gravitational waveforms (i.e. potential GW signatures of supermassive black holes). We begin by exploring time-like periodic geodesic orbits around the BH. Using the Lagrangian formalism, we derive the equations of motion for test particles. We then analyze how the DM halo affects the effective potential of these test particles. Furthermore, we analyze marginally bound orbits (MBOs) and innermost stable circular orbits (ISCOs) using the corresponding required conditions \cite{Dadhich22a}. We investigate the dependence of the rational number $q$ on the energy and angular momentum of the test particles for various spacetime parameters. Next, we plot the periodic orbits around the Schwarzschild BH embedded in the Dehnen-type DM halo, characterized by integers $(z,w,v)$. Finally, we employ the numerical kludge method to investigate the gravitational waveforms generated by these periodic orbits, considering a stellar-mass object orbiting the SMBH \cite{Babak07PRD}.

The paper is organized as follows. In Section~\ref{sec2}, we briefly describe a Schwarzschild BH surrounded by a Dehnen-type DM halo, followed by the study of the test particle dynamics, including the ISCOs and MBOs. In Section~\ref{sec3}, we investigate the rational number $q$ of periodic orbits and explore the effect of the DM halo on the energy and orbital angular momentum of periodic orbits. Furthermore, we examine the gravitational waveforms of the EMRI system from the periodic orbits around the Schwarzschild BH surrounded by a Dehnen-type DM halo in Section~\ref{sec4}. Section~\ref{summary} contains the conclusions and discussion of our findings. 

\section{Spacetime metric and timelike geodesics}\label{sec2}

\begin{figure*}[!htb]
    \centering
    \includegraphics[scale=0.5]{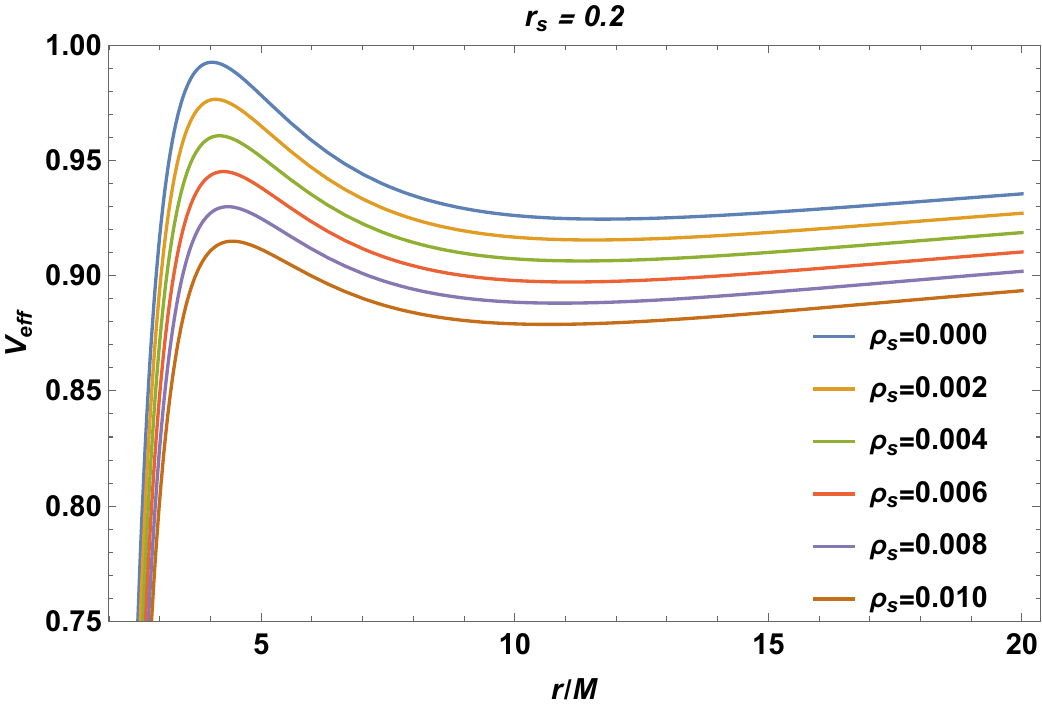}
    \includegraphics[scale=0.5]{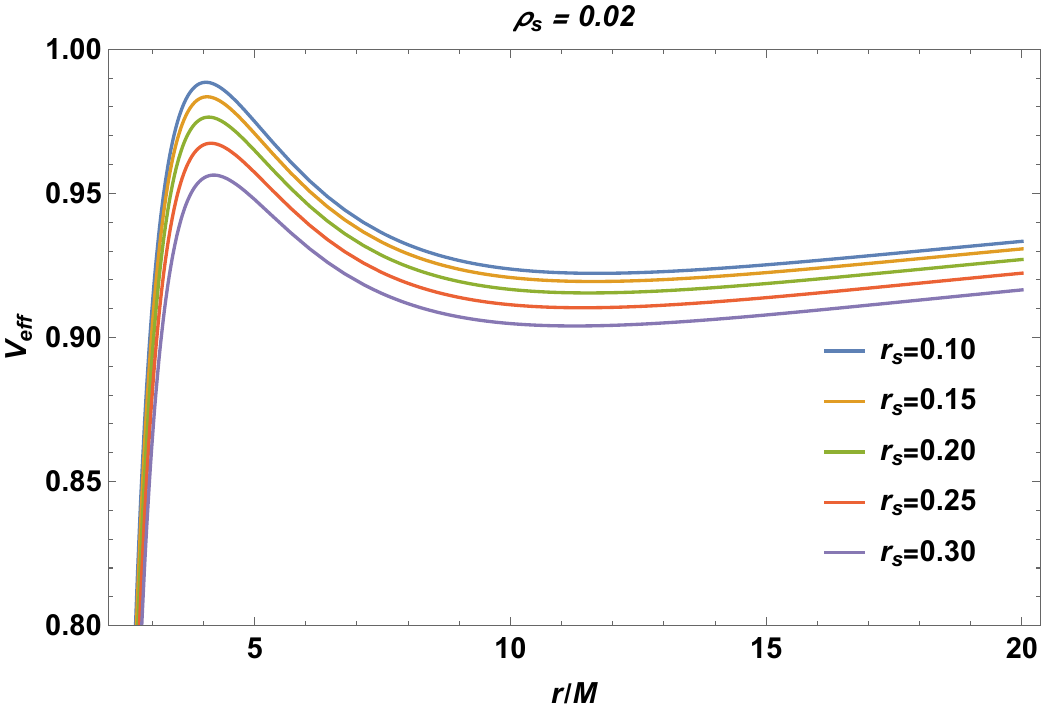}
    \includegraphics[scale=0.5]{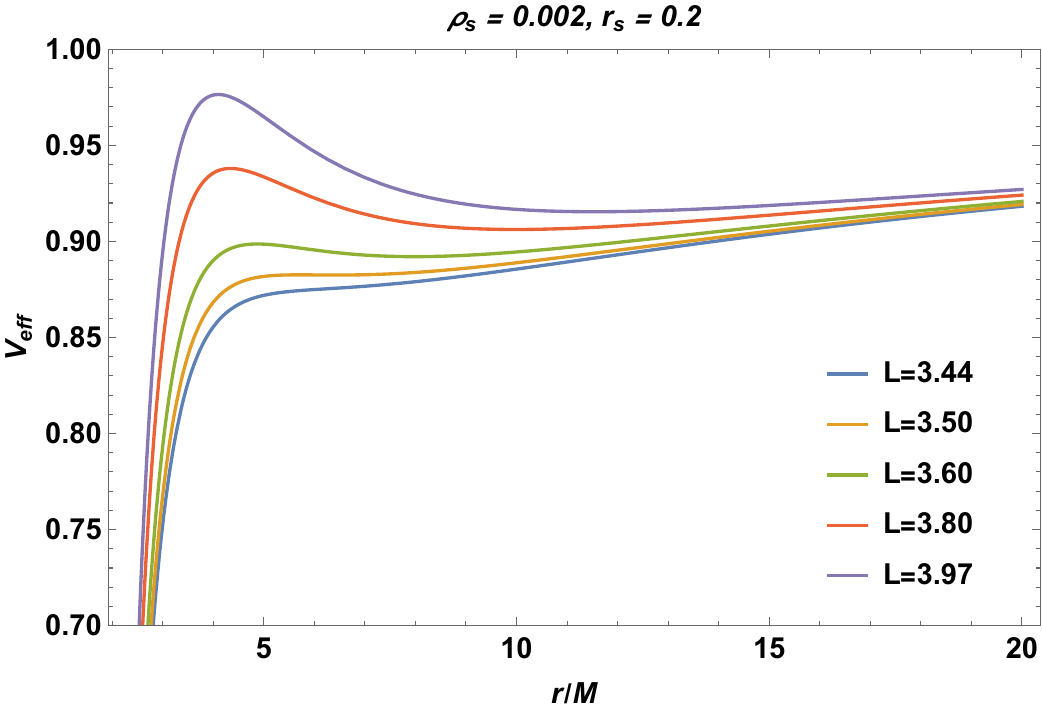}
    \caption{The radial dependence of the effective potential for different values of the spacetime parameters $\rho_s$, $r_s$ and the orbital angular momentum. }
    \label{fig:effpotential}
\end{figure*}
\begin{figure*}[!htb]
    \centering
    \includegraphics[scale=0.5]{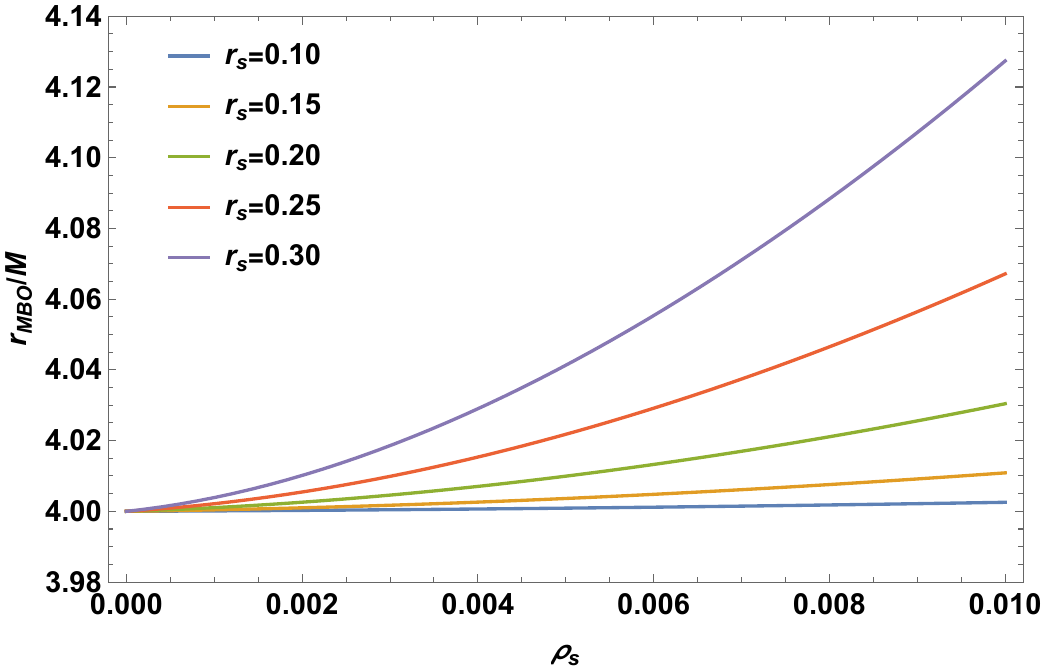}
    \includegraphics[scale=0.5]{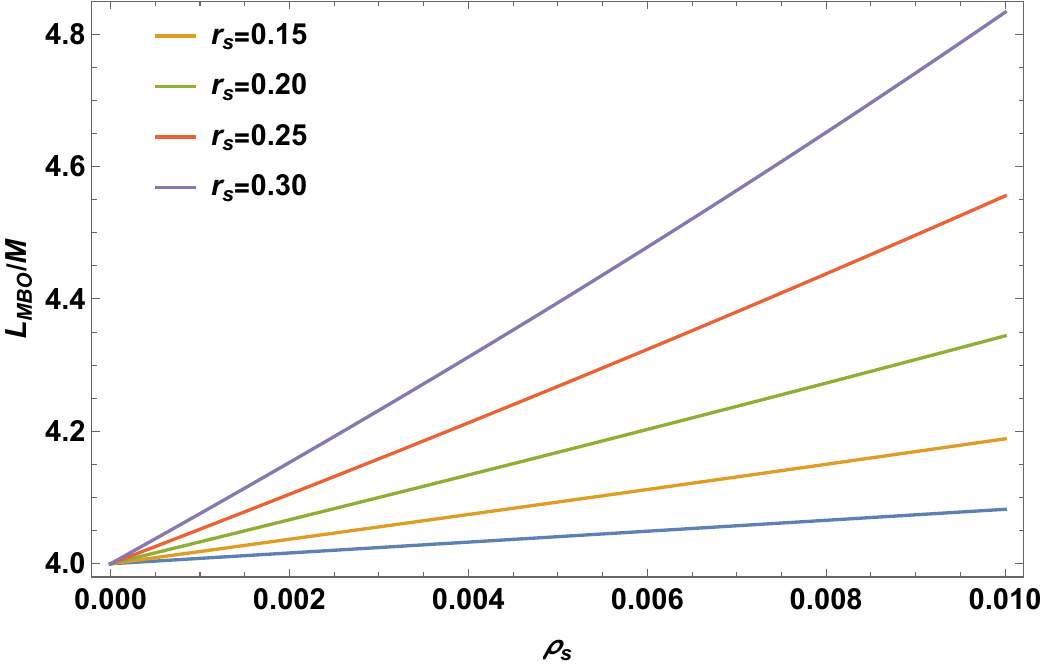}
    \caption{\cblue{Left panel: The dependence of the MBO radius on the DM halo's density $\rho_s$ for different values of the core radius $r_s$. Right panel: The angular momentum of the MBO as a function of the DM halo's density $\rho_s$ for different values of the core radius $r_s$. }}
    \label{fig:mbo}
\end{figure*}
\begin{figure*}[htbp]
    \centering
    \includegraphics[scale=0.5]{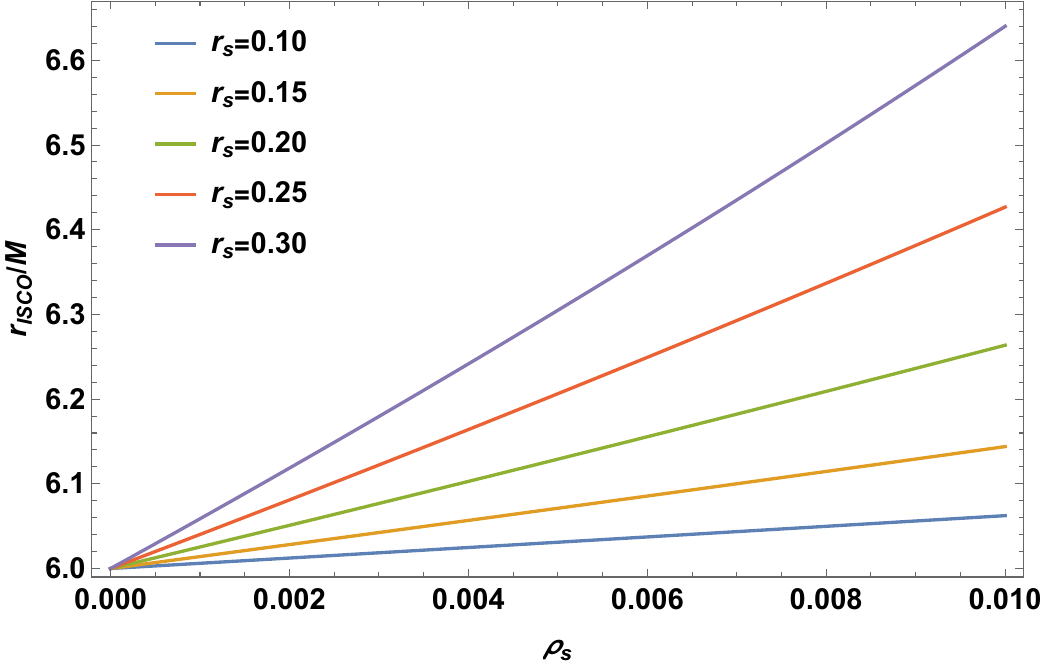}
    \includegraphics[scale=0.5]{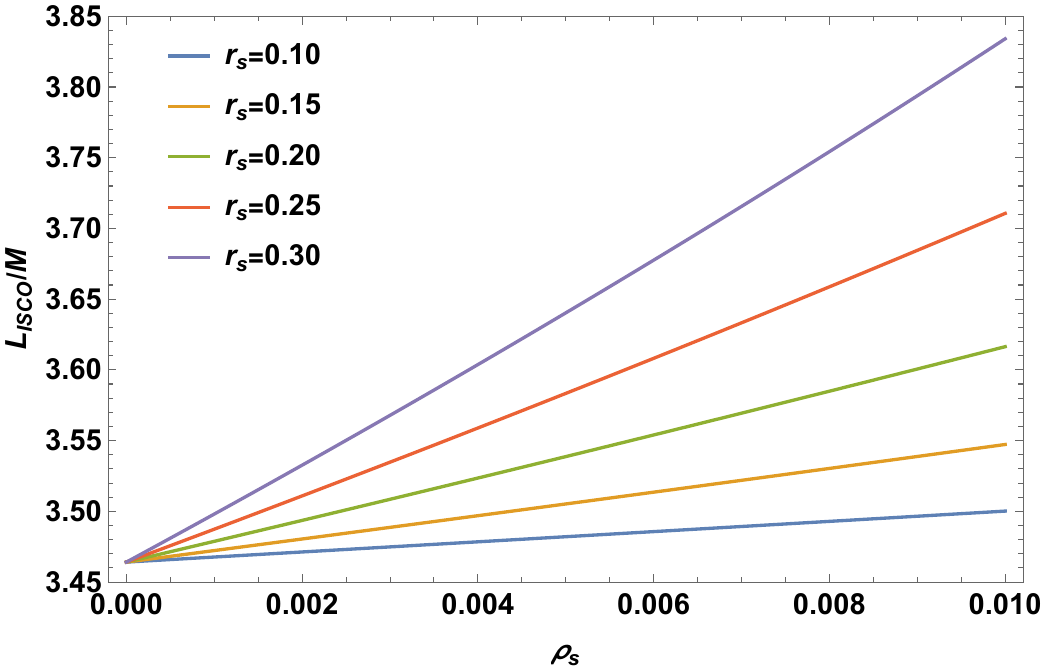}
    \includegraphics[scale=0.5]{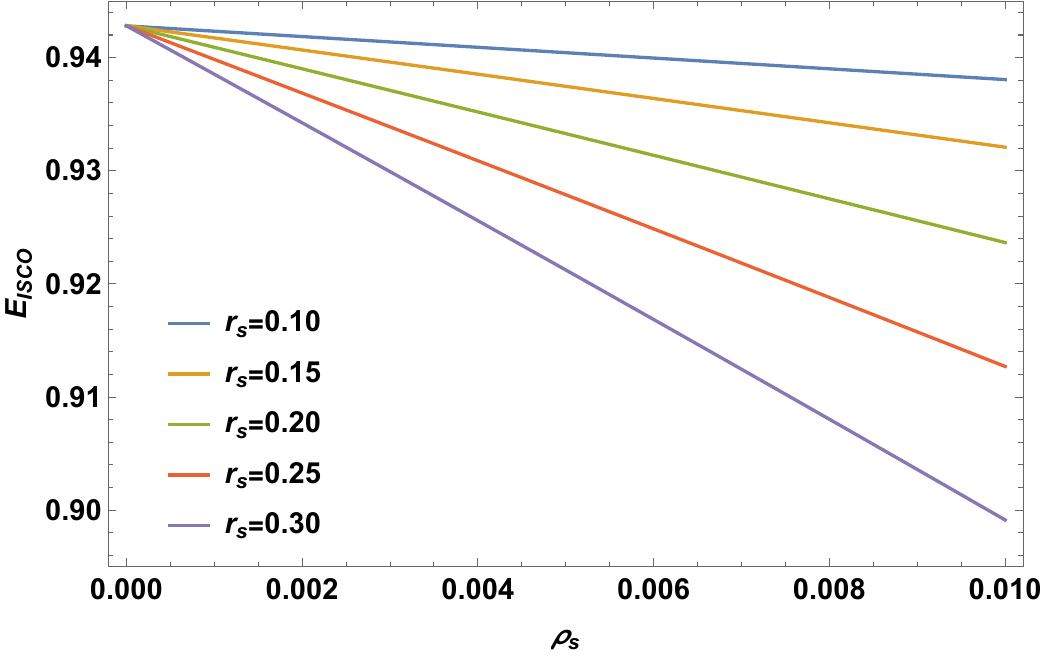}
    \caption{\cblue{Top left panel: The ISCO radius as a function of the DM halo's density $\rho_s$ for different values of the core radius $r_s$. Top right panel: The dependence of the orbital angular momentum of the ISCO on the DM halo's density $\rho_s$ for different values of the core radius $r_s$. Bottom panel illustrates the energy of the ISCO as a function of the DM halo's density for different values of the core radius $r_s$.}}
    \label{fig:isco}
\end{figure*}
\begin{figure*}[htbp]
\centering
 \includegraphics[scale=0.5]{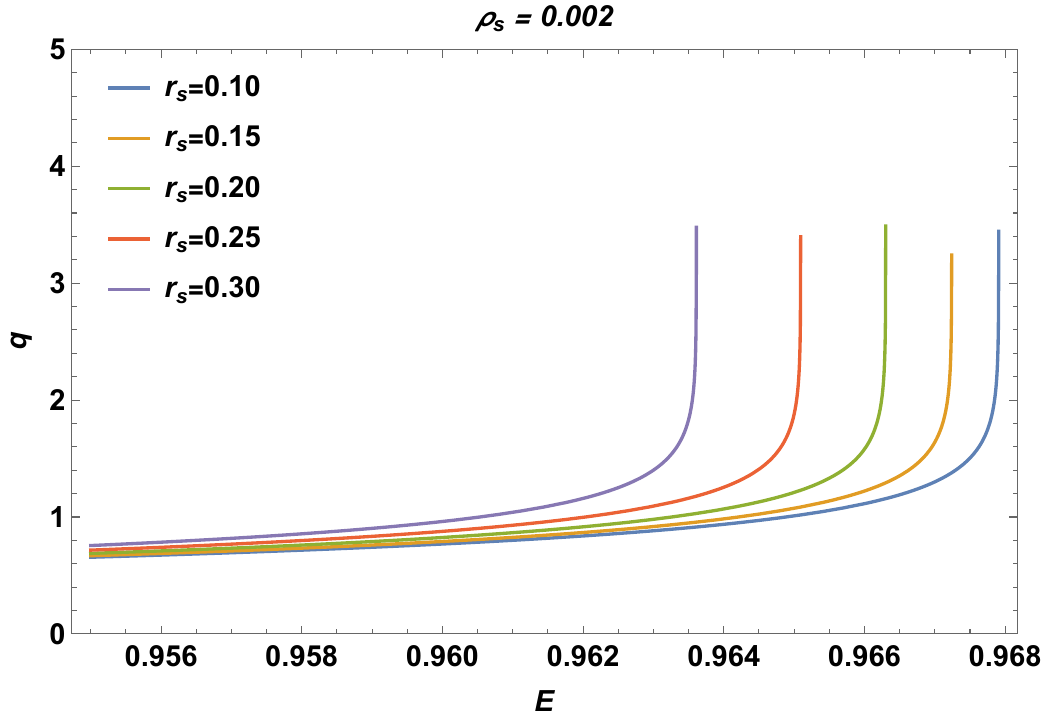}
 \includegraphics[scale=0.49]{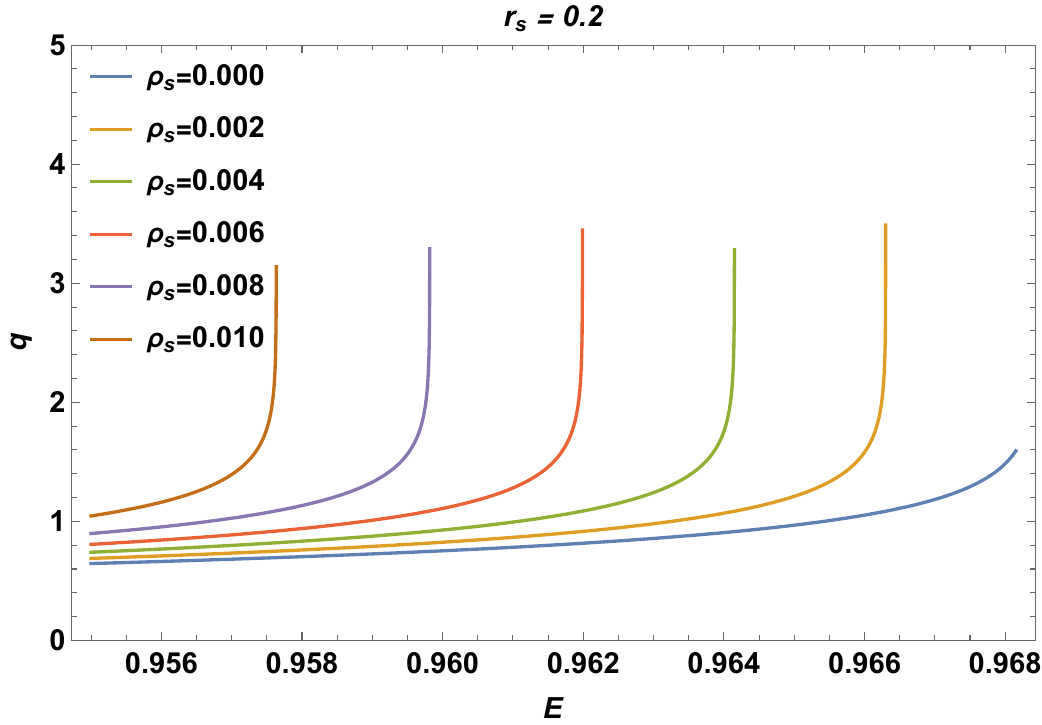}
  \includegraphics[scale=0.5]{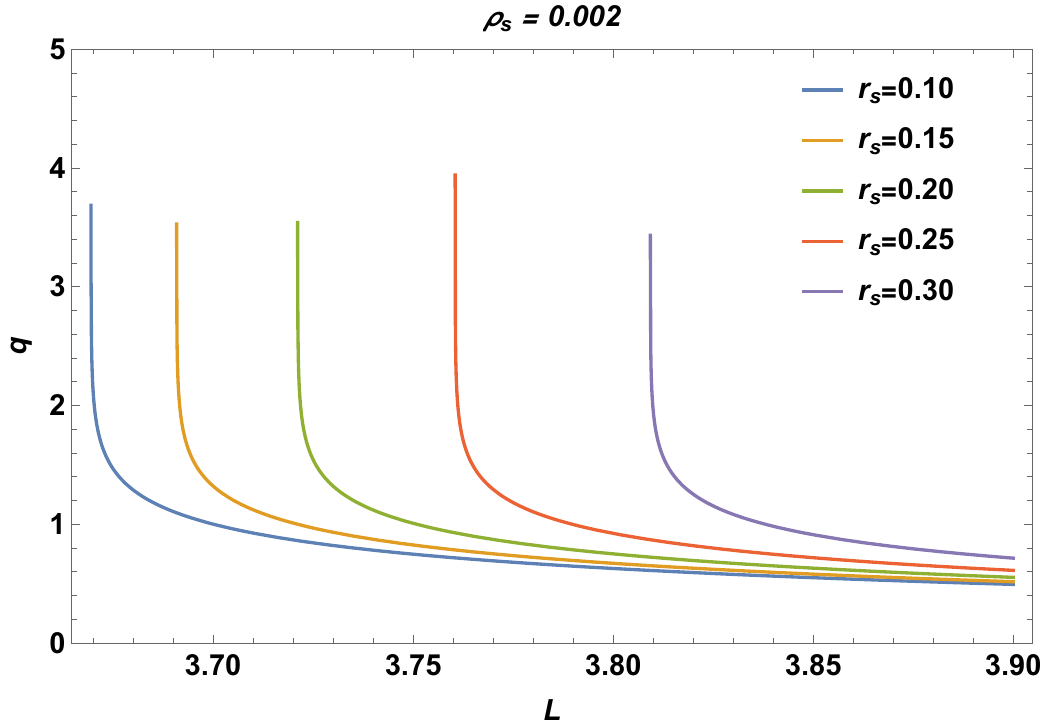}
 \includegraphics[scale=0.49]{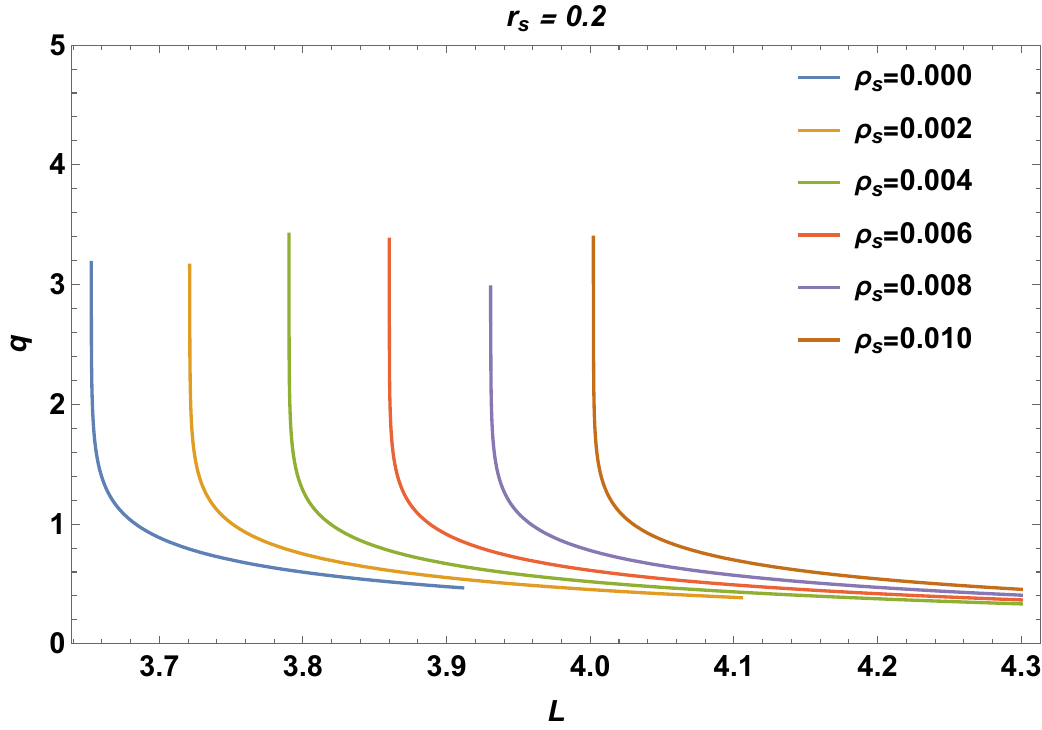}
 \caption{\cblue{The dependence of the rational number $q$ on the energy of periodic orbits around the Schwarzschild BH surrounded by Dehnen-type DM halo for different values of the parameters $\rho_s$ (top-left panel) and $r_s$ (top-right panel). Here, the orbital angular momentum is $L=\frac{1}{2}(L_{MBO}+L_{ISCO})$. The bottom panel depicts the dependence of the rational number $q$ on the orbital angular momentum of periodic orbits around the Schwarzschild BH surrounded by a Dehnen-type DM halo for different values of the parameters $\rho_s$ and $r_s$. Here, we set the energy as $E=0.96$.}}
 \label{fig:q}
\end{figure*}

In this section, we review the geodesic orbits around the Schwarzschild BH surrounded by a Dehnen-type DM halo~\cite{Al-Badawi_2025}. One can write the metric as follows
\begin{equation}
ds^{2}=-f(r)dt^{2}+f(r)^{-1}dr^{2}+r^{2}d\theta^{2}+r^{2}\sin^{2}\theta d\phi^{2},
\end{equation}
with
\begin{equation}
    f(r)=1-\frac{2M}{r}-32\pi \rho_s r^3_s\sqrt{\frac{r+r_s}{r^2_s r}}\, ,
\end{equation}
where $r_s$ and $\rho_s$ refer to the central halo density and radius, respectively. It is worth noting that we consider the test particle motion around the Schwarzschild BH surrounded by a Dehnen-type DM halo. One can write the Lagrangian of the particle as~\cite{1983mtbh.book.....C}
\begin{equation}
\mathcal{L}=\frac{1}{2}m\,g_{\mu\nu}\,\frac{dx^{\mu}}{d\tau}\,\frac{dx^{\nu}}{d\tau},
\end{equation}
where $\tau$ and $m$ are the proper time and the mass of the test particle, respectively. One can write the generalized momentum of the particle by setting $m=1$ for simplicity, as follows
\begin{equation}
    p_{\mu}=\frac{\partial {\cal L}}{\partial \dot{x}^{\mu}}=g_{\mu \nu}\dot{x}^{\nu}\, .
\end{equation}
After that, we can obtain the equations of motion for the test particles in the following form
\begin{eqnarray}\label{eq:eqmotion}
p_{t} &=& -\left(1-\frac{2M}{r}-32\pi \rho_s r^3_s\sqrt{\frac{r+r_s}{r^2_s r}}\right)\dot{t}=-E \, , \nonumber\\
p_{\phi} &=& r^{2}\sin^{2}\theta\dot{\phi}=L \, , \nonumber \\
p_{r} &=& \left(1-\frac{2M}{r}-32\pi \rho_s r^3_s\sqrt{\frac{r+r_s}{r^2_s r}}\right)^{-1}\dot{r} \, , \nonumber\\
p_{\theta} &=& r^{2}\dot{\theta} \, ,
\end{eqnarray}
where $E$ and $L$ refer to the energy and orbital angular momentum of the particle, respectively.
Note that we consider the motion in the equatorial plane ($\theta=\pi/2$). With the usage of the normalization condition we can write the following expression.
\begin{widetext}
\begin{equation}
\frac{\dot{r}^{2}}{1-\frac{2M}{r}-32\pi \rho_s r^3_s\sqrt{\frac{r+r_s}{r^2_s r}}}+\frac{L^{2}}{r^{2}}-\frac{E^{2}}{1-\frac{2M}{r}-32\pi \rho_s r^3_s\sqrt{\frac{r+r_s}{r^2_s r}}}=-1.
\end{equation}
\end{widetext}
We can rewrite the above equation as
\begin{equation}
\dot{r}^{2}+V_{\rm eff}=E^{2},
\end{equation}
with
\begin{equation}
V_{\rm eff}=\left(1-\frac{2M}{r}-32\pi \rho_s r^3_s\sqrt{\frac{r+r_s}{r^2_s r}}\right)\left(1+\frac{L^{2}}{r^{2}}\right)\, ,
\end{equation}
where $V_{eff}$ is the effective potential for the motion of the test particle. To be more informative, we demonstrate the radial dependence of the effective potential for different values of the central halo density and radius in Fig.~\ref{fig:effpotential}. We can see from this figure that the values of the effective potential decrease with the increase of halo density and radius. Moreover, it can be seen from the bottom panel of Fig.~\ref{fig:effpotential} that there is an increase under the influence of the particle's orbital angular momentum. 
\renewcommand{\arraystretch}{1.2}
\begin{table*}[]
\centering
\resizebox{1.0\textwidth}{!}{
\begin{tabular}{|c|c|c|c|c|c|c|c|c|c|}
\hline
$\rho_s$ & $L$ & $E_{(1,1,0)}$ & $E_{(1,2,0)}$ & $E_{(2,1,1)}$ & $E_{(2,2,1)}$ & $E_{(3,1,2)}$ & $E_{(3,2,2)}$ & $E_{(4,1,3)}$ & $E_{(4,2,3)}$ \\ \hline
0.000    & 3.73205 & 0.965425   & 0.968383   & 0.968026   & 0.968435   & 0.968225   & 0.968438   & 0.968285   & 0.968440   \\ \hline
0.002    & 3.78003 & 0.963261   & 0.966243   & 0.965888   & 0.966294   & 0.966087   & 0.966298   & 0.966146   & 0.966299   \\ \hline
0.004    & 3.82881 & 0.961081   & 0.964093   & 0.963739   & 0.964142   & 0.963937   & 0.964146   & 0.963996   & 0.964148   \\ \hline
0.006    & 3.87844 & 0.958886   & 0.961932   & 0.961578   & 0.961981   & 0.961777   & 0.961985   & 0.961836   & 0.961986   \\ \hline
0.008    & 3.92894 & 0.956677   & 0.959761   & 0.959405   & 0.959809   & 0.959606   & 0.959813   & 0.959665   & 0.959814   \\ \hline
0.010    & 3.98034 & 0.954453   & 0.957579   & 0.957222   & 0.957627   & 0.957424   & 0.957631   & 0.957484   & 0.957632   \\ \hline
\end{tabular}
}
\caption{The values of the energy $E$ are tabulated for different periodic orbits characterized by $(z,w,v)$. The density of the DM halo varies from $\rho_s=0$ to $\rho_s=0.01$ with an interval of $0.002$. Here, we set $L=\frac{1}{2}(L_{MBO}+L_{ISCO})$ and $r_s=0.2$.}
\label{table1}
\end{table*}
\begin{table*}[]
\resizebox{1.0\textwidth}{!}{
\begin{tabular}{|c|c|c|c|c|c|c|c|c|}
\hline
$\rho_s$ & $L_{(1,1,0)}$ & $L_{(1,2,0)}$ & $L_{(2,1,1)}$ & $L_{(2,2,1)}$ & $L_{(3,1,2)}$ & $L_{(3,2,2)}$ & $L_{(4,1,3)}$ & $L_{(4,2,3)}$ \\ \hline
0.000    & 3.683588    & 3.653440    & 3.657596    & 3.652701    & 3.655335    & 3.652636    & 3.654621    & 3.652616    \\ \hline
0.002    & 3.750959    & 3.721847    & 3.725695    & 3.721240    & 3.723592    & 3.721187    & 3.722939    & 3.721171    \\ \hline
0.004    & 3.819185    & 3.790859    & 3.794447    & 3.790322    & 3.792465    & 3.790277    & 3.791858    & 3.790264    \\ \hline
0.006    & 3.888351    & 3.860590    & 3.863974    & 3.860099    & 3.862085    & 3.860060    & 3.861513    & 3.860048    \\ \hline
0.008    & 3.958534    & 3.931141    & 3.934379    & 3.930696    & 3.932560    & 3.930661    & 3.932015    & 3.930651    \\ \hline
0.010    & 4.029808    & 4.002633    & 4.005754    & 4.002218    & 4.003990    & 4.002187    & 4.003466    & 4.002177    \\ \hline
\end{tabular}
}
\caption{The values of the orbital angular momentum $L$ are tabulated for different periodic orbits characterized by $(z,w,v)$. The density of the DM halo varies from $\rho_s=0$ to $\rho_s=0.01$ with an interval of $0.002$. Here, we set $E = 0.96$ and $r_s = 0.2$.
}
\label{table2}
\end{table*}
The main focus of our work is the periodic orbits around the Schwarzschild BH surrounded by a Dehnen-type DM halo. Periodic orbits are a type of bound orbits. In the case of the particle moving on a general bound orbit, its energy and orbital angular momentum must meet the following conditions:
\begin{equation}
    L_{ISCO}\leq L \quad\mbox{and}\quad E_{ISCO}\leq E \leq E_{MBO}=1\, ,
\end{equation}
where $E_{ISCO}$ and $ L_{ISCO}$ refer to the energy and the orbital angular momentum of the particle that is moving on ISCO, respectively, and $E_{MBO}$ is the energy of the particle along MBO. We can explore the effect of the DM halo on the properties of the periodic orbits by investigating the MBO and ISCO. One can express the following condition for determining the MBO
\begin{equation}
    V_{eff}=1\quad\mbox{and}\quad \frac{d V_{eff}}{dr}=0.
\end{equation}
Using the above conditions, we numerically investigate the radius and orbital angular momentum of the MBO. We demonstrate the radius and orbital angular momentum as a function of the DM halo's density for the different values of the DM halo's radius in Fig.~\ref{fig:mbo}. It can be seen from this figure that there is an increase with the increase of the DM halo's radius. We then turn to another important kind of bound orbit, which is ISCO. One can determine the ISCO by the following conditions
\begin{equation}
    \dot{r}=0, \quad \frac{d V_{eff}}{dr}=0 \quad\mbox{and}\quad \frac{d^2V_{eff}}{dr^2}=0.
\end{equation}
We also numerically explore the ISCO parameters, which are the radius, orbital angular momentum, and energy of the ISCO as a function of the DM halo's density for different values of the DM halo's density in Fig.~\ref{fig:isco}. We can see from this figure that the values of the radius and orbital angular momentum increase with the increase of the DM halo's density. In contrast, there is a decrease with the increasing density of the DM halo as demonstrated in Fig.~\ref{fig:isco}.

\begin{figure*}[htbp]
    \centering
    \includegraphics[width=0.32\textwidth]{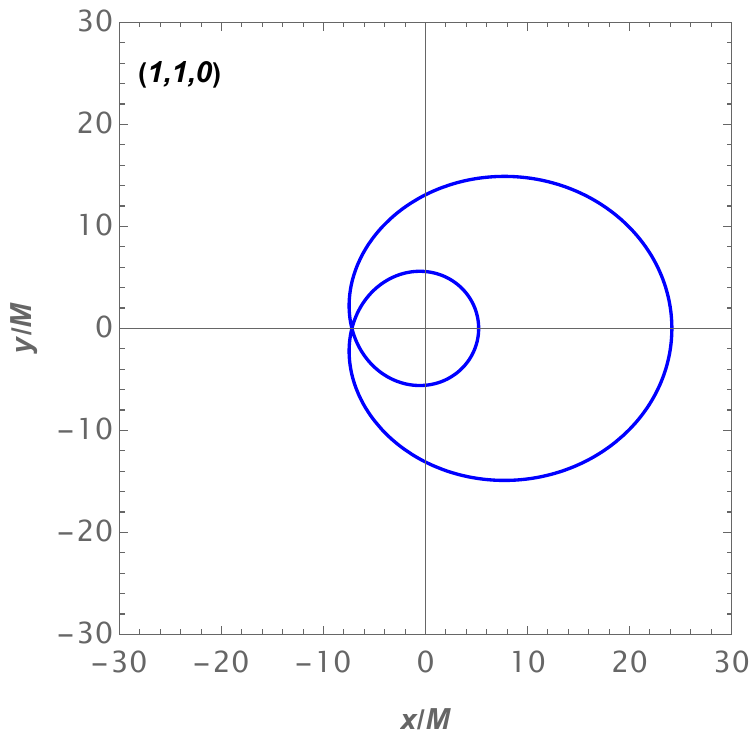} \hfill
    \includegraphics[width=0.32\textwidth]{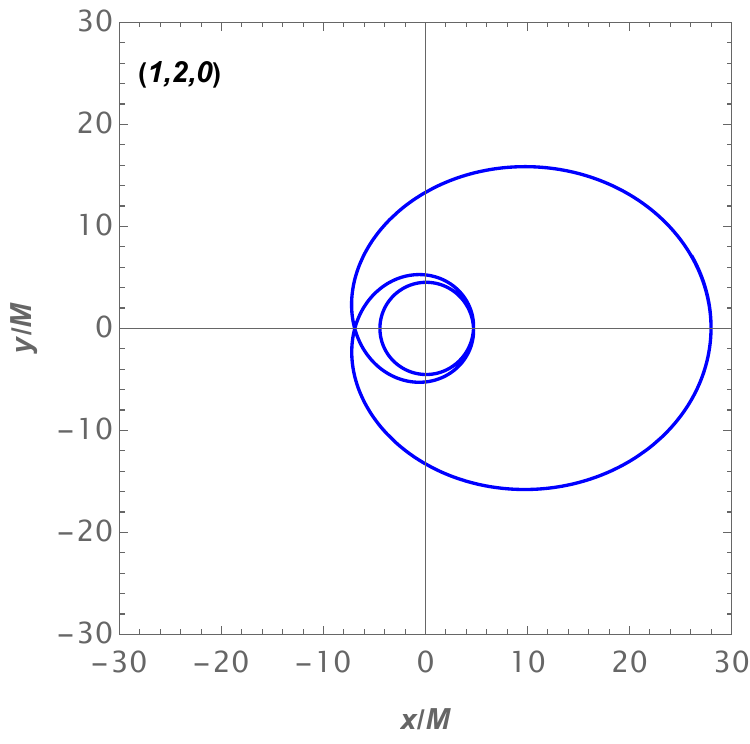} \hfill
    \includegraphics[width=0.32\textwidth]{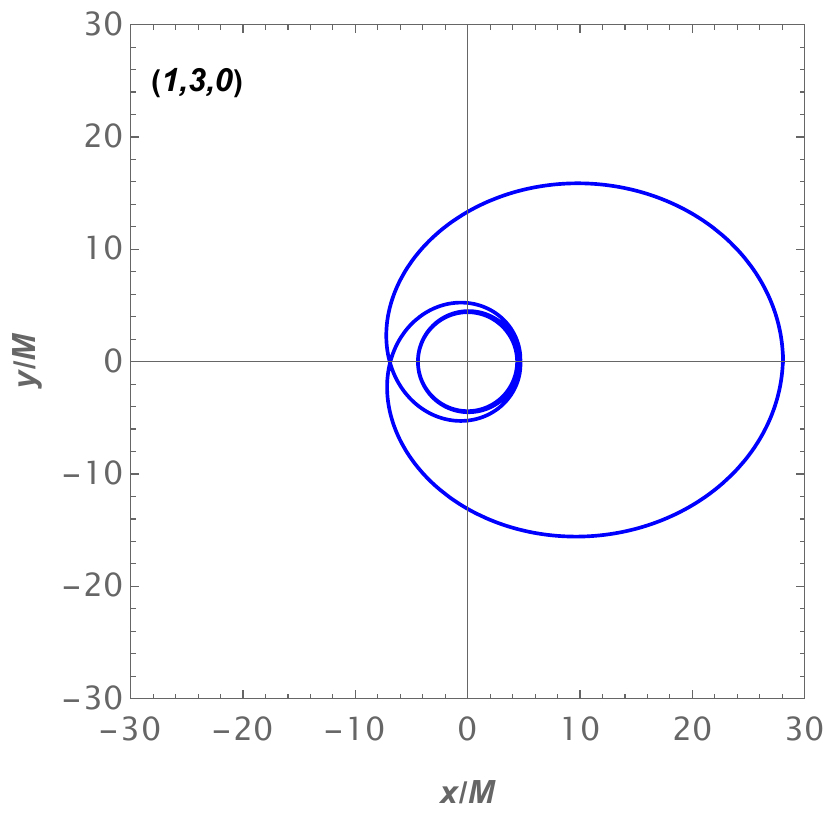} \\
    
    \vspace{0.2cm} 
    \includegraphics[width=0.32\textwidth]{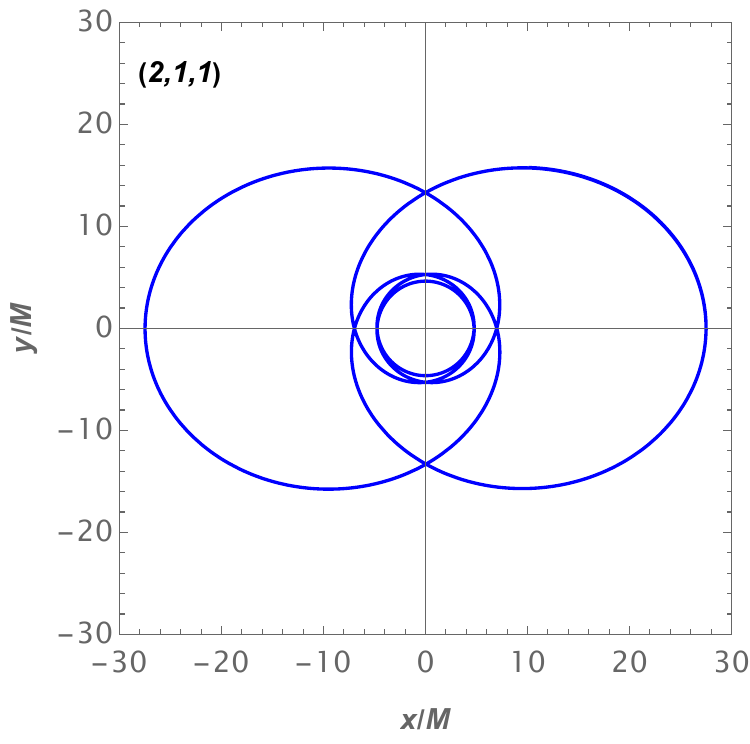} \hfill
    \includegraphics[width=0.32\textwidth]{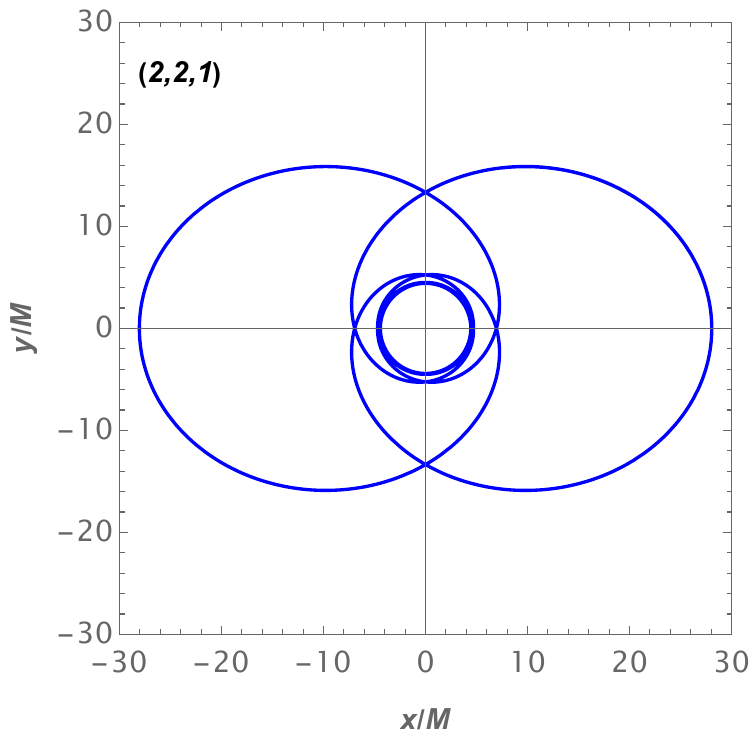} \hfill
    \includegraphics[width=0.32\textwidth]{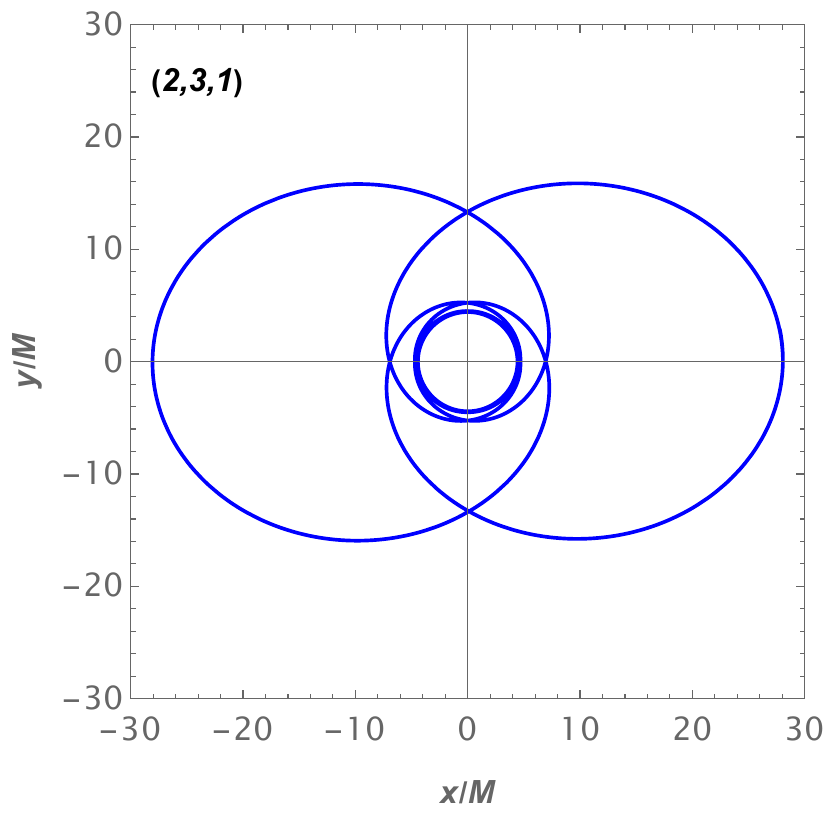} \\
    
    \vspace{0.2cm} 
    \includegraphics[width=0.32\textwidth]{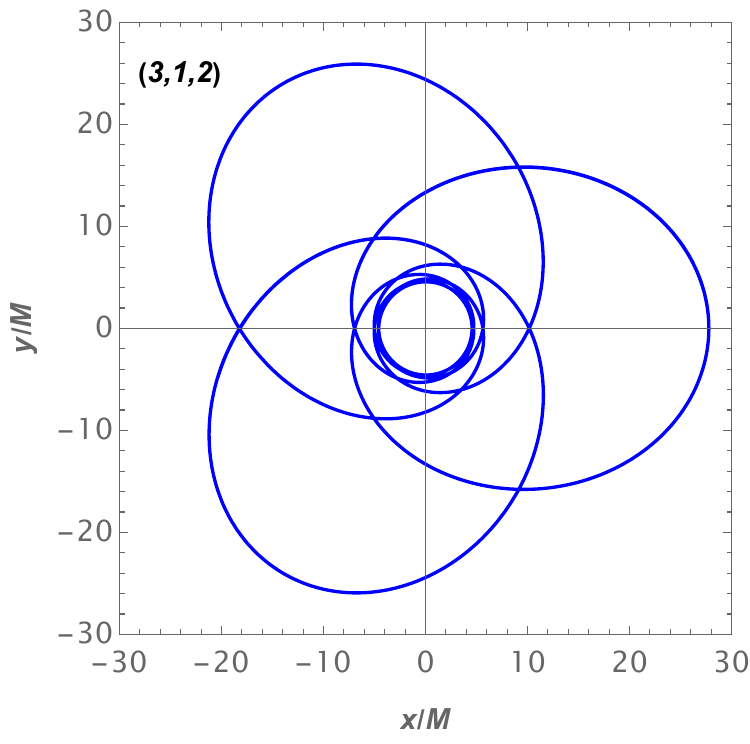} \hfill
    \includegraphics[width=0.32\textwidth]{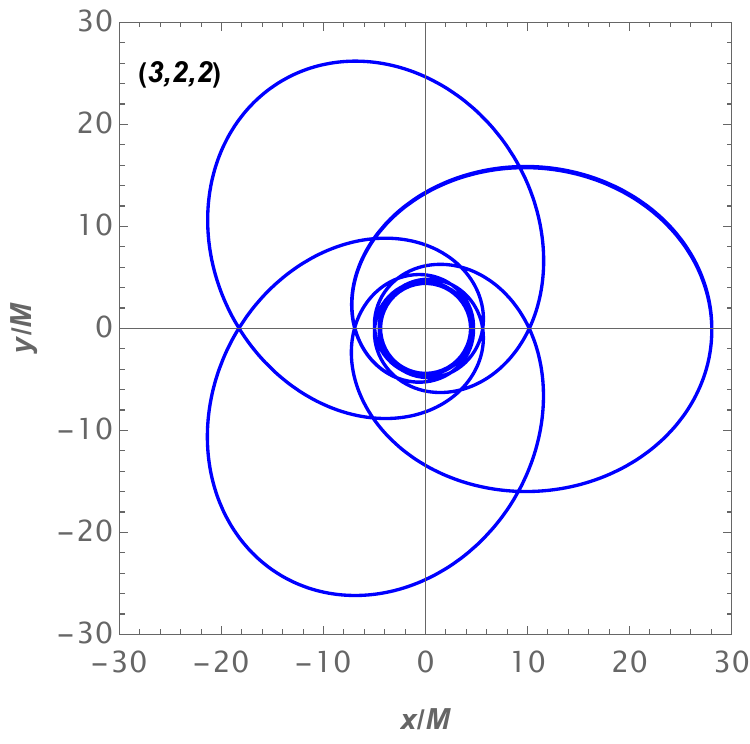} \hfill
    \includegraphics[width=0.32\textwidth]{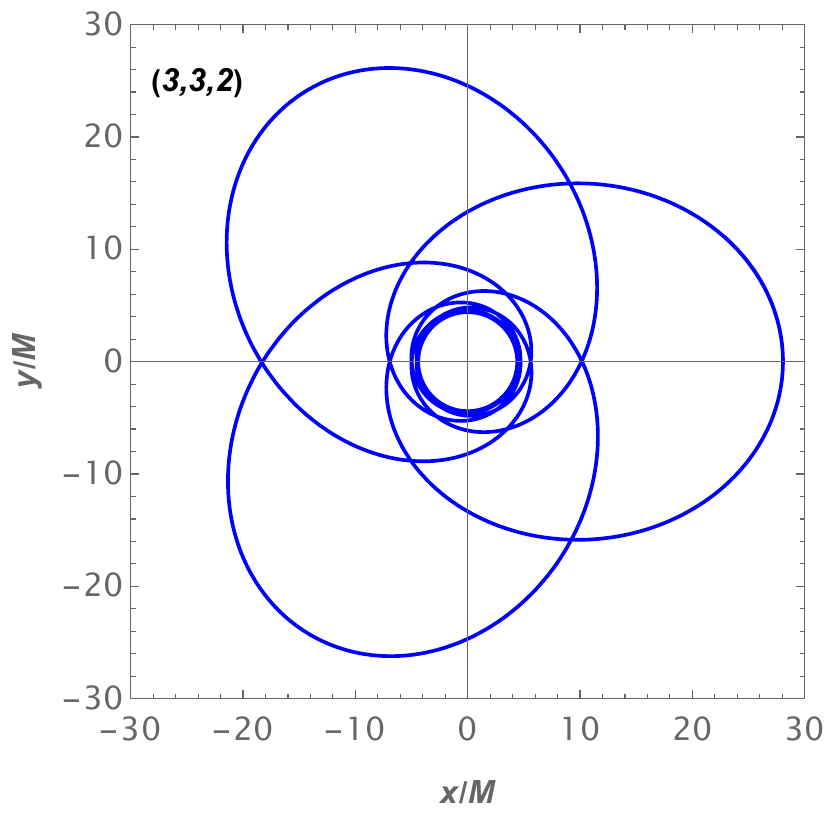} \\
    
    \vspace{0.2cm} 
    \includegraphics[width=0.32\textwidth]{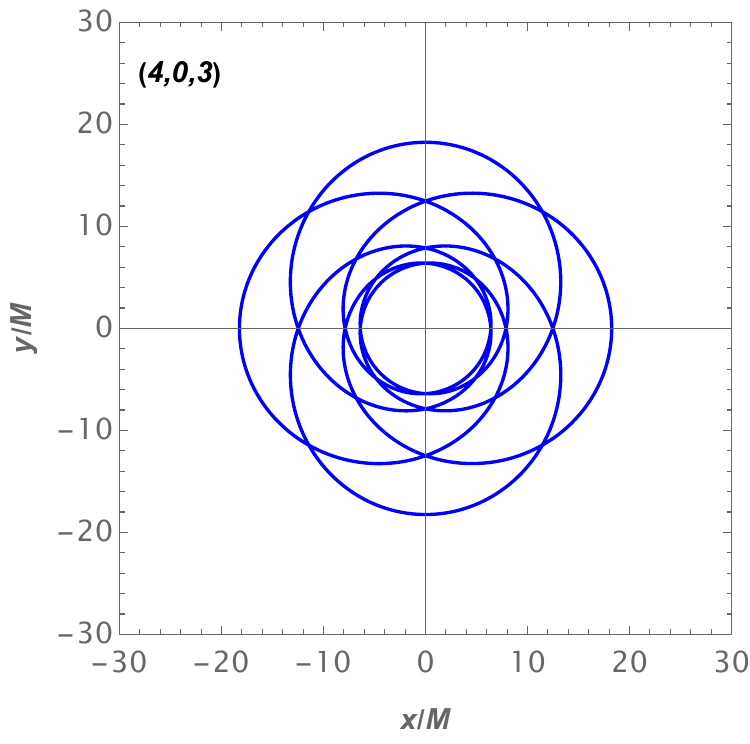} \hfill
    \includegraphics[width=0.32\textwidth]{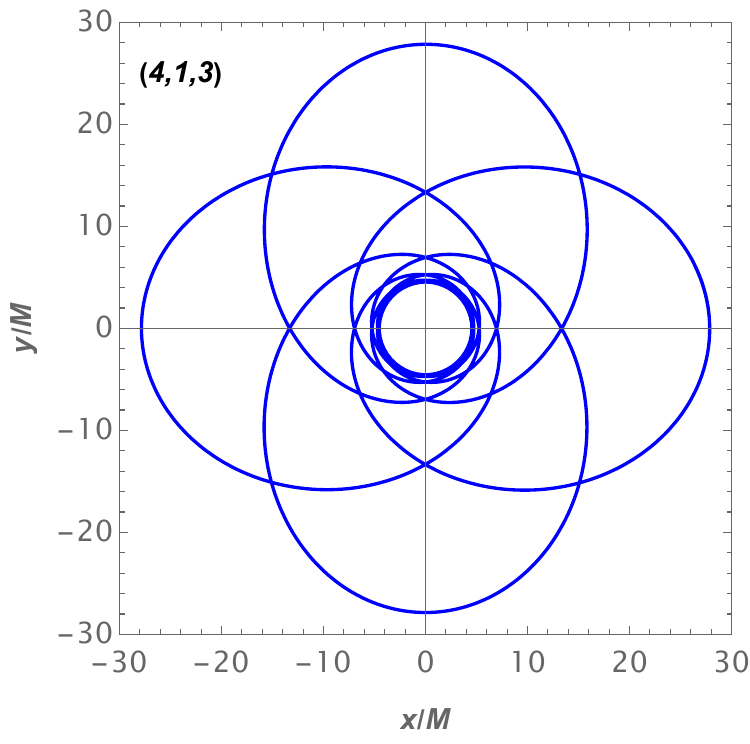} \hfill
    \includegraphics[width=0.32\textwidth]{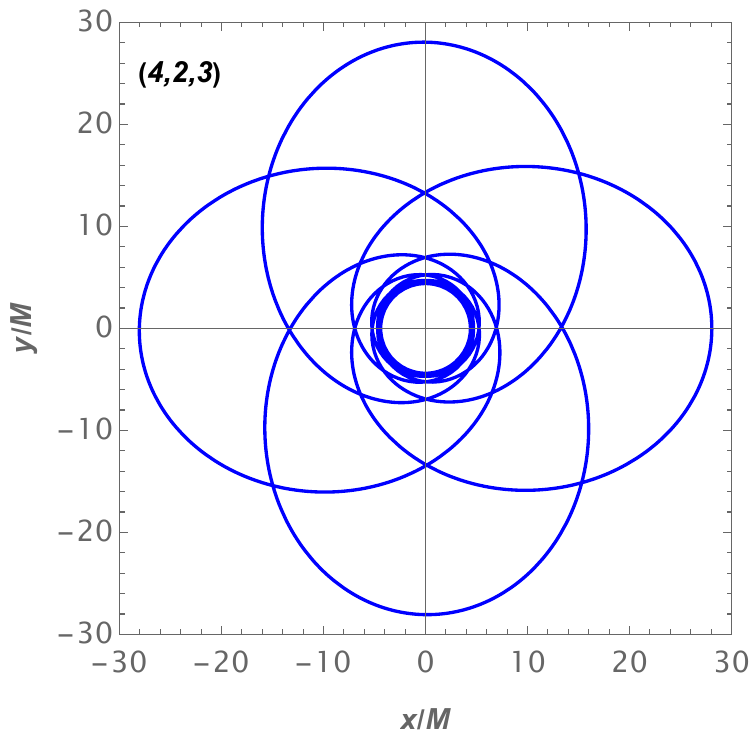}
    
    \caption{\cblue{The figure demonstrates the periodic orbits for different $(z,w,v)$ around the Schwarzschild BH surrounded by a Dehnen-type DM halo. Here, $\rho_s=0.004$, $r_s=0.2$ and $L=\frac{1}{2}(L_{MBO}+L_{ISCO})$.}}
    \label{fig:periodic}
\end{figure*}
\begin{figure*}[htbp]
    \centering
    \includegraphics[width=0.32\textwidth]{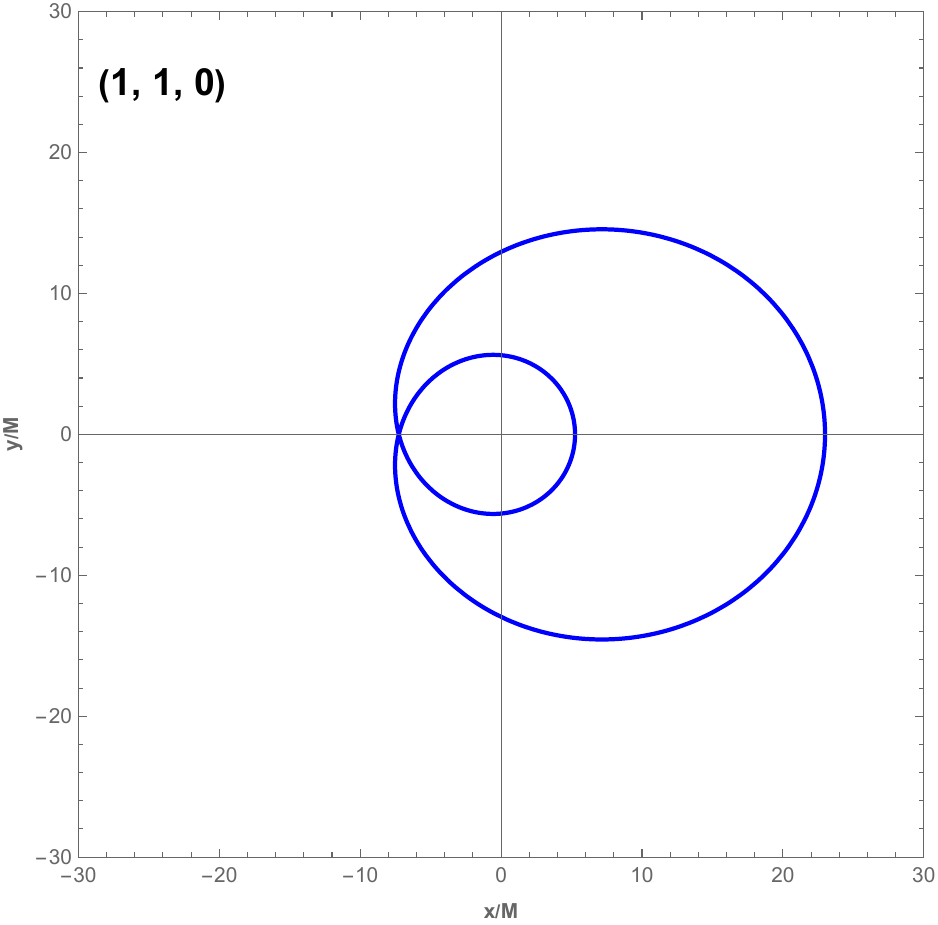} \hfill
    \includegraphics[width=0.32\textwidth]{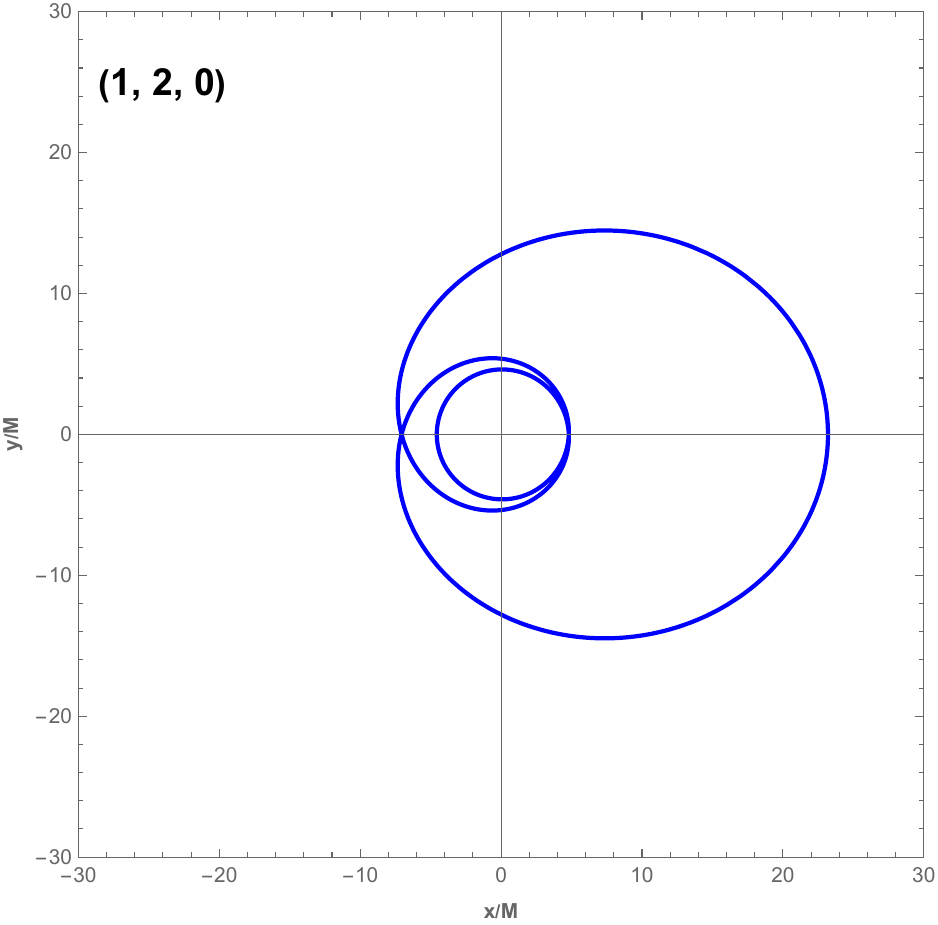} \hfill
    \includegraphics[width=0.32\textwidth]{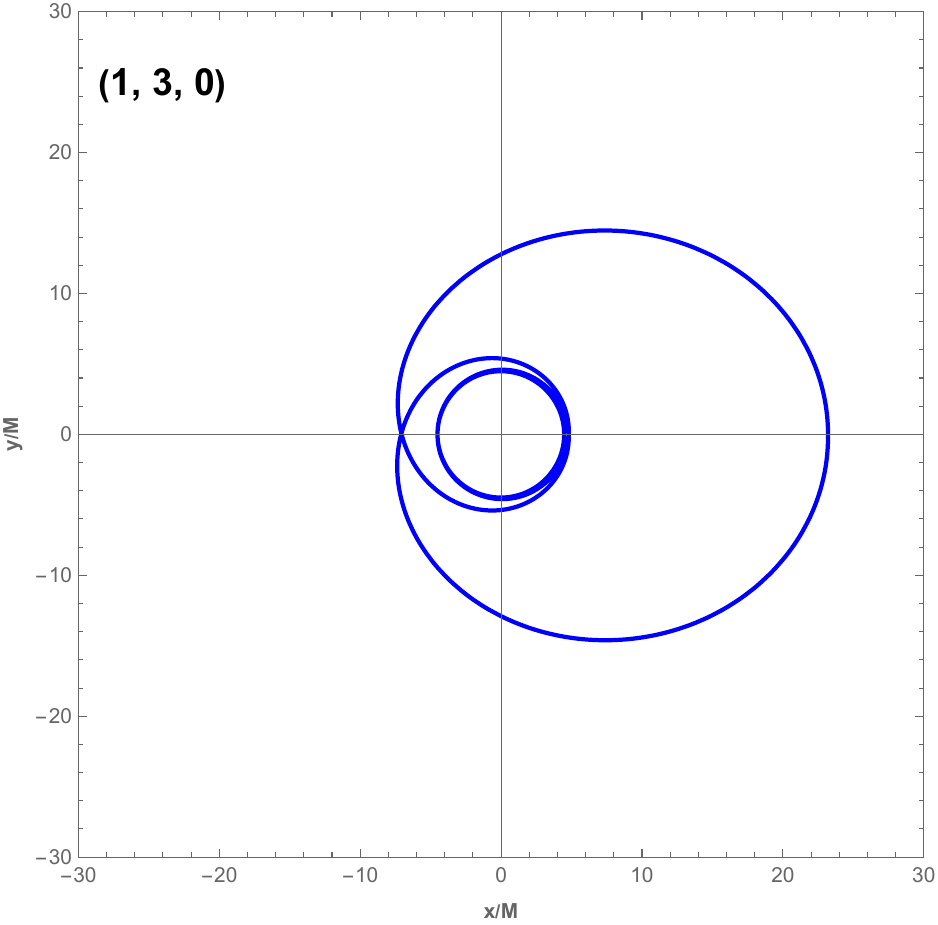} \\
    
    \vspace{0.2cm} 
    \includegraphics[width=0.32\textwidth]{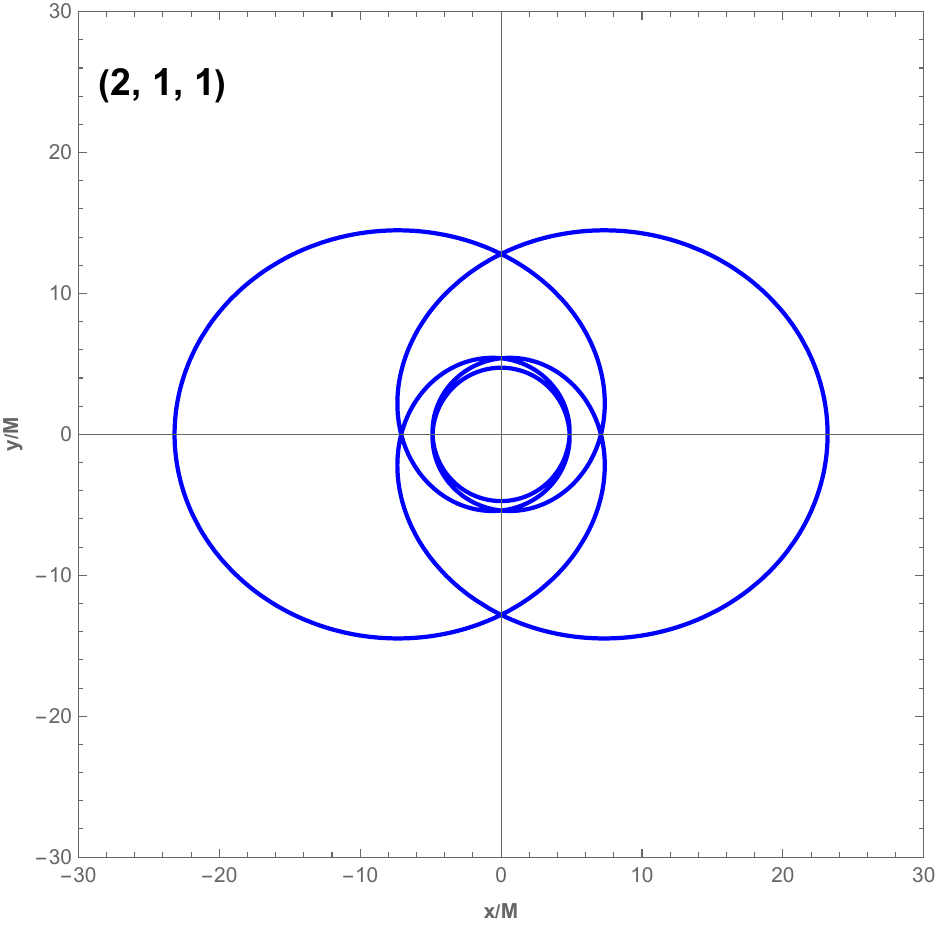} \hfill
    \includegraphics[width=0.32\textwidth]{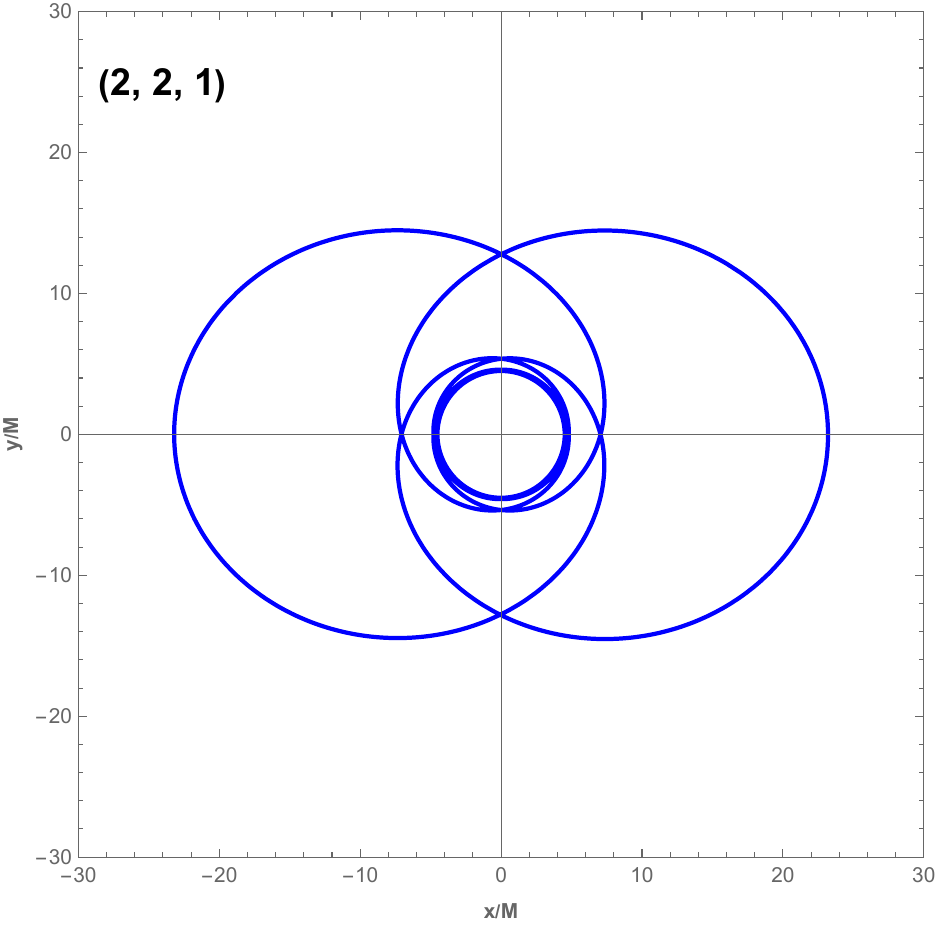} \hfill
    \includegraphics[width=0.32\textwidth]{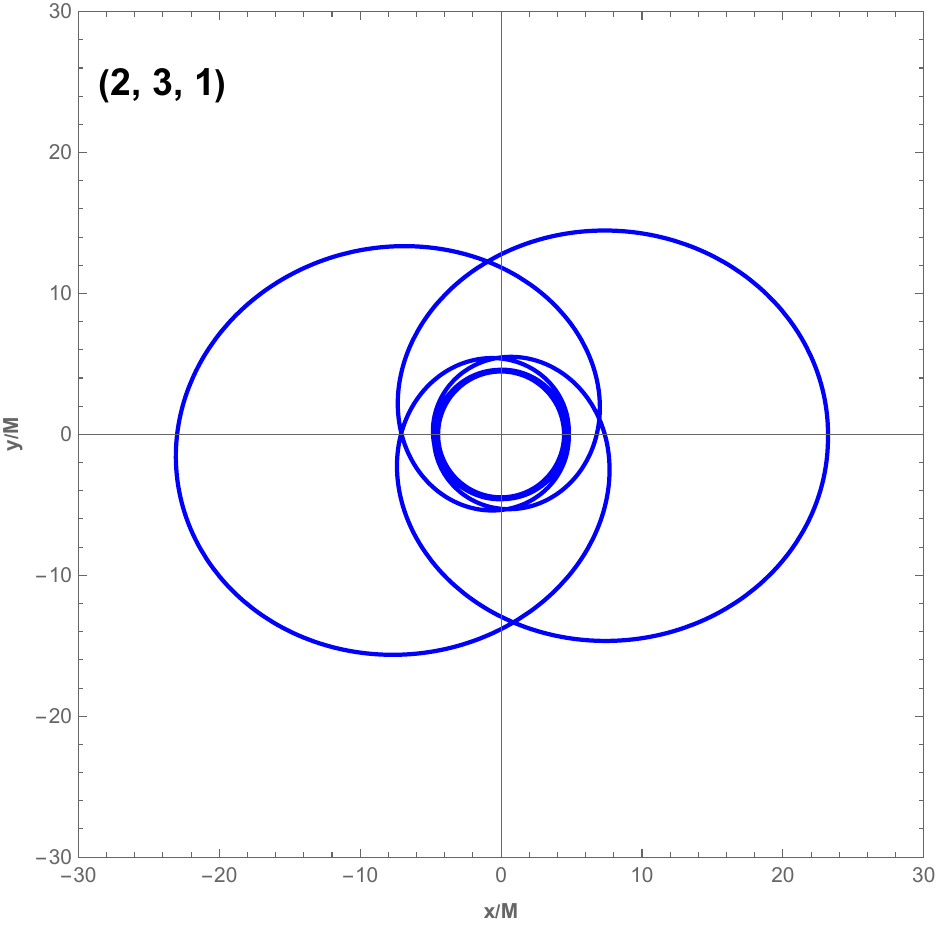} \\
    
    \vspace{0.2cm} 
    \includegraphics[width=0.32\textwidth]{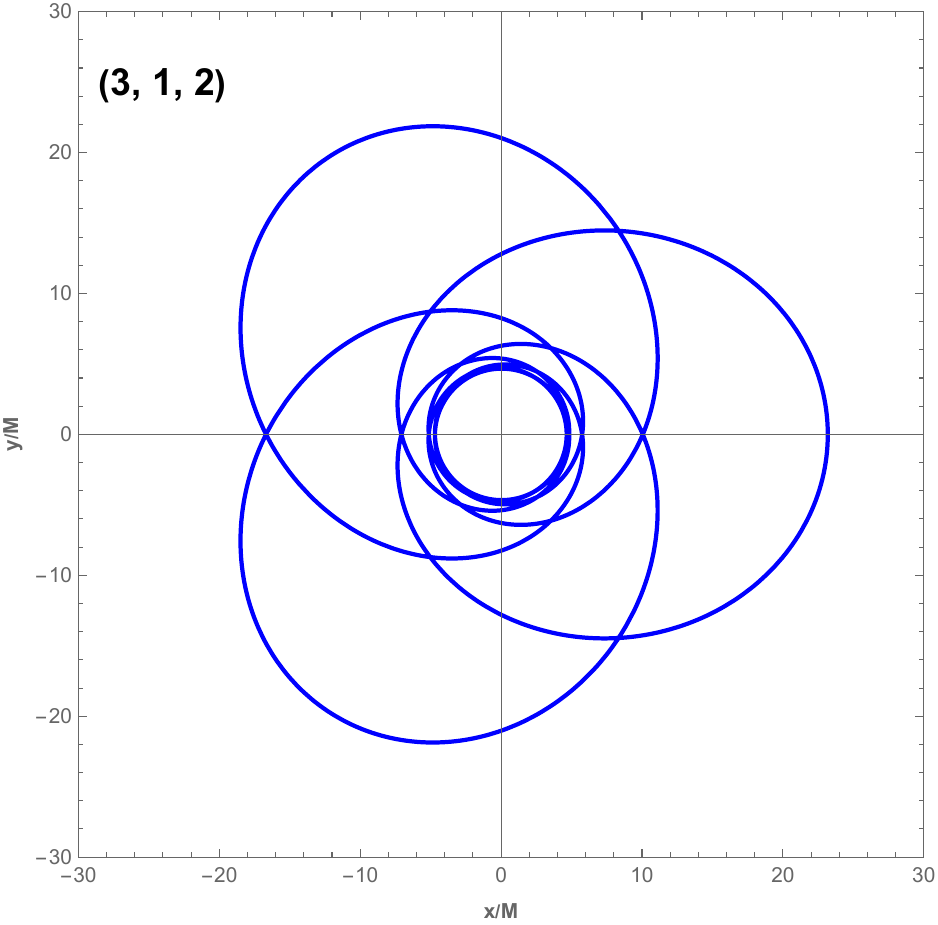} \hfill
    \includegraphics[width=0.32\textwidth]{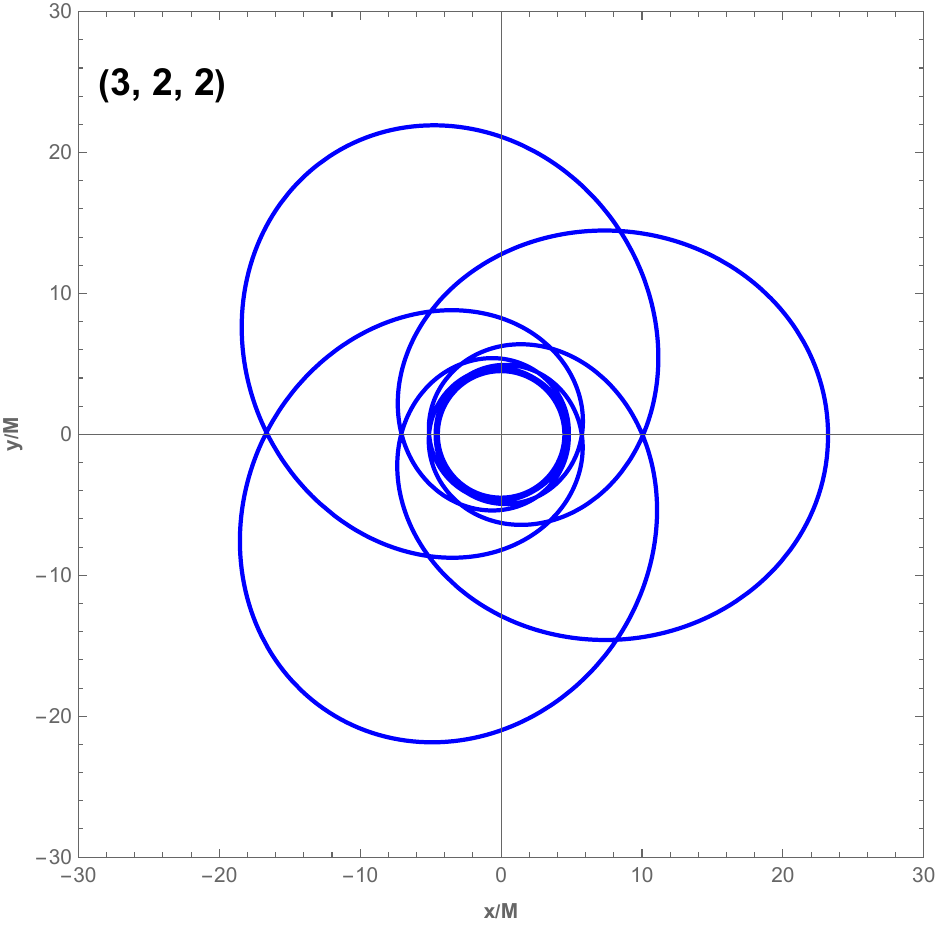} \hfill
    \includegraphics[width=0.32\textwidth]{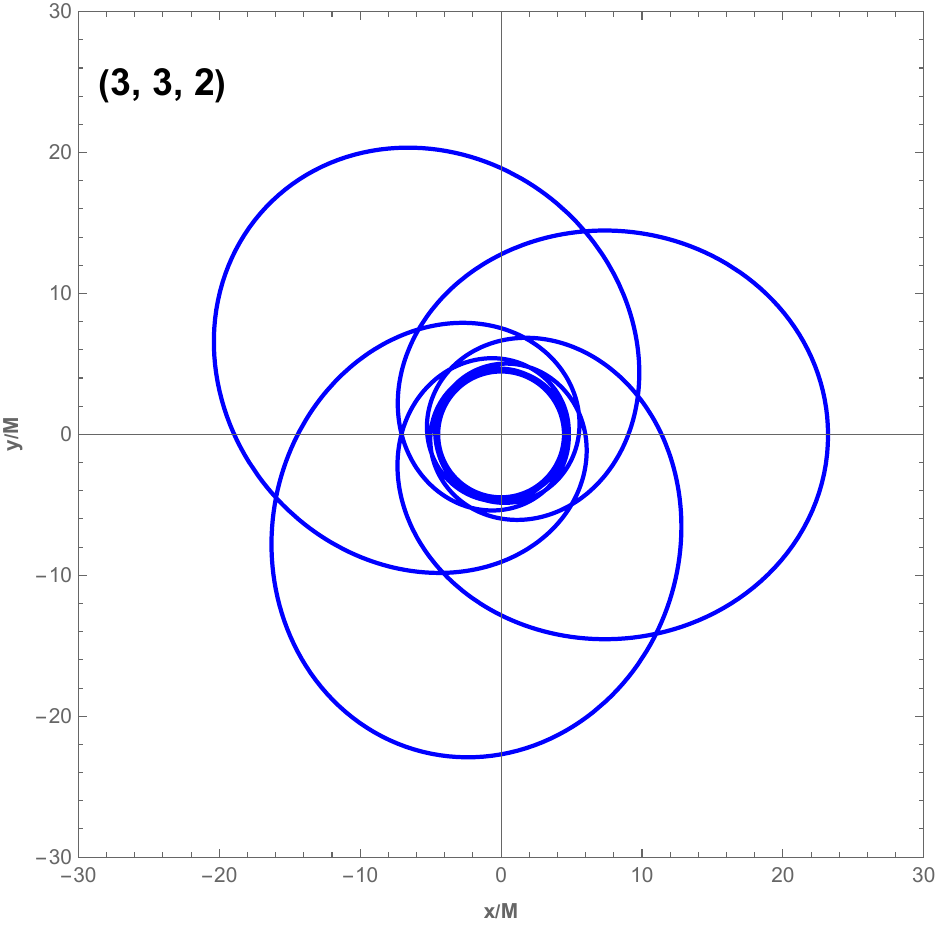} \\
    
    \vspace{0.2cm} 
    \includegraphics[width=0.32\textwidth]{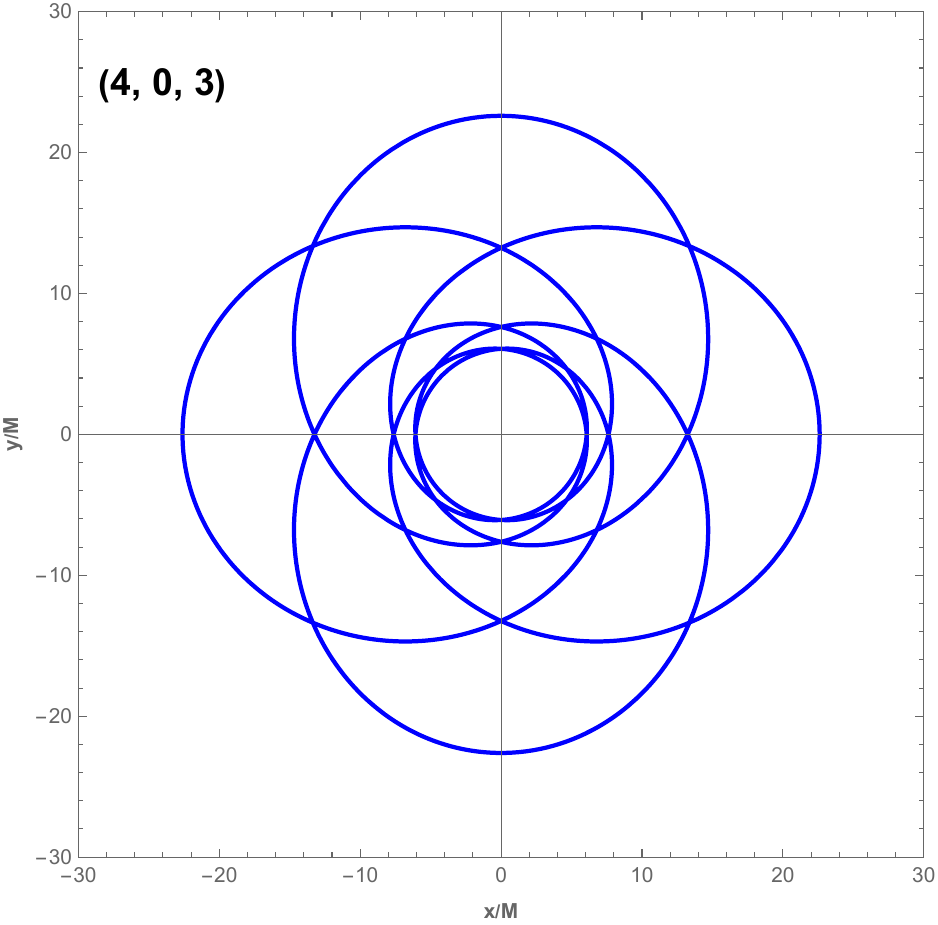} \hfill
    \includegraphics[width=0.32\textwidth]{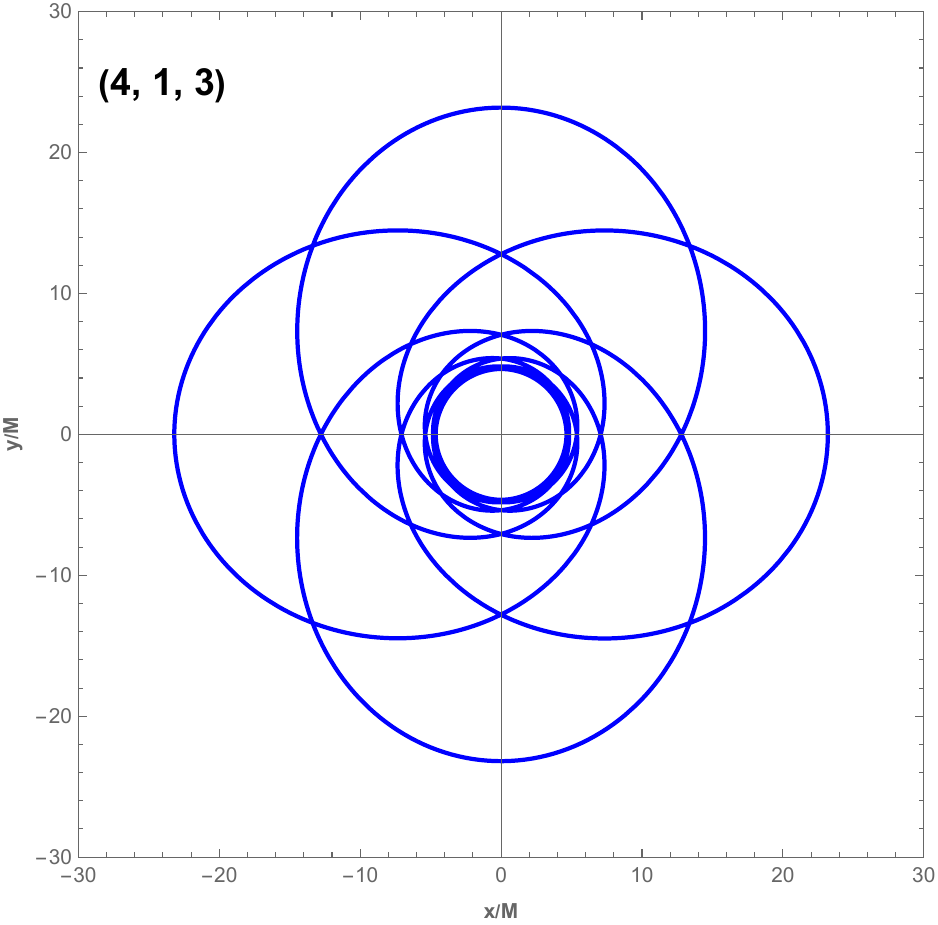} \hfill
    \includegraphics[width=0.32\textwidth]{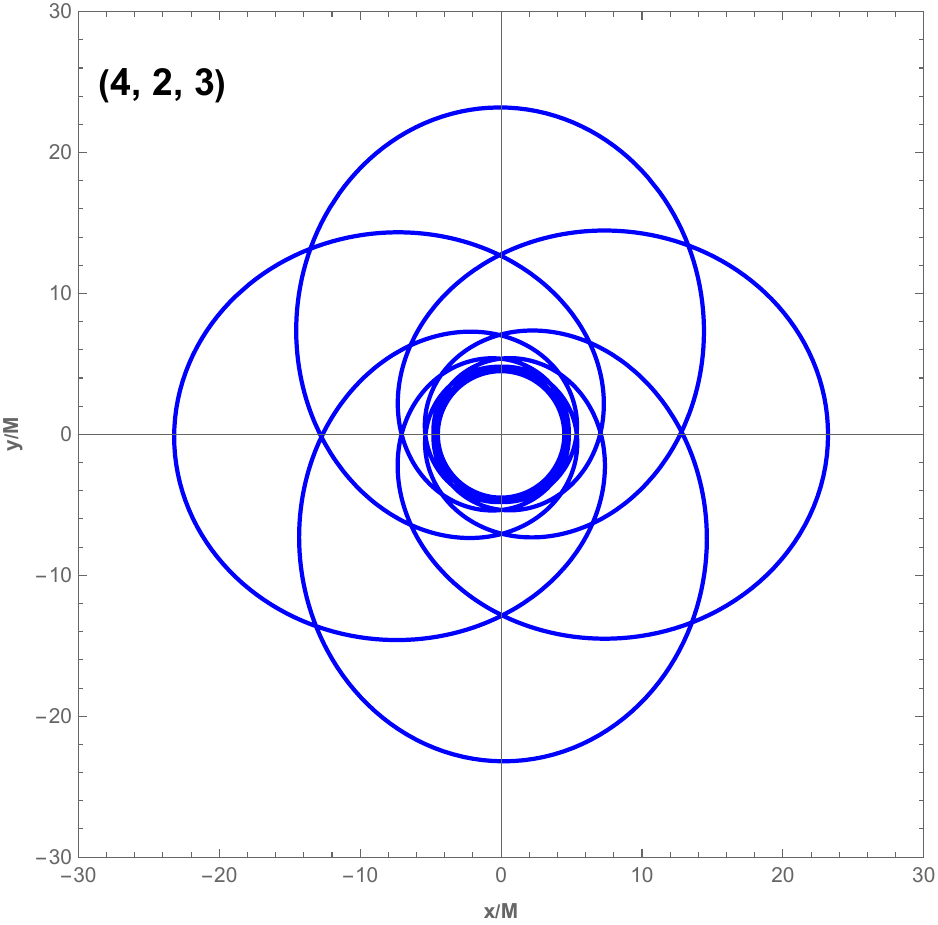}
    
    \caption{\cblue{The figure demonstrates the periodic orbits for different $(z,w,v)$ around the Schwarzschild BH surrounded by a Dehnen-type DM halo. Here, $\rho_s=0.004$, $r_s=0.2$ and $E=0.96$.}}
    \label{fig:periodicL}
\end{figure*}
\begin{figure*}[htbp]
\begin{subfigure}[b]{0.45\textwidth}
\includegraphics[width=\textwidth]{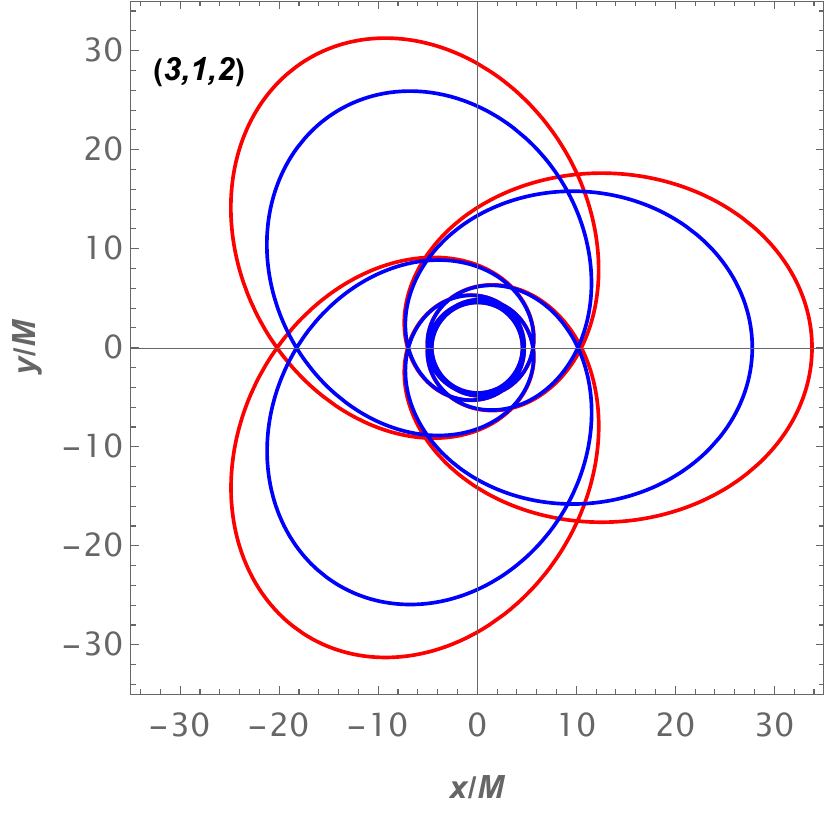}
\end{subfigure}
\begin{subfigure}[b]{0.52\textwidth}
\includegraphics[width=\textwidth]{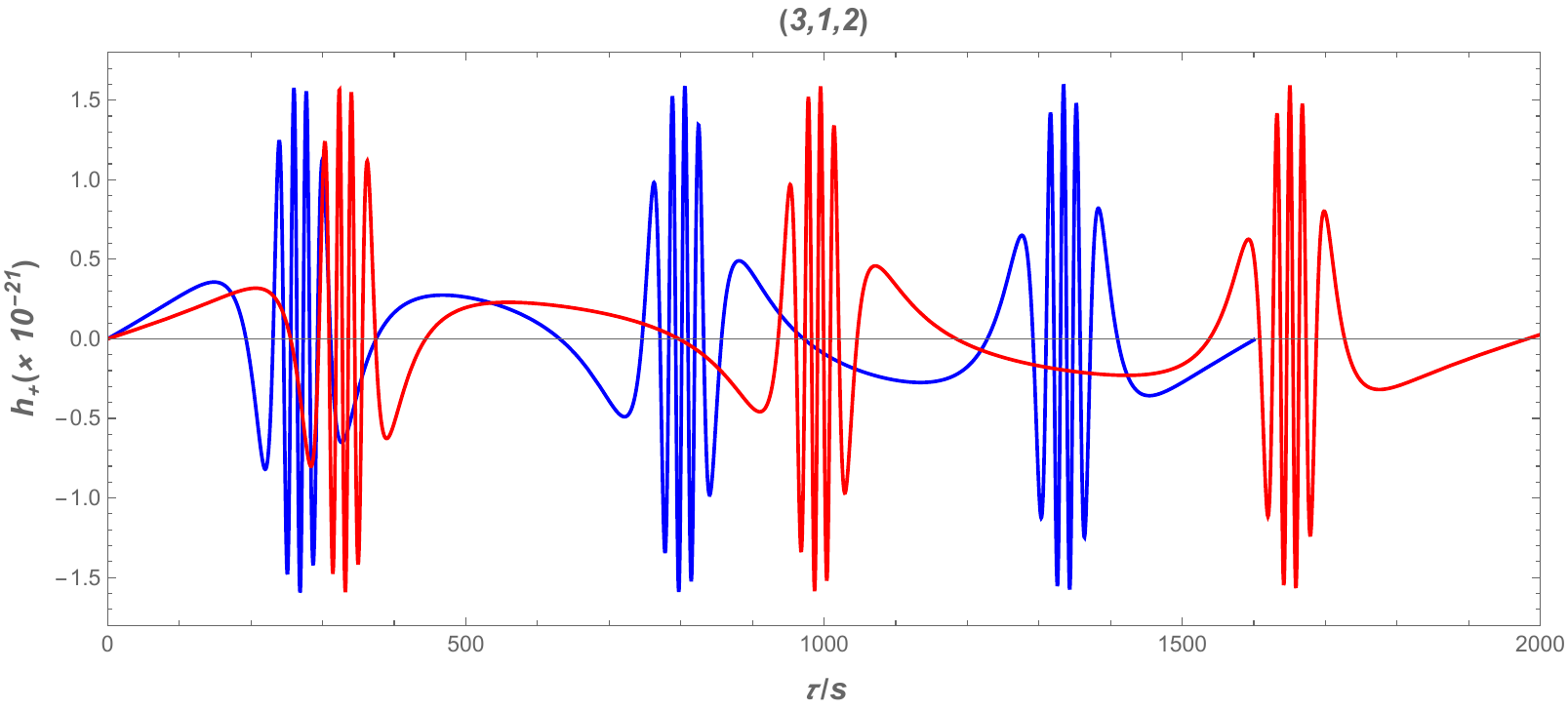}
\includegraphics[width=\textwidth]{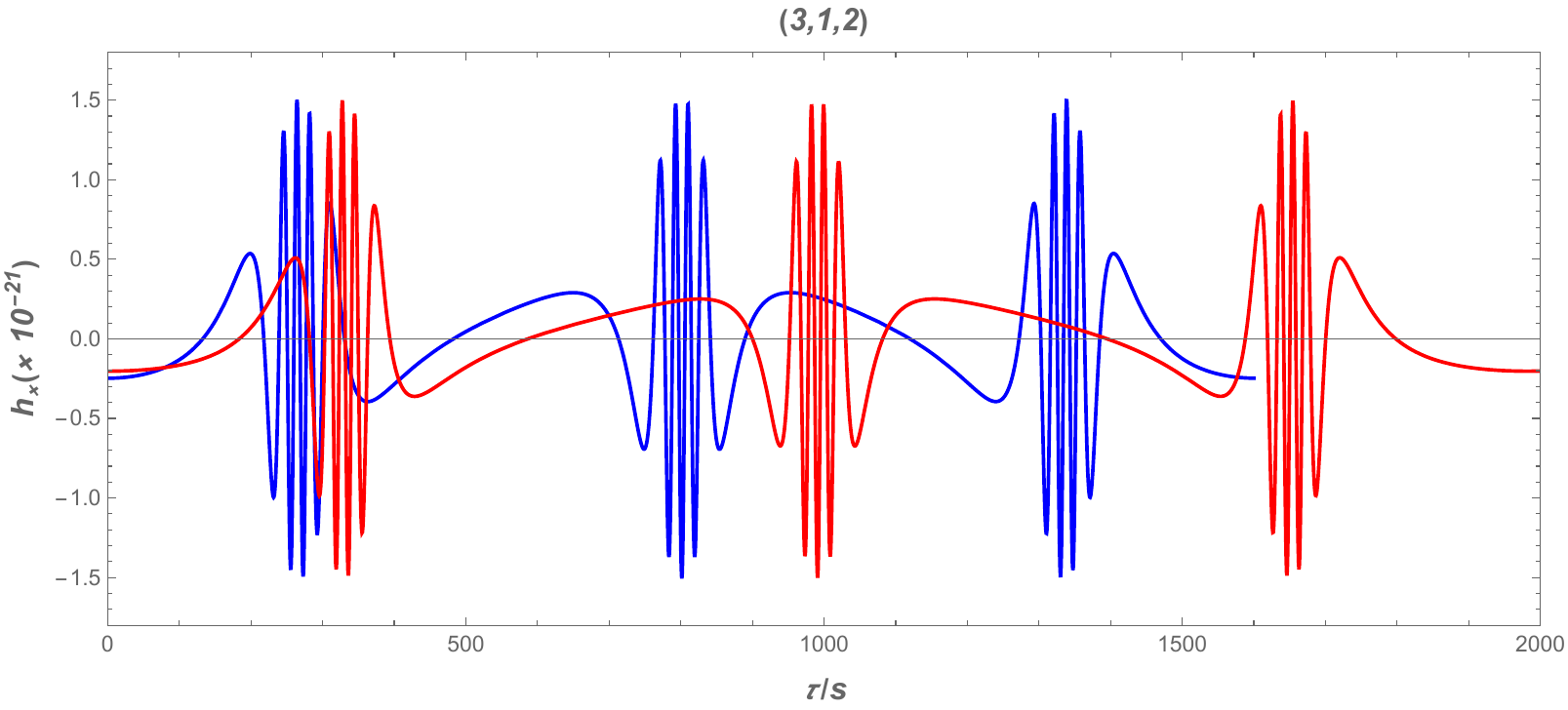}    
\end{subfigure} 
\caption{\cblue{The $(3,1,2)$ periodic orbit and the corresponding gravitational waveform for the EMRI system, which consists of the small object around a supermassive Schwarzschild BH surrounded by a Dehnen-type DM halo. The mass of SMBH and the small object are $M\sim 10^7 M_{\odot}$ and $m\sim 10M_{\odot}$, respectively. The blue and red lines correspond to the $\rho_s=0.04$ and $\rho_s=0.08$, respectively. Here, we set $r_s=0.2$.}}
\label{fig:combination}
\end{figure*}
\begin{figure*}[htbp]
\begin{subfigure}[b]{0.45\textwidth}
\includegraphics[width=\textwidth]{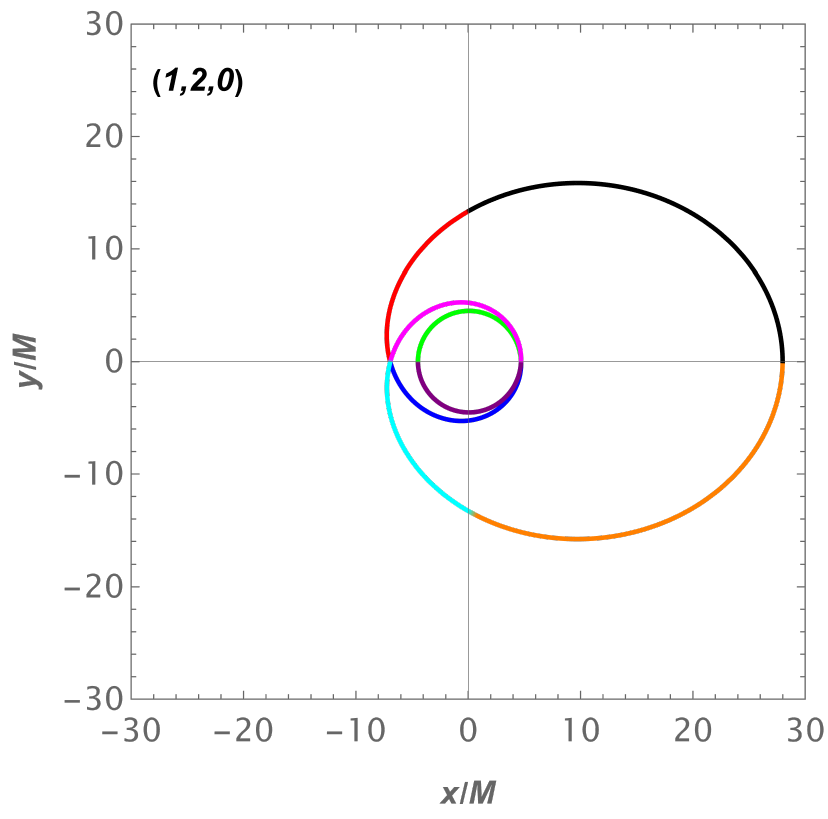}
\end{subfigure}
\begin{subfigure}[b]{0.52\textwidth}
\includegraphics[width=\textwidth]{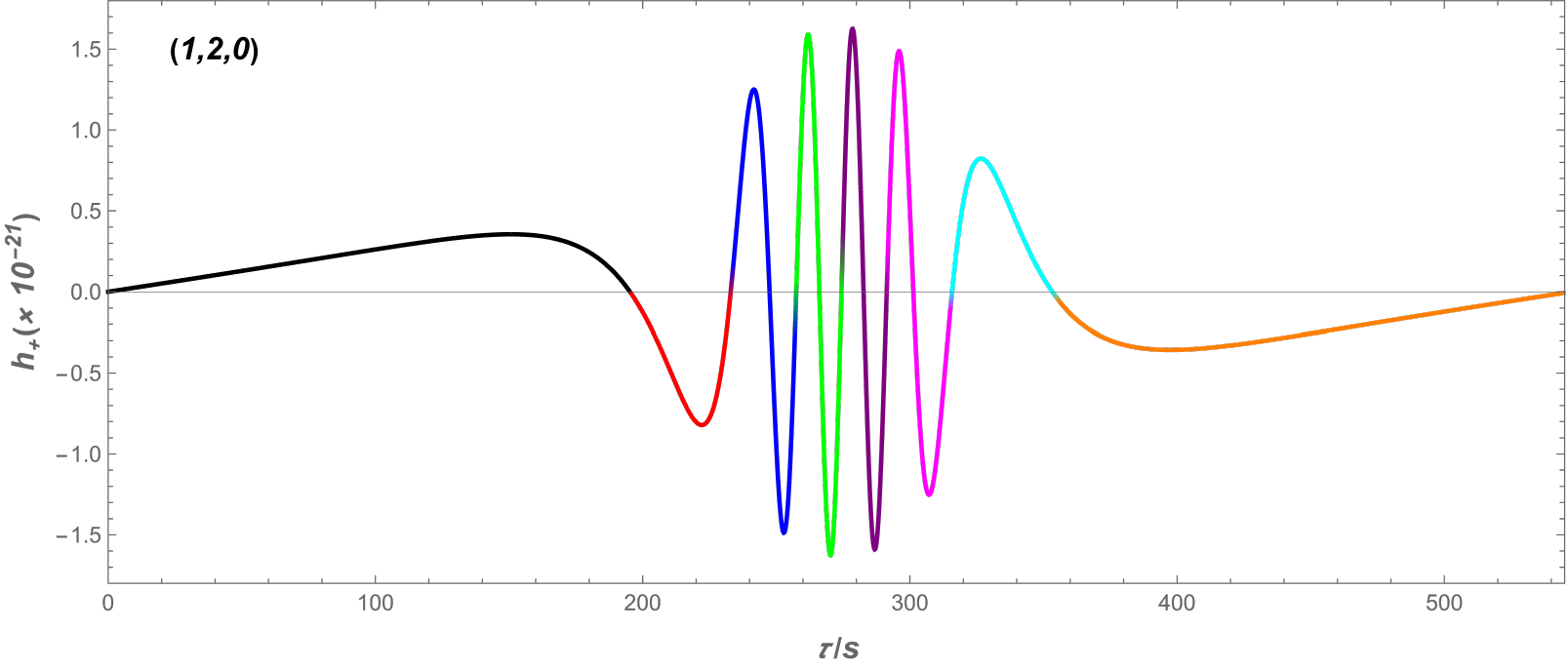}
\includegraphics[width=\textwidth]{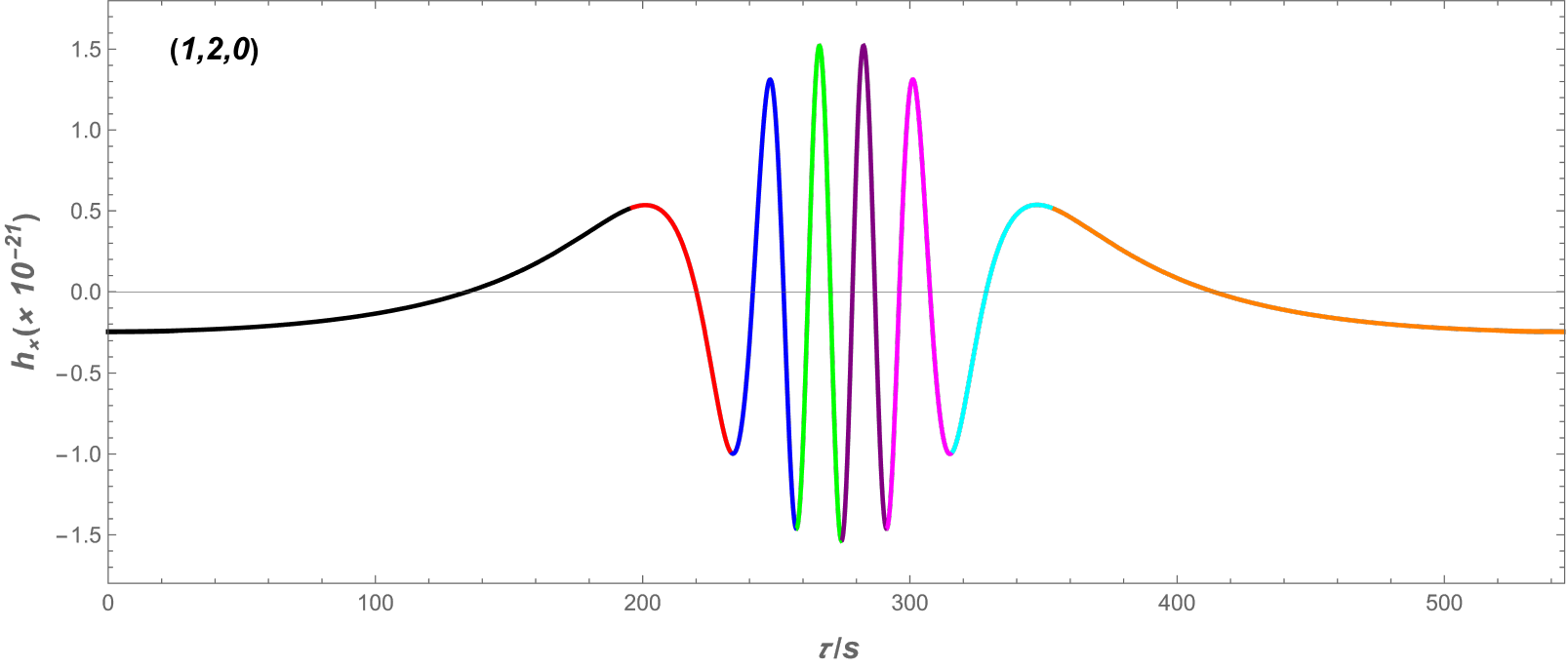}    
\end{subfigure} 
\caption{\cblue{The $(1,2,0)$ periodic orbit and the corresponding gravitational waveform for the EMRI system, which consists of the small object around a supermassive Schwarzschild BH surrounded by a Dehnen-type DM halo. The mass of SMBH and the small object are $M\sim 10^7 M_{\odot}$ and $m\sim 10M_{\odot}$, respectively. Different segments are marked using different colors.} }
\label{fig:combination2}
\end{figure*}
\begin{figure*}[htbp]
\includegraphics[width=0.85\textwidth]{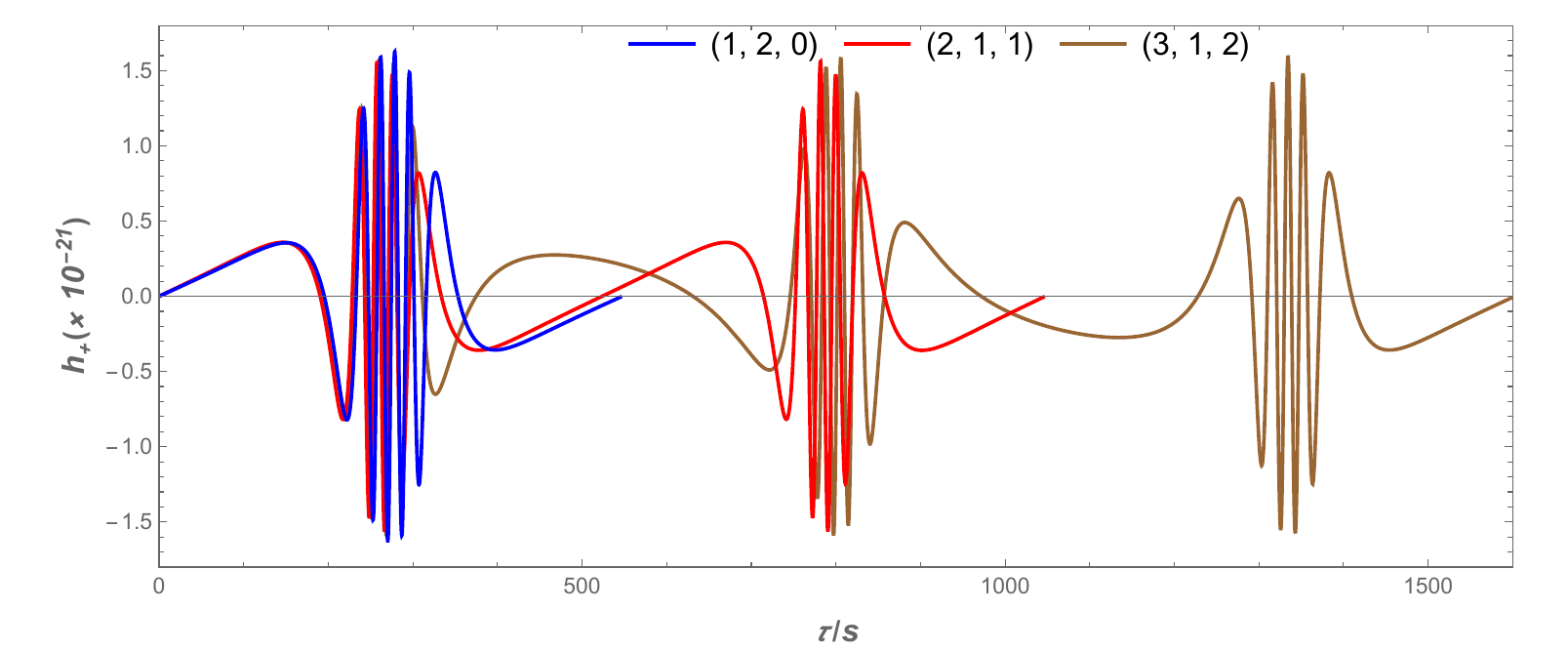}
\includegraphics[width=0.85\textwidth]{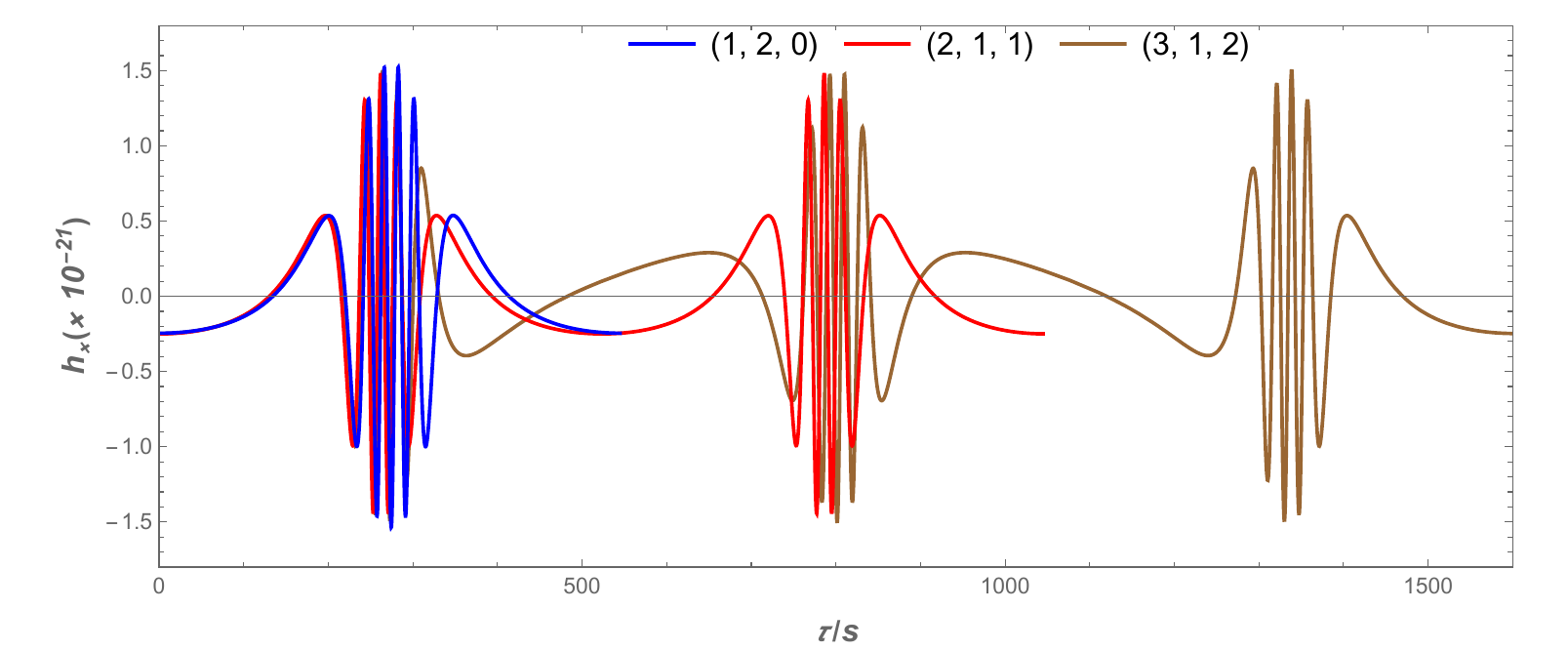}
\caption{\cblue{The figure demonstrates the difference in the gravitational waveforms between various periodic orbits. The top and bottom panels refer to the $h_{+}$ and $h_{\times}$, respectively. The mass of SMBH and the small object are $M\sim 10^7 M_{\odot}$ and $m\sim 10M_{\odot}$, respectively. The parameters of the spacetime are $\rho_s=0.004$ and $r_s=0.2$. }}
\label{fig:GW}
\end{figure*}

\section{Periodic orbits around the Schwarzschild black hole surrounded by dark matter halo}\label{sec3}

In this part, we investigate the periodic orbits around the Schwarzschild BH surrounded by a Dehnen-type DM halo. The fundamental orbital frequencies of the periodic orbits are related by rational ratios. Note that every periodic orbit is defined by three integers: the ($z,w,v$). $z$, $w$, and $v$ refer to the zoom, whirl, and vertex numbers, respectively. The rational number can be written as~\cite{Levin_2008}
\begin{equation}
    q=\frac{\omega_{\phi}}{\omega_r}-1=w+\frac{v}{z}\, ,
\end{equation}
where $\omega_{\phi}$ and $\omega_r$ are the angular and radial frequencies, respectively. If we consider the equations of motion, the rational number takes the following form~\cite{2025JCAP...01..091Y,SHABBIR2025101816,JIANG2024101627}
\begin{widetext}
\begin{equation}
    q=\frac{1}{\pi}\int^{r_2}_{r_1} \frac{\dot{\phi}}{\dot{r}}-1=\frac{1}{\pi} \int^{r_2}_{r_1}\frac{1}{r^2\sqrt{E^2-(1-\frac{2M}{r}-32\pi \rho_s r^3_s\sqrt{\frac{r+r_s}{r^2_s r}})(1+\frac{L^2}{r^2})}}\, ,
\end{equation}
\end{widetext}
where $r_1$ and $r_2$ refer to the radius of periapsis and apoapsis of the periodic orbits, respectively.
We show the dependence of the rational number $q$ on the energy and orbital angular momentum of the particle for different values of the spacetime parameters in Fig.~\ref{fig:q}. It can be seen from the top panel of this figure that the rational number $q$ increases with increasing energy $E$, and then there is a rapid increase as $E$ approaches its maximum value. Moreover, the values of the rational number $q$ shift toward the lower side of energy with the increase of the spacetime parameters. The bottom panel illustrates the rational number $q$ as a function of the orbital angular momentum. As the angular momentum $L$ approaches its minimum value, the rational number $q$ increases dramatically, but gradually decreases as $L$ grows larger. Furthermore, the values of the rational number $q$ shift to the lower side of energy as the spacetime parameters increase.
Furthermore, we numerically calculate the values of the energy $E$ of periodic orbits that are characterized by different $(z,w,v)$ for fixed values of the orbital angular momentum $L=\frac{1}{2}(L_{MBO}+L_{ISCO})$ in Table~\ref{table1}. Using the results of Table~\ref{table1}, we plot the periodic orbits around the Schwarzschild BH surrounded by a Dehnen-type DM halo for different values of the $(z,w,v)$ in Fig.~\ref{fig:periodic}. The spacetime parameters are fixed as $\rho_s=0.004$ and $r_s=0.2$ in this figure. In addition, we tabulated the values of the orbital angular momentum $L$ of periodic orbits characterized by different $(z,w,v)$ in Table~\ref{table2}. We set $E=0.96$ and $r_s=0.2$ for this table. To provide more information, we demonstrate the periodic orbits for different $(z,w,v)$ in Fig.~\ref{fig:periodicL} by using the results in Table~\ref{table2}. It is evident that periodic orbits with a higher zoom number $z$ display more intricate structural features, whereas those with a larger whirl number $w$ complete more revolutions around the central black hole between successive apoapses.

\section{NUMERICAL KLUDGE GRAVITATIONAL WAVEFORMS FROM PERIODIC ORBITS}\label{sec4}

In this section, we investigate the gravitational waveforms from periodic orbits around the Schwarzschild BH surrounded by a Dehnen-type DM halo. It is worth noting that we consider an extreme mass-ratio inspiral (EMRI) system. The EMRI system consists of a stellar-mass object along the periodic orbit around a supermassive Schwarzschild BH surrounded by a Dehnen-type DM halo. The gravitational waves emitted by this EMRI system could carry information about the periodic orbit and the supermassive Schwarzschild BH surrounded by a Dehnen-type DM halo. We use the adiabatic approximation method, which is widely recognized for calculating the gravitational waveforms from an EMRI system. The energy and orbital angular momentum can be considered as a constant in the short term because the small object's orbital parameters change over a time scale significantly longer than the orbital periods. Therefore, the orbit can be approximated as a geodesic over a few orbital cycles, and we define the gravitational waveforms generated by the periodic orbits over one full cycle. Note that we can neglect the effects of the gravitational radiation on the motion of the small object for the given brief time frame. We investigate the gravitational waveforms from the EMRI system with the usage of the numerical kludge waveform model. First, we numerically solve the equations of motion, which are written in Eq.~(\ref{eq:eqmotion}), to explore the periodic orbits of the small object around a supermassive Schwarzschild BH surrounded by a Dehnen-type DM halo. After that, we can generate the gravitational waveforms using the symmetric and trace-free (STF) mass quadrupole equation of gravitational radiation (see \cite{2025JCAP...01..091Y}). The quadrupole formula can be written by calculating the second-order term as~\cite{2025EPJC...85...36Z}
\begin{equation}\label{eq:perturbation}
h_{ij}=\frac{4\mu M}{D_{L}}\left(v_{i}v_{j}-\frac{m}{r}n_{i}n_{j}\right).
\end{equation}
where $m$ and $M$ refer to the mass of the small object and the BH, respectively. It is worth noting that \(m\ll M\) because we are exploring the EMRI system. $D_{L}$ and \(\mu=\frac{Mm}{(M+m)^{2}}\) are the luminosity distance of the EMRI system and the symmetric mass ratio of the system, respectively. Moreover, $n_{i}$ is the unit pointing vector and $v_{i}$ is the component of the velocity vector of the small object. This waveform illustrates the fundamental distribution properties of gravitational waves in space. To be compatible with real detectors, the computed gravitational wave signal must be adjusted to align with the detector's framework. This requires projecting the gravitational wave onto a coordinate system compatible with the detector~\cite{Poisson_Will_2014,2025JCAP...01..091Y,2025EPJC...85...36Z}, and its coordinate basis can be expressed in the following form
\begin{equation}
\begin{array}{l}
e_{X}=[\cos\zeta,-\sin\zeta,0]\,,\\
e_{Y}=[\cos\iota\sin\zeta,\cos\iota\cos\zeta,-\sin\iota]\,,\\
e_{Z}=[\sin\iota\sin\zeta,\sin\iota\cos\zeta,\cos\iota]\,.
\end{array}
\end{equation}
Here, \(\zeta\) is the longitude of the periastron, while \(\iota\) represents the inclination angle of the test particle orbit relative to the observation direction. \(e_{X}\), \(e_{Y}\), \(e_{Z}\) refer to the orthogonal coordinate basis aligned with the detector, used to decompose the gravitational wave signal into its distinct polarization modes. The polarization components \(h_{+}\) and \(h_{\times}\) can be written within this coordinate basis as~\cite{2025EPJC...85...36Z,2024arXiv241101858M}
\begin{eqnarray}
&& h_{+}=-\frac{2\mu M^{2}}{D_{L}r}\left(1+\cos^{2}\iota\right)\cos\left(2\phi+2\zeta\right), \\
&& h_{\times}=-\frac{4\mu M^{2}}{D_{L}r}\cos\iota\sin\left(2\phi+2\zeta\right),
\end{eqnarray}
where \(\phi\) is the phase angle. To visualize the gravitational waves radiated by the small objects along periodic orbital motion we assume the parameters of the EMRI system as follows: the supermassive mass of the BH is $M \sim 10^7 M_{\odot}$, and the mass of the small object is $m \sim 10M_{\odot}$. We also set the latitude and the angle of inclination as $\zeta=\iota=\pi/4$, and the luminosity distance is $D_{L}=200$ \text{Mpc}. We show the gravitational waveforms ($h_{+}$ and $h_{\times}$) radiated by the $(3,1,2)$ periodic orbit in Fig.~\ref{fig:combination}. It can be seen from this figure that there are zoom and whirl stages. The calm section of the waveform aligns with the zoom stage, which occurs when the small object enters a highly elliptical orbit far from the black hole. In contrast, the section with rapid oscillations corresponds to the whirl stage, where the small object nears the black hole and begins to exhibit circular, whirl motion. In this stage, the gravitational wave frequency rises sharply, leading to the intense oscillations. Moreover, we compare $(3,1,2)$ orbits for two different values of $\rho_s$ in Fig.~\ref{fig:combination}. The blue and red lines correspond to $\rho_s=0.04$ and $\rho_s=0.08$, respectively. From Fig.~\ref{fig:combination}, the zoom area expands as $\rho_s$ increases, thus shifting the gravitational waveform toward larger values of time. To provide more information, different colors are used to emphasize the correspondence between the $(1,2,0)$ periodic orbit and the gravitational waveform, facilitating a detailed analysis of how the orbit influences the resulting gravitational radiation. One can see from Fig.~\ref{fig:combination2} which part of the gravitational waveform corresponds to which part of the periodic orbit. Furthermore, we compare the gravitational waveforms radiated by the different periodic orbits in Fig.~\ref{fig:GW}. It is worth noting that we consider only one period of the small object around the Schwarzschild BH surrounded by a Dehnen-type DM halo. Therefore, gravitational waveforms are different for every periodic orbit.

\section{Summary}\label{summary}

In this work, we investigated the geodesic motion of test particles around a Schwarzschild black hole embedded in a Dehnen-type dark matter halo. Employing the Lagrangian formalism, we derived the effective potential governing the particle motion. The effective potential clearly shows a slight rightward shift of the curves, suggesting larger radial distances (r), while the potential maximum decreases with increasing $\rho_s$ (DM halo density) and $r_s$ (core radius). This weakening of the gravitational barrier is a direct consequence of the increased DM density $\rho_s$. The properties of marginally bound orbits (MBOs) and innermost stable circular orbits (ISCOs) were determined using the corresponding conditions. Both the radius and orbital angular momentum of MBOs increase with increasing $\rho_s$ and $r_s$. Similarly, for ISCOs, both the radius and orbital angular momentum increase with increasing $\rho_s$ and $r_s$. However, the energy of the ISCO decreases as the DM halo density and core radius increase.

We also investigated the influence of parameters $\rho_s$ and $r_s$ on the relationship between the rational number $q$ and both the energy $E$ and orbital angular momentum $L$. We found that $q$ increases with increasing $E$, exhibiting a rapid rise as $E$ approaches its maximum value. The presence of the DM halo, parameterized by $\rho_s$ and $r_s$, shifts $q$ towards lower energies. Similarly, as $L$ approaches its minimum, $q$ increases dramatically and then gradually decreases with increasing $L$. Again, larger values of $\rho_s$ and $r_s$ shift $q$ towards lower angular momenta. For periodic orbits, we numerically determined the energy and orbital angular momentum for various periodic orbits $(z,w,v)$, presenting the results in tabular form. These tables were then used to illustrate the periodic orbits characterized by the three integers $z$, $w$, and $v$; see Tables \ref{table1} and \ref{table2}.

Finally, we investigated the gravitational waveforms generated by the periodic orbits. We considered an extreme mass ratio inspiral (EMRI) system consisting of a $m \sim 10M_{\odot}$ test object moving along a periodic orbit around a $M \sim 10^7 M_{\odot}$ supermassive Schwarzschild black hole embedded in a Dehnen-type dark matter halo. Using the numerical kludge method, we generated the corresponding gravitational waveforms for the periodic orbits $(3,1,2)$, as seen in Fig.~\ref{fig:combination}. For a more detailed visualization, we showed the waveforms emitted at each stage of the $(1,2,0)$ orbit, as depicted in Fig.~\ref{fig:combination2}. A comparison of waveforms for different periodic orbits is presented in Fig.~\ref{fig:GW}. Note that only one complete period is considered for each orbit, resulting in distinct waveforms for the orbits that are compared.

Given the importance of gravitational waveforms from EMRIs as potential gravitational wave signatures of SMBHs, we emphasize that our findings, while based on a simplified model, have significant astrophysical relevance. This model demonstrates the potential influence of a DM halo on the background spacetime and the resulting gravitational waveforms. As noted previously, we considered only one complete period for each orbit and neglected the impact of gravitational radiation on the orbital parameters of the smaller object. However, this effect is crucial in realistic astrophysical scenarios and will be addressed in our future works, along with an analysis of detectable frequency ranges for ground-based detectors. This study highlights the significant role of the Dehnen-type DM halo properties in shaping GW signals, presenting promising prospects for future observations to help detect and constrain the influence of dark matter. Our theoretical findings may provide key insights for interpreting astrophysical observations and predicting gravitational wave signatures arising from the interaction between BH gravity and the DM halo.

\section*{Acknowledgements}

The research is supported by the National Natural Science Foundation of China under Grant No. W2433018.

\bibliographystyle{apsrev4-1}
\bibliography{ref}

\begin{thebibliography}{69}%
\makeatletter
\providecommand \@ifxundefined [1]{%
 \@ifx{#1\undefined}
}%
\providecommand \@ifnum [1]{%
 \ifnum #1\expandafter \@firstoftwo
 \else \expandafter \@secondoftwo
 \fi
}%
\providecommand \@ifx [1]{%
 \ifx #1\expandafter \@firstoftwo
 \else \expandafter \@secondoftwo
 \fi
}%
\providecommand \natexlab [1]{#1}%
\providecommand \enquote  [1]{``#1''}%
\providecommand \bibnamefont  [1]{#1}%
\providecommand \bibfnamefont [1]{#1}%
\providecommand \citenamefont [1]{#1}%
\providecommand \href@noop [0]{\@secondoftwo}%
\providecommand \href [0]{\begingroup \@sanitize@url \@href}%
\providecommand \@href[1]{\@@startlink{#1}\@@href}%
\providecommand \@@href[1]{\endgroup#1\@@endlink}%
\providecommand \@sanitize@url [0]{\catcode `\\12\catcode `\$12\catcode `\&12\catcode `\#12\catcode `\^12\catcode `\_12\catcode `\%12\relax}%
\providecommand \@@startlink[1]{}%
\providecommand \@@endlink[0]{}%
\providecommand \url  [0]{\begingroup\@sanitize@url \@url }%
\providecommand \@url [1]{\endgroup\@href {#1}{\urlprefix }}%
\providecommand \urlprefix  [0]{URL }%
\providecommand \Eprint [0]{\href }%
\providecommand \doibase [0]{http://dx.doi.org/}%
\providecommand \selectlanguage [0]{\@gobble}%
\providecommand \bibinfo  [0]{\@secondoftwo}%
\providecommand \bibfield  [0]{\@secondoftwo}%
\providecommand \translation [1]{[#1]}%
\providecommand \BibitemOpen [0]{}%
\providecommand \bibitemStop [0]{}%
\providecommand \bibitemNoStop [0]{.\EOS\space}%
\providecommand \EOS [0]{\spacefactor3000\relax}%
\providecommand \BibitemShut  [1]{\csname bibitem#1\endcsname}%
\let\auto@bib@innerbib\@empty
%</preamble>
\bibitem [{\citenamefont {{Schwarzschild}}(1916)}]{1916SPAW.......189S}%
  \BibitemOpen
  \bibfield  {author} {\bibinfo {author} {\bibfnamefont {K.}~\bibnamefont {{Schwarzschild}}},\ }\href@noop {} {\bibfield  {journal} {\bibinfo  {journal} {Sitzungsberichte der K\&ouml;niglich Preussischen Akademie der Wissenschaften}\ ,\ \bibinfo {pages} {189}} (\bibinfo {year} {1916})}\BibitemShut {NoStop}%
\bibitem [{\citenamefont {{Blinder}}(2015)}]{2015arXiv151202061B}%
  \BibitemOpen
  \bibfield  {author} {\bibinfo {author} {\bibfnamefont {S.~M.}\ \bibnamefont {{Blinder}}},\ }\href {\doibase 10.48550/arXiv.1512.02061} {\bibfield  {journal} {\bibinfo  {journal} {arXiv e-prints}\ ,\ \bibinfo {eid} {arXiv:1512.02061}} (\bibinfo {year} {2015})},\ \Eprint {http://arxiv.org/abs/1512.02061} {arXiv:1512.02061 [physics.pop-ph]} \BibitemShut {NoStop}%
\bibitem [{\citenamefont {{Kerr}}(1963)}]{1963PhRvL..11..237K}%
  \BibitemOpen
  \bibfield  {author} {\bibinfo {author} {\bibfnamefont {R.~P.}\ \bibnamefont {{Kerr}}},\ }\href {\doibase 10.1103/PhysRevLett.11.237} {\bibfield  {journal} {\bibinfo  {journal} {Phys. Rev. Lett.}\ }\textbf {\bibinfo {volume} {11}},\ \bibinfo {pages} {237} (\bibinfo {year} {1963})}\BibitemShut {NoStop}%
\bibitem [{\citenamefont {{Rubin}}\ and\ \citenamefont {{Ford}}(1970)}]{Rubin70ApJ}%
  \BibitemOpen
  \bibfield  {author} {\bibinfo {author} {\bibfnamefont {V.~C.}\ \bibnamefont {{Rubin}}}\ and\ \bibinfo {author} {\bibfnamefont {J.}~\bibnamefont {{Ford}}, \bibfnamefont {W.~Kent}},\ }\href {\doibase 10.1086/150317} {\bibfield  {journal} {\bibinfo  {journal} {Astrophys. J.}\ }\textbf {\bibinfo {volume} {159}},\ \bibinfo {pages} {379} (\bibinfo {year} {1970})}\BibitemShut {NoStop}%
\bibitem [{\citenamefont {{Bertone}}\ and\ \citenamefont {{Tait}}(2018)}]{Bertone18Nature}%
  \BibitemOpen
  \bibfield  {author} {\bibinfo {author} {\bibfnamefont {G.}~\bibnamefont {{Bertone}}}\ and\ \bibinfo {author} {\bibfnamefont {T.~M.~P.}\ \bibnamefont {{Tait}}},\ }\href {\doibase 10.1038/s41586-018-0542-z} {\bibfield  {journal} {\bibinfo  {journal} {Nature}\ }\textbf {\bibinfo {volume} {562}},\ \bibinfo {pages} {51} (\bibinfo {year} {2018})},\ \Eprint {http://arxiv.org/abs/1810.01668} {arXiv:1810.01668 [astro-ph.CO]} \BibitemShut {NoStop}%
\bibitem [{\citenamefont {{Corbelli}}\ and\ \citenamefont {{Salucci}}(2000)}]{Corbelli00MNRAS}%
  \BibitemOpen
  \bibfield  {author} {\bibinfo {author} {\bibfnamefont {E.}~\bibnamefont {{Corbelli}}}\ and\ \bibinfo {author} {\bibfnamefont {P.}~\bibnamefont {{Salucci}}},\ }\href {\doibase 10.1046/j.1365-8711.2000.03075.x} {\bibfield  {journal} {\bibinfo  {journal} {Mon. Not. Roy. Astron. Soc.}\ }\textbf {\bibinfo {volume} {311}},\ \bibinfo {pages} {441} (\bibinfo {year} {2000})},\ \Eprint {http://arxiv.org/abs/astro-ph/9909252} {arXiv:astro-ph/9909252 [astro-ph]} \BibitemShut {NoStop}%
\bibitem [{\citenamefont {{Clowe}}\ \emph {et~al.}(2006)\citenamefont {{Clowe}}, \citenamefont {{Brada{\v{c}}}}, \citenamefont {{Gonzalez}}, \citenamefont {{Markevitch}}, \citenamefont {{Randall}}, \citenamefont {{Jones}},\ and\ \citenamefont {{Zaritsky}}}]{Clowe06ApJL}%
  \BibitemOpen
  \bibfield  {author} {\bibinfo {author} {\bibfnamefont {D.}~\bibnamefont {{Clowe}}}, \bibinfo {author} {\bibfnamefont {M.}~\bibnamefont {{Brada{\v{c}}}}}, \bibinfo {author} {\bibfnamefont {A.~H.}\ \bibnamefont {{Gonzalez}}}, \bibinfo {author} {\bibfnamefont {M.}~\bibnamefont {{Markevitch}}}, \bibinfo {author} {\bibfnamefont {S.~W.}\ \bibnamefont {{Randall}}}, \bibinfo {author} {\bibfnamefont {C.}~\bibnamefont {{Jones}}}, \ and\ \bibinfo {author} {\bibfnamefont {D.}~\bibnamefont {{Zaritsky}}},\ }\href {\doibase 10.1086/508162} {\bibfield  {journal} {\bibinfo  {journal} {Astrophys. J. Lett.}\ }\textbf {\bibinfo {volume} {648}},\ \bibinfo {pages} {L109} (\bibinfo {year} {2006})},\ \Eprint {http://arxiv.org/abs/astro-ph/0608407} {arXiv:astro-ph/0608407 [astro-ph]} \BibitemShut {NoStop}%
\bibitem [{\citenamefont {{Bertone}}\ \emph {et~al.}(2005)\citenamefont {{Bertone}}, \citenamefont {{Hooper}},\ and\ \citenamefont {{Silk}}}]{Bertone05}%
  \BibitemOpen
  \bibfield  {author} {\bibinfo {author} {\bibfnamefont {G.}~\bibnamefont {{Bertone}}}, \bibinfo {author} {\bibfnamefont {D.}~\bibnamefont {{Hooper}}}, \ and\ \bibinfo {author} {\bibfnamefont {J.}~\bibnamefont {{Silk}}},\ }\href {\doibase 10.1016/j.physrep.2004.08.031} {\bibfield  {journal} {\bibinfo  {journal} {Phys. Rep.}\ }\textbf {\bibinfo {volume} {405}},\ \bibinfo {pages} {279} (\bibinfo {year} {2005})},\ \Eprint {http://arxiv.org/abs/hep-ph/0404175} {arXiv:hep-ph/0404175 [hep-ph]} \BibitemShut {NoStop}%
\bibitem [{\citenamefont {{de Swart}}\ \emph {et~al.}(2017)\citenamefont {{de Swart}}, \citenamefont {{Bertone}},\ and\ \citenamefont {{van Dongen}}}]{deSwart17Nat}%
  \BibitemOpen
  \bibfield  {author} {\bibinfo {author} {\bibfnamefont {J.~G.}\ \bibnamefont {{de Swart}}}, \bibinfo {author} {\bibfnamefont {G.}~\bibnamefont {{Bertone}}}, \ and\ \bibinfo {author} {\bibfnamefont {J.}~\bibnamefont {{van Dongen}}},\ }\href {\doibase 10.1038/s41550-017-0059} {\bibfield  {journal} {\bibinfo  {journal} {Nature Astron.}\ }\textbf {\bibinfo {volume} {1}},\ \bibinfo {eid} {0059} (\bibinfo {year} {2017})},\ \Eprint {http://arxiv.org/abs/1703.00013} {arXiv:1703.00013 [astro-ph.CO]} \BibitemShut {NoStop}%
\bibitem [{\citenamefont {{Wechsler}}\ and\ \citenamefont {{Tinker}}(2018)}]{Wechsler18}%
  \BibitemOpen
  \bibfield  {author} {\bibinfo {author} {\bibfnamefont {R.~H.}\ \bibnamefont {{Wechsler}}}\ and\ \bibinfo {author} {\bibfnamefont {J.~L.}\ \bibnamefont {{Tinker}}},\ }\href {\doibase 10.1146/annurev-astro-081817-051756} {\bibfield  {journal} {\bibinfo  {journal} {Annu. Rev. Astron. Astrophys.}\ }\textbf {\bibinfo {volume} {56}},\ \bibinfo {pages} {435} (\bibinfo {year} {2018})},\ \Eprint {http://arxiv.org/abs/1804.03097} {arXiv:1804.03097 [astro-ph.GA]} \BibitemShut {NoStop}%
\bibitem [{\citenamefont {{Valluri}}\ \emph {et~al.}(2004)\citenamefont {{Valluri}}, \citenamefont {{Merritt}},\ and\ \citenamefont {{Emsellem}}}]{Valluri04ApJ}%
  \BibitemOpen
  \bibfield  {author} {\bibinfo {author} {\bibfnamefont {M.}~\bibnamefont {{Valluri}}}, \bibinfo {author} {\bibfnamefont {D.}~\bibnamefont {{Merritt}}}, \ and\ \bibinfo {author} {\bibfnamefont {E.}~\bibnamefont {{Emsellem}}},\ }\href {\doibase 10.1086/380896} {\bibfield  {journal} {\bibinfo  {journal} {Astrophys. J.}\ }\textbf {\bibinfo {volume} {602}},\ \bibinfo {pages} {66} (\bibinfo {year} {2004})},\ \Eprint {http://arxiv.org/abs/astro-ph/0210379} {arXiv:astro-ph/0210379 [astro-ph]} \BibitemShut {NoStop}%
\bibitem [{\citenamefont {{Akiyama}}\ and\ \citenamefont {et~al. {(Event Horizon Telescope Collaboration)}}(2019{\natexlab{a}})}]{Akiyama19L1}%
  \BibitemOpen
  \bibfield  {author} {\bibinfo {author} {\bibfnamefont {K.}~\bibnamefont {{Akiyama}}}\ and\ \bibinfo {author} {\bibnamefont {et~al. {(Event Horizon Telescope Collaboration)}}},\ }\href {\doibase 10.3847/2041-8213/ab0ec7} {\bibfield  {journal} {\bibinfo  {journal} {Astrophys. J.}\ }\textbf {\bibinfo {volume} {875}},\ \bibinfo {eid} {L1} (\bibinfo {year} {2019}{\natexlab{a}})},\ \Eprint {http://arxiv.org/abs/1906.11238} {arXiv:1906.11238 [astro-ph.GA]} \BibitemShut {NoStop}%
\bibitem [{\citenamefont {{Akiyama}}\ and\ \citenamefont {et~al. {(Event Horizon Telescope Collaboration)}}(2019{\natexlab{b}})}]{Akiyama19L6}%
  \BibitemOpen
  \bibfield  {author} {\bibinfo {author} {\bibfnamefont {K.}~\bibnamefont {{Akiyama}}}\ and\ \bibinfo {author} {\bibnamefont {et~al. {(Event Horizon Telescope Collaboration)}}},\ }\href {\doibase 10.3847/2041-8213/ab1141} {\bibfield  {journal} {\bibinfo  {journal} {Astrophys. J.}\ }\textbf {\bibinfo {volume} {875}},\ \bibinfo {eid} {L6} (\bibinfo {year} {2019}{\natexlab{b}})},\ \Eprint {http://arxiv.org/abs/1906.11243} {arXiv:1906.11243 [astro-ph.GA]} \BibitemShut {NoStop}%
\bibitem [{\citenamefont {{Akiyama}}\ and\ \citenamefont {et~al. {(Event Horizon Telescope Collaboration)}}(2022)}]{Akiyama22L12}%
  \BibitemOpen
  \bibfield  {author} {\bibinfo {author} {\bibfnamefont {K.}~\bibnamefont {{Akiyama}}}\ and\ \bibinfo {author} {\bibnamefont {et~al. {(Event Horizon Telescope Collaboration)}}},\ }\href {\doibase 10.3847/2041-8213/ac6674} {\bibfield  {journal} {\bibinfo  {journal} {Astrophys. J. Lett.}\ }\textbf {\bibinfo {volume} {930}},\ \bibinfo {eid} {L12} (\bibinfo {year} {2022})}\BibitemShut {NoStop}%
\bibitem [{\citenamefont {{Persic}}\ \emph {et~al.}(1996)\citenamefont {{Persic}}, \citenamefont {{Salucci}},\ and\ \citenamefont {{Stel}}}]{Persic96}%
  \BibitemOpen
  \bibfield  {author} {\bibinfo {author} {\bibfnamefont {M.}~\bibnamefont {{Persic}}}, \bibinfo {author} {\bibfnamefont {P.}~\bibnamefont {{Salucci}}}, \ and\ \bibinfo {author} {\bibfnamefont {F.}~\bibnamefont {{Stel}}},\ }\href {\doibase 10.1093/mnras/278.1.27} {\bibfield  {journal} {\bibinfo  {journal} {Mon. Not. R. Astron. Soc.}\ }\textbf {\bibinfo {volume} {281}},\ \bibinfo {pages} {27} (\bibinfo {year} {1996})},\ \Eprint {http://arxiv.org/abs/astro-ph/9506004} {arXiv:astro-ph/9506004 [astro-ph]} \BibitemShut {NoStop}%
\bibitem [{\citenamefont {Merritt}\ \emph {et~al.}(2006)\citenamefont {Merritt}, \citenamefont {Graham}, \citenamefont {Moore}, \citenamefont {Diemand},\ and\ \citenamefont {Terzić}}]{Merritt_2006}%
  \BibitemOpen
  \bibfield  {author} {\bibinfo {author} {\bibfnamefont {D.}~\bibnamefont {Merritt}}, \bibinfo {author} {\bibfnamefont {A.}~\bibnamefont {Graham}}, \bibinfo {author} {\bibfnamefont {B.}~\bibnamefont {Moore}}, \bibinfo {author} {\bibfnamefont {J.}~\bibnamefont {Diemand}}, \ and\ \bibinfo {author} {\bibfnamefont {B.}~\bibnamefont {Terzić}},\ }\href {\doibase 10.1086/508988} {\bibfield  {journal} {\bibinfo  {journal} {Astron. J.}\ }\textbf {\bibinfo {volume} {132}},\ \bibinfo {pages} {2685–2700} (\bibinfo {year} {2006})}\BibitemShut {NoStop}%
\bibitem [{\citenamefont {Dutton}\ and\ \citenamefont {Macciò}(2014)}]{Dutton_2014}%
  \BibitemOpen
  \bibfield  {author} {\bibinfo {author} {\bibfnamefont {A.~A.}\ \bibnamefont {Dutton}}\ and\ \bibinfo {author} {\bibfnamefont {A.~V.}\ \bibnamefont {Macciò}},\ }\href {\doibase 10.1093/mnras/stu742} {\bibfield  {journal} {\bibinfo  {journal} {Mon. Not. R. Astron. Soc.}\ }\textbf {\bibinfo {volume} {441}},\ \bibinfo {pages} {3359–3374} (\bibinfo {year} {2014})}\BibitemShut {NoStop}%
\bibitem [{\citenamefont {Navarro}\ \emph {et~al.}(1996)\citenamefont {Navarro}, \citenamefont {Frenk},\ and\ \citenamefont {White}}]{Navarro_1996}%
  \BibitemOpen
  \bibfield  {author} {\bibinfo {author} {\bibfnamefont {J.~F.}\ \bibnamefont {Navarro}}, \bibinfo {author} {\bibfnamefont {C.~S.}\ \bibnamefont {Frenk}}, \ and\ \bibinfo {author} {\bibfnamefont {S.~D.~M.}\ \bibnamefont {White}},\ }\href {\doibase 10.1086/177173} {\bibfield  {journal} {\bibinfo  {journal} {Astrophys. J.}\ }\textbf {\bibinfo {volume} {462}},\ \bibinfo {pages} {563} (\bibinfo {year} {1996})}\BibitemShut {NoStop}%
\bibitem [{\citenamefont {Burkert}(1995)}]{Burkert_1995}%
  \BibitemOpen
  \bibfield  {author} {\bibinfo {author} {\bibfnamefont {A.}~\bibnamefont {Burkert}},\ }\href {\doibase 10.1086/309560} {\bibfield  {journal} {\bibinfo  {journal} {Astrophys. J.}\ }\textbf {\bibinfo {volume} {447}} (\bibinfo {year} {1995}),\ 10.1086/309560}\BibitemShut {NoStop}%
\bibitem [{\citenamefont {{Dehnen}}(1993)}]{Dehnen93}%
  \BibitemOpen
  \bibfield  {author} {\bibinfo {author} {\bibfnamefont {W.}~\bibnamefont {{Dehnen}}},\ }\href {\doibase 10.1093/mnras/265.1.250} {\bibfield  {journal} {\bibinfo  {journal} {Mon. Not. R. Astron. Soc.}\ }\textbf {\bibinfo {volume} {265}},\ \bibinfo {pages} {250} (\bibinfo {year} {1993})}\BibitemShut {NoStop}%
\bibitem [{\citenamefont {{Shukirgaliyev, B.}}\ \emph {et~al.}(2021)\citenamefont {{Shukirgaliyev, B.}}, \citenamefont {{Otebay, A.}}, \citenamefont {{Sobolenko, M.}}, \citenamefont {{Ishchenko, M.}}, \citenamefont {{Borodina, O.}}, \citenamefont {{Panamarev, T.}}, \citenamefont {{Myrzakul, S.}}, \citenamefont {{Kalambay, M.}}, \citenamefont {{Naurzbayeva, A.}}, \citenamefont {{Abdikamalov, E.}}, \citenamefont {{Polyachenko, E.}}, \citenamefont {{Banerjee, S.}}, \citenamefont {{Berczik, P.}}, \citenamefont {{Spurzem, R.}},\ and\ \citenamefont {{Just, A.}}}]{refId0}%
  \BibitemOpen
  \bibfield  {author} {\bibinfo {author} {\bibnamefont {{Shukirgaliyev, B.}}}, \bibinfo {author} {\bibnamefont {{Otebay, A.}}}, \bibinfo {author} {\bibnamefont {{Sobolenko, M.}}}, \bibinfo {author} {\bibnamefont {{Ishchenko, M.}}}, \bibinfo {author} {\bibnamefont {{Borodina, O.}}}, \bibinfo {author} {\bibnamefont {{Panamarev, T.}}}, \bibinfo {author} {\bibnamefont {{Myrzakul, S.}}}, \bibinfo {author} {\bibnamefont {{Kalambay, M.}}}, \bibinfo {author} {\bibnamefont {{Naurzbayeva, A.}}}, \bibinfo {author} {\bibnamefont {{Abdikamalov, E.}}}, \bibinfo {author} {\bibnamefont {{Polyachenko, E.}}}, \bibinfo {author} {\bibnamefont {{Banerjee, S.}}}, \bibinfo {author} {\bibnamefont {{Berczik, P.}}}, \bibinfo {author} {\bibnamefont {{Spurzem, R.}}}, \ and\ \bibinfo {author} {\bibnamefont {{Just, A.}}},\ }\href {\doibase 10.1051/0004-6361/202141299} {\bibfield  {journal} {\bibinfo  {journal} {A\&A}\ }\textbf {\bibinfo {volume} {654}},\ \bibinfo {pages} {A53} (\bibinfo {year} {2021})}\BibitemShut {NoStop}%
\bibitem [{\citenamefont {Gohain}\ \emph {et~al.}(2024)\citenamefont {Gohain}, \citenamefont {Phukon},\ and\ \citenamefont {Bhuyan}}]{Gohain_2024}%
  \BibitemOpen
  \bibfield  {author} {\bibinfo {author} {\bibfnamefont {M.~M.}\ \bibnamefont {Gohain}}, \bibinfo {author} {\bibfnamefont {P.}~\bibnamefont {Phukon}}, \ and\ \bibinfo {author} {\bibfnamefont {K.}~\bibnamefont {Bhuyan}},\ }\href {\doibase 10.1016/j.dark.2024.101683} {\bibfield  {journal} {\bibinfo  {journal} {Phys. Dark Universe}\ }\textbf {\bibinfo {volume} {46}},\ \bibinfo {pages} {101683} (\bibinfo {year} {2024})}\BibitemShut {NoStop}%
\bibitem [{\citenamefont {Pantig}\ and\ \citenamefont {Övgün}(2022)}]{Pantig_2022}%
  \BibitemOpen
  \bibfield  {author} {\bibinfo {author} {\bibfnamefont {R.~C.}\ \bibnamefont {Pantig}}\ and\ \bibinfo {author} {\bibfnamefont {A.}~\bibnamefont {Övgün}},\ }\href {\doibase 10.1088/1475-7516/2022/08/056} {\bibfield  {journal} {\bibinfo  {journal} {J. Cosmol. Astropart. Phys.}\ }\textbf {\bibinfo {volume} {2022}},\ \bibinfo {pages} {056} (\bibinfo {year} {2022})}\BibitemShut {NoStop}%
\bibitem [{\citenamefont {{Li}}\ and\ \citenamefont {{Yang}}(2012)}]{Li-Yang12}%
  \BibitemOpen
  \bibfield  {author} {\bibinfo {author} {\bibfnamefont {M.-H.}\ \bibnamefont {{Li}}}\ and\ \bibinfo {author} {\bibfnamefont {K.-C.}\ \bibnamefont {{Yang}}},\ }\href {\doibase 10.1103/PhysRevD.86.123015} {\bibfield  {journal} {\bibinfo  {journal} {Phys. Rev. D}\ }\textbf {\bibinfo {volume} {86}},\ \bibinfo {eid} {123015} (\bibinfo {year} {2012})},\ \Eprint {http://arxiv.org/abs/1204.3178} {arXiv:1204.3178 [astro-ph.CO]} \BibitemShut {NoStop}%
\bibitem [{\citenamefont {{Shaymatov}}\ \emph {et~al.}(2021{\natexlab{a}})\citenamefont {{Shaymatov}}, \citenamefont {{Ahmedov}},\ and\ \citenamefont {{Jamil}}}]{Shaymatov21d}%
  \BibitemOpen
  \bibfield  {author} {\bibinfo {author} {\bibfnamefont {S.}~\bibnamefont {{Shaymatov}}}, \bibinfo {author} {\bibfnamefont {B.}~\bibnamefont {{Ahmedov}}}, \ and\ \bibinfo {author} {\bibfnamefont {M.}~\bibnamefont {{Jamil}}},\ }\href {\doibase 10.1140/epjc/s10052-021-09398-w} {\bibfield  {journal} {\bibinfo  {journal} {Eur. Phys. J. C}\ }\textbf {\bibinfo {volume} {81}},\ \bibinfo {eid} {588} (\bibinfo {year} {2021}{\natexlab{a}})}\BibitemShut {NoStop}%
\bibitem [{\citenamefont {{Shaymatov}}\ \emph {et~al.}(2021{\natexlab{b}})\citenamefont {{Shaymatov}}, \citenamefont {{Malafarina}},\ and\ \citenamefont {{Ahmedov}}}]{Shaymatov21pdu}%
  \BibitemOpen
  \bibfield  {author} {\bibinfo {author} {\bibfnamefont {S.}~\bibnamefont {{Shaymatov}}}, \bibinfo {author} {\bibfnamefont {D.}~\bibnamefont {{Malafarina}}}, \ and\ \bibinfo {author} {\bibfnamefont {B.}~\bibnamefont {{Ahmedov}}},\ }\href {\doibase 10.1016/j.dark.2021.100891} {\bibfield  {journal} {\bibinfo  {journal} {Phys. Dark Universe}\ }\textbf {\bibinfo {volume} {34}},\ \bibinfo {eid} {100891} (\bibinfo {year} {2021}{\natexlab{b}})},\ \Eprint {http://arxiv.org/abs/2004.06811} {arXiv:2004.06811 [gr-qc]} \BibitemShut {NoStop}%
\bibitem [{\citenamefont {{Shaymatov}}\ \emph {et~al.}(2022)\citenamefont {{Shaymatov}}, \citenamefont {{Sheoran}},\ and\ \citenamefont {{Siwach}}}]{Shaymatov22a}%
  \BibitemOpen
  \bibfield  {author} {\bibinfo {author} {\bibfnamefont {S.}~\bibnamefont {{Shaymatov}}}, \bibinfo {author} {\bibfnamefont {P.}~\bibnamefont {{Sheoran}}}, \ and\ \bibinfo {author} {\bibfnamefont {S.}~\bibnamefont {{Siwach}}},\ }\href {\doibase 10.1103/PhysRevD.105.104059} {\bibfield  {journal} {\bibinfo  {journal} {Phys. Rev. D}\ }\textbf {\bibinfo {volume} {105}},\ \bibinfo {eid} {104059} (\bibinfo {year} {2022})},\ \Eprint {http://arxiv.org/abs/2110.10610} {arXiv:2110.10610 [gr-qc]} \BibitemShut {NoStop}%
\bibitem [{\citenamefont {{Cardoso}}\ \emph {et~al.}(2022{\natexlab{a}})\citenamefont {{Cardoso}}, \citenamefont {{Destounis}}, \citenamefont {{Duque}}, \citenamefont {{Macedo}},\ and\ \citenamefont {{Maselli}}}]{Cardoso22DM}%
  \BibitemOpen
  \bibfield  {author} {\bibinfo {author} {\bibfnamefont {V.}~\bibnamefont {{Cardoso}}}, \bibinfo {author} {\bibfnamefont {K.}~\bibnamefont {{Destounis}}}, \bibinfo {author} {\bibfnamefont {F.}~\bibnamefont {{Duque}}}, \bibinfo {author} {\bibfnamefont {R.~P.}\ \bibnamefont {{Macedo}}}, \ and\ \bibinfo {author} {\bibfnamefont {A.}~\bibnamefont {{Maselli}}},\ }\href {\doibase 10.1103/PhysRevD.105.L061501} {\bibfield  {journal} {\bibinfo  {journal} {Phys. Rev. D}\ }\textbf {\bibinfo {volume} {105}},\ \bibinfo {eid} {L061501} (\bibinfo {year} {2022}{\natexlab{a}})},\ \Eprint {http://arxiv.org/abs/2109.00005} {arXiv:2109.00005 [gr-qc]} \BibitemShut {NoStop}%
\bibitem [{\citenamefont {{Hou}}\ \emph {et~al.}(2018)\citenamefont {{Hou}}, \citenamefont {{Xu}}, \citenamefont {{Zhou}},\ and\ \citenamefont {{Wang}}}]{Hou18-dm}%
  \BibitemOpen
  \bibfield  {author} {\bibinfo {author} {\bibfnamefont {X.}~\bibnamefont {{Hou}}}, \bibinfo {author} {\bibfnamefont {Z.}~\bibnamefont {{Xu}}}, \bibinfo {author} {\bibfnamefont {M.}~\bibnamefont {{Zhou}}}, \ and\ \bibinfo {author} {\bibfnamefont {J.}~\bibnamefont {{Wang}}},\ }\href {\doibase 10.1088/1475-7516/2018/07/015} {\bibfield  {journal} {\bibinfo  {journal} {J. Cosmol. Astropart. Phys.}\ }\textbf {\bibinfo {volume} {2018}},\ \bibinfo {eid} {015} (\bibinfo {year} {2018})},\ \Eprint {http://arxiv.org/abs/1804.08110} {arXiv:1804.08110 [gr-qc]} \BibitemShut {NoStop}%
\bibitem [{\citenamefont {{Shen}}\ \emph {et~al.}(2024)\citenamefont {{Shen}}, \citenamefont {{Wang}}, \citenamefont {{Gong}},\ and\ \citenamefont {{Yin}}}]{Shen24PLB}%
  \BibitemOpen
  \bibfield  {author} {\bibinfo {author} {\bibfnamefont {Z.}~\bibnamefont {{Shen}}}, \bibinfo {author} {\bibfnamefont {A.}~\bibnamefont {{Wang}}}, \bibinfo {author} {\bibfnamefont {Y.}~\bibnamefont {{Gong}}}, \ and\ \bibinfo {author} {\bibfnamefont {S.}~\bibnamefont {{Yin}}},\ }\href {\doibase 10.1016/j.physletb.2024.138797} {\bibfield  {journal} {\bibinfo  {journal} {Phys. Lett. B}\ }\textbf {\bibinfo {volume} {855}},\ \bibinfo {eid} {138797} (\bibinfo {year} {2024})},\ \Eprint {http://arxiv.org/abs/2311.12259} {arXiv:2311.12259 [gr-qc]} \BibitemShut {NoStop}%
\bibitem [{\citenamefont {{Abbott}}\ and\ \citenamefont {et~al. {(Virgo and LIGO Scientific Collaborations)}}(2016{\natexlab{a}})}]{Abbott16a}%
  \BibitemOpen
  \bibfield  {author} {\bibinfo {author} {\bibfnamefont {B.~P.}\ \bibnamefont {{Abbott}}}\ and\ \bibinfo {author} {\bibnamefont {et~al. {(Virgo and LIGO Scientific Collaborations)}}},\ }\href {\doibase 10.1103/PhysRevLett.116.061102} {\bibfield  {journal} {\bibinfo  {journal} {Phys. Rev. Lett.}\ }\textbf {\bibinfo {volume} {116}},\ \bibinfo {eid} {061102} (\bibinfo {year} {2016}{\natexlab{a}})},\ \Eprint {http://arxiv.org/abs/1602.03837} {arXiv:1602.03837 [gr-qc]} \BibitemShut {NoStop}%
\bibitem [{\citenamefont {{Abbott}}\ and\ \citenamefont {et~al. {(Virgo and LIGO Scientific Collaborations)}}(2016{\natexlab{b}})}]{Abbott16b}%
  \BibitemOpen
  \bibfield  {author} {\bibinfo {author} {\bibfnamefont {B.~P.}\ \bibnamefont {{Abbott}}}\ and\ \bibinfo {author} {\bibnamefont {et~al. {(Virgo and LIGO Scientific Collaborations)}}},\ }\href {\doibase 10.1103/PhysRevLett.116.241102} {\bibfield  {journal} {\bibinfo  {journal} {Phys. Rev. Lett.}\ }\textbf {\bibinfo {volume} {116}},\ \bibinfo {eid} {241102} (\bibinfo {year} {2016}{\natexlab{b}})},\ \Eprint {http://arxiv.org/abs/1602.03840} {arXiv:1602.03840 [gr-qc]} \BibitemShut {NoStop}%
\bibitem [{\citenamefont {{Amaro-Seoane}}\ and\ \citenamefont {et~al. {(Laser Interferometer Space Antenna)}}(2017)}]{Amaro-Seoane2017LISA}%
  \BibitemOpen
  \bibfield  {author} {\bibinfo {author} {\bibfnamefont {P.}~\bibnamefont {{Amaro-Seoane}}}\ and\ \bibinfo {author} {\bibnamefont {et~al. {(Laser Interferometer Space Antenna)}}},\ }\href {https://arxiv.org/abs/1702.00786} {\enquote {\bibinfo {title} {Laser interferometer space antenna},}\ } (\bibinfo {year} {2017}),\ \Eprint {http://arxiv.org/abs/1702.00786} {arXiv:1702.00786 [astro-ph.IM]} \BibitemShut {NoStop}%
\bibitem [{\citenamefont {Hu}\ and\ \citenamefont {Wu}(2017)}]{10.1093/nsr/nwx116}%
  \BibitemOpen
  \bibfield  {author} {\bibinfo {author} {\bibfnamefont {W.-R.}\ \bibnamefont {Hu}}\ and\ \bibinfo {author} {\bibfnamefont {Y.-L.}\ \bibnamefont {Wu}},\ }\href {\doibase 10.1093/nsr/nwx116} {\bibfield  {journal} {\bibinfo  {journal} {Natl. Sci. Rev.}\ }\textbf {\bibinfo {volume} {4}},\ \bibinfo {pages} {685} (\bibinfo {year} {2017})},\ \Eprint {http://arxiv.org/abs/https://academic.oup.com/nsr/article-pdf/4/5/685/31566708/nwx116.pdf} {https://academic.oup.com/nsr/article-pdf/4/5/685/31566708/nwx116.pdf} \BibitemShut {NoStop}%
\bibitem [{\citenamefont {Hughes}(2001)}]{Hughes_2001}%
  \BibitemOpen
  \bibfield  {author} {\bibinfo {author} {\bibfnamefont {S.~A.}\ \bibnamefont {Hughes}},\ }\href {\doibase 10.1088/0264-9381/18/19/314} {\bibfield  {journal} {\bibinfo  {journal} {Class. Quantum Gravity}\ }\textbf {\bibinfo {volume} {18}},\ \bibinfo {pages} {4067–4073} (\bibinfo {year} {2001})}\BibitemShut {NoStop}%
\bibitem [{\citenamefont {{Amaro-Seoane}}(2018)}]{Amaro-Seoane18LRR}%
  \BibitemOpen
  \bibfield  {author} {\bibinfo {author} {\bibfnamefont {P.}~\bibnamefont {{Amaro-Seoane}}},\ }\href {\doibase 10.1007/s41114-018-0013-8} {\bibfield  {journal} {\bibinfo  {journal} {Living Rev. Rel.}\ }\textbf {\bibinfo {volume} {21}},\ \bibinfo {eid} {4} (\bibinfo {year} {2018})},\ \Eprint {http://arxiv.org/abs/1205.5240} {arXiv:1205.5240 [astro-ph.CO]} \BibitemShut {NoStop}%
\bibitem [{\citenamefont {{Babak}}\ \emph {et~al.}(2017)\citenamefont {{Babak}}, \citenamefont {{Gair}}, \citenamefont {{Sesana}}, \citenamefont {{Barausse}}, \citenamefont {{Sopuerta}}, \citenamefont {{Berry}}, \citenamefont {{Berti}}, \citenamefont {{Amaro-Seoane}}, \citenamefont {{Petiteau}},\ and\ \citenamefont {{Klein}}}]{Babak17PRD}%
  \BibitemOpen
  \bibfield  {author} {\bibinfo {author} {\bibfnamefont {S.}~\bibnamefont {{Babak}}}, \bibinfo {author} {\bibfnamefont {J.}~\bibnamefont {{Gair}}}, \bibinfo {author} {\bibfnamefont {A.}~\bibnamefont {{Sesana}}}, \bibinfo {author} {\bibfnamefont {E.}~\bibnamefont {{Barausse}}}, \bibinfo {author} {\bibfnamefont {C.~F.}\ \bibnamefont {{Sopuerta}}}, \bibinfo {author} {\bibfnamefont {C.~P.~L.}\ \bibnamefont {{Berry}}}, \bibinfo {author} {\bibfnamefont {E.}~\bibnamefont {{Berti}}}, \bibinfo {author} {\bibfnamefont {P.}~\bibnamefont {{Amaro-Seoane}}}, \bibinfo {author} {\bibfnamefont {A.}~\bibnamefont {{Petiteau}}}, \ and\ \bibinfo {author} {\bibfnamefont {A.}~\bibnamefont {{Klein}}},\ }\href {\doibase 10.1103/PhysRevD.95.103012} {\bibfield  {journal} {\bibinfo  {journal} {Phys. Rev. D}\ }\textbf {\bibinfo {volume} {95}},\ \bibinfo {eid} {103012} (\bibinfo {year} {2017})},\ \Eprint {http://arxiv.org/abs/1703.09722} {arXiv:1703.09722 [gr-qc]} \BibitemShut {NoStop}%
\bibitem [{\citenamefont {{Yue}}\ \emph {et~al.}(2019)\citenamefont {{Yue}}, \citenamefont {{Han}},\ and\ \citenamefont {{Chen}}}]{Yue19ApJ}%
  \BibitemOpen
  \bibfield  {author} {\bibinfo {author} {\bibfnamefont {X.-J.}\ \bibnamefont {{Yue}}}, \bibinfo {author} {\bibfnamefont {W.-B.}\ \bibnamefont {{Han}}}, \ and\ \bibinfo {author} {\bibfnamefont {X.}~\bibnamefont {{Chen}}},\ }\href {\doibase 10.3847/1538-4357/ab06f6} {\bibfield  {journal} {\bibinfo  {journal} {Astrophys. J.}\ }\textbf {\bibinfo {volume} {874}},\ \bibinfo {eid} {34} (\bibinfo {year} {2019})},\ \Eprint {http://arxiv.org/abs/1802.03739} {arXiv:1802.03739 [gr-qc]} \BibitemShut {NoStop}%
\bibitem [{\citenamefont {{Duque}}\ \emph {et~al.}(2024)\citenamefont {{Duque}}, \citenamefont {{Macedo}}, \citenamefont {{Vicente}},\ and\ \citenamefont {{Cardoso}}}]{Duque24PRL}%
  \BibitemOpen
  \bibfield  {author} {\bibinfo {author} {\bibfnamefont {F.}~\bibnamefont {{Duque}}}, \bibinfo {author} {\bibfnamefont {C.~F.~B.}\ \bibnamefont {{Macedo}}}, \bibinfo {author} {\bibfnamefont {R.}~\bibnamefont {{Vicente}}}, \ and\ \bibinfo {author} {\bibfnamefont {V.}~\bibnamefont {{Cardoso}}},\ }\href {\doibase 10.1103/PhysRevLett.133.121404} {\bibfield  {journal} {\bibinfo  {journal} {Phys. Rev. Lett.}\ }\textbf {\bibinfo {volume} {133}},\ \bibinfo {eid} {121404} (\bibinfo {year} {2024})},\ \Eprint {http://arxiv.org/abs/2312.06767} {arXiv:2312.06767 [gr-qc]} \BibitemShut {NoStop}%
\bibitem [{\citenamefont {{Dai}}\ \emph {et~al.}(2024)\citenamefont {{Dai}}, \citenamefont {{Gong}}, \citenamefont {{Zhao}},\ and\ \citenamefont {{Jiang}}}]{Dai24PRD}%
  \BibitemOpen
  \bibfield  {author} {\bibinfo {author} {\bibfnamefont {N.}~\bibnamefont {{Dai}}}, \bibinfo {author} {\bibfnamefont {Y.}~\bibnamefont {{Gong}}}, \bibinfo {author} {\bibfnamefont {Y.}~\bibnamefont {{Zhao}}}, \ and\ \bibinfo {author} {\bibfnamefont {T.}~\bibnamefont {{Jiang}}},\ }\href {\doibase 10.1103/PhysRevD.110.084080} {\bibfield  {journal} {\bibinfo  {journal} {Phys. Rev. D}\ }\textbf {\bibinfo {volume} {110}},\ \bibinfo {eid} {084080} (\bibinfo {year} {2024})},\ \Eprint {http://arxiv.org/abs/2301.05088} {arXiv:2301.05088 [gr-qc]} \BibitemShut {NoStop}%
\bibitem [{\citenamefont {Levin}\ and\ \citenamefont {Perez-Giz}(2008)}]{Levin_2008}%
  \BibitemOpen
  \bibfield  {author} {\bibinfo {author} {\bibfnamefont {J.}~\bibnamefont {Levin}}\ and\ \bibinfo {author} {\bibfnamefont {G.}~\bibnamefont {Perez-Giz}},\ }\href {\doibase 10.1103/physrevd.77.103005} {\bibfield  {journal} {\bibinfo  {journal} {Phys. Rev. D}\ }\textbf {\bibinfo {volume} {77}} (\bibinfo {year} {2008}),\ 10.1103/physrevd.77.103005}\BibitemShut {NoStop}%
\bibitem [{\citenamefont {Grossman}\ and\ \citenamefont {Levin}(2009)}]{Grossman_2009}%
  \BibitemOpen
  \bibfield  {author} {\bibinfo {author} {\bibfnamefont {R.}~\bibnamefont {Grossman}}\ and\ \bibinfo {author} {\bibfnamefont {J.}~\bibnamefont {Levin}},\ }\href {\doibase 10.1103/physrevd.79.043017} {\bibfield  {journal} {\bibinfo  {journal} {Phys. Rev. D}\ }\textbf {\bibinfo {volume} {79}} (\bibinfo {year} {2009}),\ 10.1103/physrevd.79.043017}\BibitemShut {NoStop}%
\bibitem [{\citenamefont {Misra}\ and\ \citenamefont {Levin}(2010{\natexlab{a}})}]{Misra_2010}%
  \BibitemOpen
  \bibfield  {author} {\bibinfo {author} {\bibfnamefont {V.}~\bibnamefont {Misra}}\ and\ \bibinfo {author} {\bibfnamefont {J.}~\bibnamefont {Levin}},\ }\href {\doibase 10.1103/physrevd.82.083001} {\bibfield  {journal} {\bibinfo  {journal} {Phys. Rev. D}\ }\textbf {\bibinfo {volume} {82}} (\bibinfo {year} {2010}{\natexlab{a}}),\ 10.1103/physrevd.82.083001}\BibitemShut {NoStop}%
\bibitem [{\citenamefont {Levin}(2009)}]{Levin_2009}%
  \BibitemOpen
  \bibfield  {author} {\bibinfo {author} {\bibfnamefont {J.}~\bibnamefont {Levin}},\ }\href {\doibase 10.1088/0264-9381/26/23/235010} {\bibfield  {journal} {\bibinfo  {journal} {Class. Quantum Gravity}\ }\textbf {\bibinfo {volume} {26}},\ \bibinfo {pages} {235010} (\bibinfo {year} {2009})}\BibitemShut {NoStop}%
\bibitem [{\citenamefont {{Glampedakis}}\ and\ \citenamefont {{Kennefick}}(2002)}]{Glampedakis02PRD}%
  \BibitemOpen
  \bibfield  {author} {\bibinfo {author} {\bibfnamefont {K.}~\bibnamefont {{Glampedakis}}}\ and\ \bibinfo {author} {\bibfnamefont {D.}~\bibnamefont {{Kennefick}}},\ }\href {\doibase 10.1103/PhysRevD.66.044002} {\bibfield  {journal} {\bibinfo  {journal} {Phys. Rev. D}\ }\textbf {\bibinfo {volume} {66}},\ \bibinfo {eid} {044002} (\bibinfo {year} {2002})},\ \Eprint {http://arxiv.org/abs/gr-qc/0203086} {arXiv:gr-qc/0203086 [gr-qc]} \BibitemShut {NoStop}%
\bibitem [{\citenamefont {Misra}\ and\ \citenamefont {Levin}(2010{\natexlab{b}})}]{Levin2010}%
  \BibitemOpen
  \bibfield  {author} {\bibinfo {author} {\bibfnamefont {V.}~\bibnamefont {Misra}}\ and\ \bibinfo {author} {\bibfnamefont {J.}~\bibnamefont {Levin}},\ }\href {\doibase 10.1103/PhysRevD.82.083001} {\bibfield  {journal} {\bibinfo  {journal} {Phys. Rev. D}\ }\textbf {\bibinfo {volume} {82}},\ \bibinfo {pages} {083001} (\bibinfo {year} {2010}{\natexlab{b}})}\BibitemShut {NoStop}%
\bibitem [{\citenamefont {{Babar}}\ \emph {et~al.}(2017)\citenamefont {{Babar}}, \citenamefont {{Babar}},\ and\ \citenamefont {{Lim}}}]{Babar17PRD}%
  \BibitemOpen
  \bibfield  {author} {\bibinfo {author} {\bibfnamefont {G.~Z.}\ \bibnamefont {{Babar}}}, \bibinfo {author} {\bibfnamefont {A.~Z.}\ \bibnamefont {{Babar}}}, \ and\ \bibinfo {author} {\bibfnamefont {Y.-K.}\ \bibnamefont {{Lim}}},\ }\href {\doibase 10.1103/PhysRevD.96.084052} {\bibfield  {journal} {\bibinfo  {journal} {Phys. Rev. D}\ }\textbf {\bibinfo {volume} {96}},\ \bibinfo {eid} {084052} (\bibinfo {year} {2017})},\ \Eprint {http://arxiv.org/abs/1710.09581} {arXiv:1710.09581 [gr-qc]} \BibitemShut {NoStop}%
\bibitem [{\citenamefont {Liu}\ \emph {et~al.}(2019)\citenamefont {Liu}, \citenamefont {Ding},\ and\ \citenamefont {Jing}}]{Liu_2019}%
  \BibitemOpen
  \bibfield  {author} {\bibinfo {author} {\bibfnamefont {C.-Q.}\ \bibnamefont {Liu}}, \bibinfo {author} {\bibfnamefont {C.-K.}\ \bibnamefont {Ding}}, \ and\ \bibinfo {author} {\bibfnamefont {J.-L.}\ \bibnamefont {Jing}},\ }\href {\doibase 10.1088/0253-6102/71/12/1461} {\bibfield  {journal} {\bibinfo  {journal} {Commun. Theor. Phys.}\ }\textbf {\bibinfo {volume} {71}},\ \bibinfo {pages} {1461} (\bibinfo {year} {2019})}\BibitemShut {NoStop}%
\bibitem [{\citenamefont {Tu}\ \emph {et~al.}(2023)\citenamefont {Tu}, \citenamefont {Zhu},\ and\ \citenamefont {Wang}}]{Tu23}%
  \BibitemOpen
  \bibfield  {author} {\bibinfo {author} {\bibfnamefont {Z.-Y.}\ \bibnamefont {Tu}}, \bibinfo {author} {\bibfnamefont {T.}~\bibnamefont {Zhu}}, \ and\ \bibinfo {author} {\bibfnamefont {A.}~\bibnamefont {Wang}},\ }\href {\doibase 10.1103/PhysRevD.108.024035} {\bibfield  {journal} {\bibinfo  {journal} {Phys. Rev. D}\ }\textbf {\bibinfo {volume} {108}},\ \bibinfo {pages} {024035} (\bibinfo {year} {2023})}\BibitemShut {NoStop}%
\bibitem [{\citenamefont {Azreg-A\"{\i}nou}\ \emph {et~al.}(2020)\citenamefont {Azreg-A\"{\i}nou}, \citenamefont {Chen}, \citenamefont {Deng}, \citenamefont {Jamil}, \citenamefont {Zhu}, \citenamefont {Wu},\ and\ \citenamefont {Lim}}]{Mustapha2020}%
  \BibitemOpen
  \bibfield  {author} {\bibinfo {author} {\bibfnamefont {M.}~\bibnamefont {Azreg-A\"{\i}nou}}, \bibinfo {author} {\bibfnamefont {Z.}~\bibnamefont {Chen}}, \bibinfo {author} {\bibfnamefont {B.}~\bibnamefont {Deng}}, \bibinfo {author} {\bibfnamefont {M.}~\bibnamefont {Jamil}}, \bibinfo {author} {\bibfnamefont {T.}~\bibnamefont {Zhu}}, \bibinfo {author} {\bibfnamefont {Q.}~\bibnamefont {Wu}}, \ and\ \bibinfo {author} {\bibfnamefont {Y.-K.}\ \bibnamefont {Lim}},\ }\href {\doibase 10.1103/PhysRevD.102.044028} {\bibfield  {journal} {\bibinfo  {journal} {Phys. Rev. D}\ }\textbf {\bibinfo {volume} {102}},\ \bibinfo {pages} {044028} (\bibinfo {year} {2020})}\BibitemShut {NoStop}%
\bibitem [{\citenamefont {Wei}\ \emph {et~al.}(2019)\citenamefont {Wei}, \citenamefont {Yang},\ and\ \citenamefont {Liu}}]{wei2019}%
  \BibitemOpen
  \bibfield  {author} {\bibinfo {author} {\bibfnamefont {S.-W.}\ \bibnamefont {Wei}}, \bibinfo {author} {\bibfnamefont {J.}~\bibnamefont {Yang}}, \ and\ \bibinfo {author} {\bibfnamefont {Y.-X.}\ \bibnamefont {Liu}},\ }\href {\doibase 10.1103/PhysRevD.99.104016} {\bibfield  {journal} {\bibinfo  {journal} {Phys. Rev. D}\ }\textbf {\bibinfo {volume} {99}},\ \bibinfo {pages} {104016} (\bibinfo {year} {2019})}\BibitemShut {NoStop}%
\bibitem [{\citenamefont {{Deng}}(2020)}]{Deng20}%
  \BibitemOpen
  \bibfield  {author} {\bibinfo {author} {\bibfnamefont {X.-M.}\ \bibnamefont {{Deng}}},\ }\href {\doibase 10.1016/j.dark.2020.100629} {\bibfield  {journal} {\bibinfo  {journal} {Phys. Dark Universe}\ }\textbf {\bibinfo {volume} {30}},\ \bibinfo {eid} {100629} (\bibinfo {year} {2020})}\BibitemShut {NoStop}%
\bibitem [{\citenamefont {Yang}\ \emph {et~al.}(2024)\citenamefont {Yang}, \citenamefont {Zhang}, \citenamefont {Zhu}, \citenamefont {Zhao},\ and\ \citenamefont {Liu}}]{yang2024}%
  \BibitemOpen
  \bibfield  {author} {\bibinfo {author} {\bibfnamefont {S.}~\bibnamefont {Yang}}, \bibinfo {author} {\bibfnamefont {Y.-P.}\ \bibnamefont {Zhang}}, \bibinfo {author} {\bibfnamefont {T.}~\bibnamefont {Zhu}}, \bibinfo {author} {\bibfnamefont {L.}~\bibnamefont {Zhao}}, \ and\ \bibinfo {author} {\bibfnamefont {Y.-X.}\ \bibnamefont {Liu}},\ }\href {https://arxiv.org/abs/2407.00283} {\enquote {\bibinfo {title} {Gravitational waveforms from periodic orbits around a quantum-corrected black hole},}\ } (\bibinfo {year} {2024}),\ \Eprint {http://arxiv.org/abs/2407.00283} {arXiv:2407.00283 [gr-qc]} \BibitemShut {NoStop}%
\bibitem [{\citenamefont {{Sadeghian}}\ \emph {et~al.}(2013)\citenamefont {{Sadeghian}}, \citenamefont {{Ferrer}},\ and\ \citenamefont {{Will}}}]{Sadeghian13PRD}%
  \BibitemOpen
  \bibfield  {author} {\bibinfo {author} {\bibfnamefont {L.}~\bibnamefont {{Sadeghian}}}, \bibinfo {author} {\bibfnamefont {F.}~\bibnamefont {{Ferrer}}}, \ and\ \bibinfo {author} {\bibfnamefont {C.~M.}\ \bibnamefont {{Will}}},\ }\href {\doibase 10.1103/PhysRevD.88.063522} {\bibfield  {journal} {\bibinfo  {journal} {Phys. Rev. D}\ }\textbf {\bibinfo {volume} {88}},\ \bibinfo {eid} {063522} (\bibinfo {year} {2013})},\ \Eprint {http://arxiv.org/abs/1305.2619} {arXiv:1305.2619 [astro-ph.GA]} \BibitemShut {NoStop}%
\bibitem [{\citenamefont {{Barausse}}\ \emph {et~al.}(2014)\citenamefont {{Barausse}}, \citenamefont {{Cardoso}},\ and\ \citenamefont {{Pani}}}]{Barausse14PRD}%
  \BibitemOpen
  \bibfield  {author} {\bibinfo {author} {\bibfnamefont {E.}~\bibnamefont {{Barausse}}}, \bibinfo {author} {\bibfnamefont {V.}~\bibnamefont {{Cardoso}}}, \ and\ \bibinfo {author} {\bibfnamefont {P.}~\bibnamefont {{Pani}}},\ }\href {\doibase 10.1103/PhysRevD.89.104059} {\bibfield  {journal} {\bibinfo  {journal} {Phys. Rev. D}\ }\textbf {\bibinfo {volume} {89}},\ \bibinfo {eid} {104059} (\bibinfo {year} {2014})},\ \Eprint {http://arxiv.org/abs/1404.7149} {arXiv:1404.7149 [gr-qc]} \BibitemShut {NoStop}%
\bibitem [{\citenamefont {{Cardoso}}\ \emph {et~al.}(2022{\natexlab{b}})\citenamefont {{Cardoso}}, \citenamefont {{Destounis}}, \citenamefont {{Duque}}, \citenamefont {{Macedo}},\ and\ \citenamefont {{Maselli}}}]{Cardoso22PRD}%
  \BibitemOpen
  \bibfield  {author} {\bibinfo {author} {\bibfnamefont {V.}~\bibnamefont {{Cardoso}}}, \bibinfo {author} {\bibfnamefont {K.}~\bibnamefont {{Destounis}}}, \bibinfo {author} {\bibfnamefont {F.}~\bibnamefont {{Duque}}}, \bibinfo {author} {\bibfnamefont {R.~P.}\ \bibnamefont {{Macedo}}}, \ and\ \bibinfo {author} {\bibfnamefont {A.}~\bibnamefont {{Maselli}}},\ }\href {\doibase 10.1103/PhysRevD.105.L061501} {\bibfield  {journal} {\bibinfo  {journal} {Phys. Rev. D}\ }\textbf {\bibinfo {volume} {105}},\ \bibinfo {eid} {L061501} (\bibinfo {year} {2022}{\natexlab{b}})},\ \Eprint {http://arxiv.org/abs/2109.00005} {arXiv:2109.00005 [gr-qc]} \BibitemShut {NoStop}%
\bibitem [{\citenamefont {Haroon}\ and\ \citenamefont {Zhu}(2025)}]{Haroon2025}%
  \BibitemOpen
  \bibfield  {author} {\bibinfo {author} {\bibfnamefont {S.}~\bibnamefont {Haroon}}\ and\ \bibinfo {author} {\bibfnamefont {T.}~\bibnamefont {Zhu}},\ }\href {https://arxiv.org/abs/2502.09171} {\enquote {\bibinfo {title} {Periodic orbits and their gravitational wave radiations in black hole with dark matter halo},}\ } (\bibinfo {year} {2025}),\ \Eprint {http://arxiv.org/abs/2502.09171} {arXiv:2502.09171 [gr-qc]} \BibitemShut {NoStop}%
\bibitem [{\citenamefont {{Navarro}}\ \emph {et~al.}(1996)\citenamefont {{Navarro}}, \citenamefont {{Frenk}},\ and\ \citenamefont {{White}}}]{Navarro96ApJ}%
  \BibitemOpen
  \bibfield  {author} {\bibinfo {author} {\bibfnamefont {J.~F.}\ \bibnamefont {{Navarro}}}, \bibinfo {author} {\bibfnamefont {C.~S.}\ \bibnamefont {{Frenk}}}, \ and\ \bibinfo {author} {\bibfnamefont {S.~D.~M.}\ \bibnamefont {{White}}},\ }\href {\doibase 10.1086/177173} {\bibfield  {journal} {\bibinfo  {journal} {Astrophys. J.}\ }\textbf {\bibinfo {volume} {462}},\ \bibinfo {pages} {563} (\bibinfo {year} {1996})},\ \Eprint {http://arxiv.org/abs/astro-ph/9508025} {arXiv:astro-ph/9508025 [astro-ph]} \BibitemShut {NoStop}%
\bibitem [{\citenamefont {{Gondolo}}\ and\ \citenamefont {{Silk}}(1999)}]{Gondolo99PRL}%
  \BibitemOpen
  \bibfield  {author} {\bibinfo {author} {\bibfnamefont {P.}~\bibnamefont {{Gondolo}}}\ and\ \bibinfo {author} {\bibfnamefont {J.}~\bibnamefont {{Silk}}},\ }\href {\doibase 10.1103/PhysRevLett.83.1719} {\bibfield  {journal} {\bibinfo  {journal} {Phys. Rev. Lett.}\ }\textbf {\bibinfo {volume} {83}},\ \bibinfo {pages} {1719} (\bibinfo {year} {1999})},\ \Eprint {http://arxiv.org/abs/astro-ph/9906391} {arXiv:astro-ph/9906391 [astro-ph]} \BibitemShut {NoStop}%
\bibitem [{\citenamefont {Al-Badawi}\ \emph {et~al.}(2025)\citenamefont {Al-Badawi}, \citenamefont {Shaymatov},\ and\ \citenamefont {Sekhmani}}]{Al-Badawi_2025}%
  \BibitemOpen
  \bibfield  {author} {\bibinfo {author} {\bibfnamefont {A.}~\bibnamefont {Al-Badawi}}, \bibinfo {author} {\bibfnamefont {S.}~\bibnamefont {Shaymatov}}, \ and\ \bibinfo {author} {\bibfnamefont {Y.}~\bibnamefont {Sekhmani}},\ }\href {\doibase 10.1088/1475-7516/2025/02/014} {\bibfield  {journal} {\bibinfo  {journal} {J. Cosmol. Astropart. Phys.}\ }\textbf {\bibinfo {volume} {2025}},\ \bibinfo {pages} {014} (\bibinfo {year} {2025})}\BibitemShut {NoStop}%
\bibitem [{\citenamefont {{Dadhich}}\ and\ \citenamefont {{Shaymatov}}(2022)}]{Dadhich22a}%
  \BibitemOpen
  \bibfield  {author} {\bibinfo {author} {\bibfnamefont {N.}~\bibnamefont {{Dadhich}}}\ and\ \bibinfo {author} {\bibfnamefont {S.}~\bibnamefont {{Shaymatov}}},\ }\href {\doibase 10.1016/j.dark.2022.100986} {\bibfield  {journal} {\bibinfo  {journal} {Phys. Dark Universe}\ }\textbf {\bibinfo {volume} {35}},\ \bibinfo {eid} {100986} (\bibinfo {year} {2022})},\ \Eprint {http://arxiv.org/abs/2104.00427} {arXiv:2104.00427 [gr-qc]} \BibitemShut {NoStop}%
\bibitem [{\citenamefont {{Babak}}\ \emph {et~al.}(2007)\citenamefont {{Babak}}, \citenamefont {{Fang}}, \citenamefont {{Gair}}, \citenamefont {{Glampedakis}},\ and\ \citenamefont {{Hughes}}}]{Babak07PRD}%
  \BibitemOpen
  \bibfield  {author} {\bibinfo {author} {\bibfnamefont {S.}~\bibnamefont {{Babak}}}, \bibinfo {author} {\bibfnamefont {H.}~\bibnamefont {{Fang}}}, \bibinfo {author} {\bibfnamefont {J.~R.}\ \bibnamefont {{Gair}}}, \bibinfo {author} {\bibfnamefont {K.}~\bibnamefont {{Glampedakis}}}, \ and\ \bibinfo {author} {\bibfnamefont {S.~A.}\ \bibnamefont {{Hughes}}},\ }\href {\doibase 10.1103/PhysRevD.75.024005} {\bibfield  {journal} {\bibinfo  {journal} {Phys. Rev. D}\ }\textbf {\bibinfo {volume} {75}},\ \bibinfo {eid} {024005} (\bibinfo {year} {2007})},\ \Eprint {http://arxiv.org/abs/gr-qc/0607007} {arXiv:gr-qc/0607007 [gr-qc]} \BibitemShut {NoStop}%
\bibitem [{\citenamefont {Chandrasekhar}(1984)}]{1983mtbh.book.....C}%
  \BibitemOpen
  \bibfield  {author} {\bibinfo {author} {\bibfnamefont {S.}~\bibnamefont {Chandrasekhar}},\ }\href {\doibase 10.1007/978-94-009-6469-3_2} {\emph {\bibinfo {title} {General Relativity and Gravitation: Invited Papers and Discussion Reports of the 10th International Conference on General Relativity and Gravitation, Padua, July 3--8, 1983}}},\ edited by\ \bibinfo {editor} {\bibfnamefont {B.}~\bibnamefont {Bertotti}}, \bibinfo {editor} {\bibfnamefont {F.}~\bibnamefont {de~Felice}}, \ and\ \bibinfo {editor} {\bibfnamefont {A.}~\bibnamefont {Pascolini}}\ (\bibinfo  {publisher} {Springer Netherlands},\ \bibinfo {address} {Dordrecht},\ \bibinfo {year} {1984})\ pp.\ \bibinfo {pages} {5--26}\BibitemShut {NoStop}%
\bibitem [{\citenamefont {{Yang}}\ \emph {et~al.}(2025)\citenamefont {{Yang}}, \citenamefont {{Zhang}}, \citenamefont {{Zhu}}, \citenamefont {{Zhao}},\ and\ \citenamefont {{Liu}}}]{2025JCAP...01..091Y}%
  \BibitemOpen
  \bibfield  {author} {\bibinfo {author} {\bibfnamefont {S.}~\bibnamefont {{Yang}}}, \bibinfo {author} {\bibfnamefont {Y.-P.}\ \bibnamefont {{Zhang}}}, \bibinfo {author} {\bibfnamefont {T.}~\bibnamefont {{Zhu}}}, \bibinfo {author} {\bibfnamefont {L.}~\bibnamefont {{Zhao}}}, \ and\ \bibinfo {author} {\bibfnamefont {Y.-X.}\ \bibnamefont {{Liu}}},\ }\href {\doibase 10.1088/1475-7516/2025/01/091} {\bibfield  {journal} {\bibinfo  {journal} {Journal of Cosmology and Astroparticle Physics}\ }\textbf {\bibinfo {volume} {2025}},\ \bibinfo {eid} {091} (\bibinfo {year} {2025})},\ \Eprint {http://arxiv.org/abs/2407.00283} {arXiv:2407.00283 [gr-qc]} \BibitemShut {NoStop}%
\bibitem [{\citenamefont {Shabbir}\ \emph {et~al.}(2025)\citenamefont {Shabbir}, \citenamefont {Jamil},\ and\ \citenamefont {Azreg-Aïnou}}]{SHABBIR2025101816}%
  \BibitemOpen
  \bibfield  {author} {\bibinfo {author} {\bibfnamefont {O.}~\bibnamefont {Shabbir}}, \bibinfo {author} {\bibfnamefont {M.}~\bibnamefont {Jamil}}, \ and\ \bibinfo {author} {\bibfnamefont {M.}~\bibnamefont {Azreg-Aïnou}},\ }\href {\doibase https://doi.org/10.1016/j.dark.2025.101816} {\bibfield  {journal} {\bibinfo  {journal} {Physics of the Dark Universe}\ }\textbf {\bibinfo {volume} {47}},\ \bibinfo {pages} {101816} (\bibinfo {year} {2025})}\BibitemShut {NoStop}%
\bibitem [{\citenamefont {Jiang}\ \emph {et~al.}(2024)\citenamefont {Jiang}, \citenamefont {Alloqulov}, \citenamefont {Wu}, \citenamefont {Shaymatov},\ and\ \citenamefont {Zhu}}]{JIANG2024101627}%
  \BibitemOpen
  \bibfield  {author} {\bibinfo {author} {\bibfnamefont {H.}~\bibnamefont {Jiang}}, \bibinfo {author} {\bibfnamefont {M.}~\bibnamefont {Alloqulov}}, \bibinfo {author} {\bibfnamefont {Q.}~\bibnamefont {Wu}}, \bibinfo {author} {\bibfnamefont {S.}~\bibnamefont {Shaymatov}}, \ and\ \bibinfo {author} {\bibfnamefont {T.}~\bibnamefont {Zhu}},\ }\href {\doibase https://doi.org/10.1016/j.dark.2024.101627} {\bibfield  {journal} {\bibinfo  {journal} {Phys. Dark Universe}\ }\textbf {\bibinfo {volume} {46}},\ \bibinfo {pages} {101627} (\bibinfo {year} {2024})}\BibitemShut {NoStop}%
\bibitem [{\citenamefont {{Zhao}}\ \emph {et~al.}(2025)\citenamefont {{Zhao}}, \citenamefont {{Tang}},\ and\ \citenamefont {{Xu}}}]{2025EPJC...85...36Z}%
  \BibitemOpen
  \bibfield  {author} {\bibinfo {author} {\bibfnamefont {L.}~\bibnamefont {{Zhao}}}, \bibinfo {author} {\bibfnamefont {M.}~\bibnamefont {{Tang}}}, \ and\ \bibinfo {author} {\bibfnamefont {Z.}~\bibnamefont {{Xu}}},\ }\href {\doibase 10.1140/epjc/s10052-025-13767-0} {\bibfield  {journal} {\bibinfo  {journal} {European Physical Journal C}\ }\textbf {\bibinfo {volume} {85}},\ \bibinfo {eid} {36} (\bibinfo {year} {2025})},\ \Eprint {http://arxiv.org/abs/2411.01979} {arXiv:2411.01979 [gr-qc]} \BibitemShut {NoStop}%
\bibitem [{\citenamefont {Poisson}\ and\ \citenamefont {Will}(2014)}]{Poisson_Will_2014}%
  \BibitemOpen
  \bibfield  {author} {\bibinfo {author} {\bibfnamefont {E.}~\bibnamefont {Poisson}}\ and\ \bibinfo {author} {\bibfnamefont {C.~M.}\ \bibnamefont {Will}},\ }\href@noop {} {\emph {\bibinfo {title} {Gravity: Newtonian, Post-Newtonian, Relativistic}}}\ (\bibinfo  {publisher} {Cambridge University Press},\ \bibinfo {year} {2014})\BibitemShut {NoStop}%
\bibitem [{\citenamefont {{Meng}}\ \emph {et~al.}(2024)\citenamefont {{Meng}}, \citenamefont {{Xu}},\ and\ \citenamefont {{Tang}}}]{2024arXiv241101858M}%
  \BibitemOpen
  \bibfield  {author} {\bibinfo {author} {\bibfnamefont {L.}~\bibnamefont {{Meng}}}, \bibinfo {author} {\bibfnamefont {Z.}~\bibnamefont {{Xu}}}, \ and\ \bibinfo {author} {\bibfnamefont {M.}~\bibnamefont {{Tang}}},\ }\href {\doibase 10.48550/arXiv.2411.01858} {\bibfield  {journal} {\bibinfo  {journal} {arXiv e-prints}\ ,\ \bibinfo {eid} {arXiv:2411.01858}} (\bibinfo {year} {2024})},\ \Eprint {http://arxiv.org/abs/2411.01858} {arXiv:2411.01858 [gr-qc]} \BibitemShut {NoStop}%
\end{thebibliography}%

\end{document}